\pdfoutput=1
\documentclass[twocolumn,showpacs,amssymb,floatfix,deluxe,prd]{revtex4}

\usepackage{graphicx}
\usepackage{dcolumn}
\usepackage{bm}
\usepackage{epsfig}
\usepackage[english]{babel}
\usepackage{amsmath}
\usepackage{amssymb}

\usepackage{verbatim}
\usepackage{alltt}

\def\D{\mathord{\rm D}}

\def\({\left(}
\def\){\right)}

\def\half{\frac{1}{2}}

\def\sp{\mspace{1mu}}

\begin{document}
CERN-PH-TH/2012-039
\preprint{CERN-PH-TH/2012-039}
\title{
Singular ways to search for the Higgs boson} 

\author{A. De R\'ujula${}^{a,b,c}$}
\affiliation{  \vspace{3mm}
${}^a$Instituto de F\'isica Te\'orica (UAM/CSIC), Univ. Aut\'onoma de Madrid, Madrid, and 
CIEMAT, Madrid, Spain,\\
${}^b$Physics Dept., Boston University, Boston, MA 02215,\\
${}^c$Physics Department, CERN, CH 1211 Geneva 23, Switzerland}

\author{A. Galindo${}^{d,e}$}
\affiliation{${}^d$Departamento de F\'isica,
Universidad Complutense, Madrid, Spain, ${}^e$CIEMAT, Madrid, Spain}%

\date{\today}

\begin{abstract}

The discovery or exclusion of the fundamental standard scalar is a hot
topic, given the data of LEP, the Tevatron and the LHC, as well as the advanced 
status of the pertinent theoretical calculations. With the current statistics at the hadron colliders,
the workhorse decay channel, at all relevant $H$ masses,  is $H\to WW,$
followed by $W\to \ell\nu$,  $\ell = e$ or $\mu$.  Using phase-space singularity
techniques, we construct and study a plethora of ``singularity variables" meant
to facilitate the difficult tasks of separating signal and backgrounds and of measuring 
the mass of a putative signal.
The simplest singularity variables are not invariant under boosts along the $pp$ or 
$p\bar p$ axes and the simulation of their distributions requires a good understanding 
of parton distribution functions, perhaps not a serious shortcoming 
during the boson hunting season. The derivation of longitudinally boost-invariant
variables, which are functions of the four charged-lepton observables that share
this invariance, is quite elaborate. But their use is simple and they are,
in a kinematical sense, optimal.
\end{abstract}

\maketitle

\begin{flushright}
{\it It is nice to know that the computer\\
understands the problem. But I\\
  would like to understand it too.}\\
 \bf{Eugene Wigner}\\
\end{flushright}

\section{Introduction}

Recent data from the LHC \cite{LHC} on a putative standard Higgs boson
exclude, at a 95\% confidence level (CL), the mass domain  127 GeV $\!<M_H\!<$ 600 GeV (CMS) 
and, with some narrow gaps, 131 GeV $<\!M_H\!<$ 453 GeV (ATLAS).
These results are obtained with full use of the standard theory, including 
radiative corrections which sometimes constitute the dominant effect. The
amplitude for Higgs boson production, for example, is largely dominated
by gluon fusion via a $t$-quark loop and so is the amplitude for $H\to\gamma\gamma$
decay.

In the ``quantum-level" setting we recounted, it would be inconsistent
not to analize the LHC data in conjunction with the constraints on $M_H$ which
follow from the profusion of high precision 
measurements that test the standard theory beyond tree level. These constraints (and the
direct searches \cite{CDFD0} at the Tevatron) result in
$M_H\!<\!161\,(156)$ GeV at a CL of 95\% \cite{LEPEW}, while the direct LEP limit is $M_H\!>\!115$ GeV.

In mass intervals akin to the one implied by the quoted constraints
 CMS finds a 1.9$\sigma$ excess of events --that could be an
indication of a Higgs signal-- at $M_H=124$ GeV 
and ATLAS a 2.5$\sigma$ one at $M_H=126$ GeV \cite{LHC}. In the current
broad mass range(s) of the searches, the corresponding ``local" significances 
are somewhat larger \cite{LHC}, but have no rigorous statistical interpretation.

For a standard Higgs boson of mass $M_H> 140$ GeV the branching ratio for the decay
$H\to WW$ is the dominant one.
Below this mass and above the LEP limit, the winner is $H\to b\bar b$, 
a process beset by terrifying backgrounds at a hadron collider. 
The branching ratio $W\to q\bar q$ 
is one order of magnitude larger than the one for $W\to \ell\nu$, $\ell=e,\mu$.
But a light-lepton signal is much ``cleaner" than that of a quark-generated jet.
This makes the chain $H\to W^+W^-$, $W^\pm\to \ell^\pm\nu$  the all-mass
workhorse at a hadron collider. In brief, we refer to this process as $H\to WW$, including
the ``off-shell" $M_H<2\,M_W$ case, often dubbed $H\to WW^*$.

The obvious problem with the $H\to WW$ channel is that $M_H$ cannot be reconstructed
event by event, as
a lot of information escapes detection
with the unobserved neutrinos and, at a hadron collider, also with the unobserved hadrons
that exit ``longitudinally" close to the beam pipe(s). This makes taming the workhorse almost an art,
not only a science. The formal and theoretically optimal {\it singularity variable}
procedure to deal with this kind of incomplete
information is summarized in \cite{Kim,AA} and  exploited
for the hadron-collider production of a single $W$ in \cite{AA}.
We shall see that, for the $H\to WW$ process, the situation is much more challenging,
mainly because two missing neutrinos are many more than one.

The other crucial obstacles in the process we study are the large backgrounds 
with kinematics akin to those of the signal. The main and irreducible one is
the direct non-resonant production of $W$ pairs by $q\bar q$ annihilation.
The next most relevant one is $t\bar t$ production, which also results in $W$ pairs.
For simplicity we shall illustrate our theoretical results only for the chain $W\to e\nu$,
$W\to \mu\nu$, for which the ``Drell-Yan" background is not a problem.

The analysis tools used to deal with the $H\to WW$ channel range from a simple ``cut and count"
approach to ``matrix-element" techniques and avant-garde neural networks or
``boosted decision trees" \cite{LHC}. There is no question that in the long range the
methods that input and utilize the largest amount of information are likely to be the 
most powerful ones. Whether this is also the case at an exploratory ``Higgs-hunting" stage
is more doubtful. 
Here we shall explore a ``copy and paste" avenue of intermediate sophistication: 
the derivation of singularity variables --functions of the observable momenta and of $M_H$--
whose measured histograms are to be compared (in one or more dimensions) with pre-prepared
templates. 

For the production of a Higgs boson at a hadron collider,
followed by the decays $H\to W^+W^-$, $W^\pm\to \ell^\pm\nu$, 
we shall limit our discussion to the distributions
of various functions of the charged lepton three-momenta
$\vec k$ and $\vec l$. The treatment of  the 2D transverse momentum of the 
final state hadrons, $\vec p_T$, deserves a separate paragraph, the next. 
The use of other observables,
such as the number of jets, is beyond our scope.

Two practical problems are that $\vec k$ and $\vec l$
are measured to much higher precision than $\vec p_T$
and that the formulae for the $H\to WW$ singularity variables are much
more complex for $\vec p_T\neq 0$ than for $\vec p_T=0$. 
We deal with both problems by setting $\vec p_T=0$ in our theoretical expressions.
This is less cavalier than it seems. The transverse momentum of a Higgs boson
in a given event is $\simeq\! -\sp \vec p_T$.
For a given ansatz $M_H$ value, its observed lepton momenta can be Lorentz boosted
closer to the $\vec p_T=0$ frame. The precise boost would require
knowledge of the boson's longitudinal momentum, $p_3$. But, typically, 
$p_3^2\ll 2\sp M_H^2$, it is a fair approximation to neglect $p_3$.
More importantly,  the singularity variables for $\vec p_T=0$ are
very useful even in the analysis of events 
with $\vec p_T\neq 0$, even if one does
not boost the events back closer to the $\vec p_T=0$ frame, and even if one
is also dealing with the quoted backgrounds, whose $W$ pairs do not have
a fixed invariant mass.

Let the lepton momenta be
$\vec k\equiv\{\vec k_T,k_3\}$ and $\vec l\equiv\{\vec l_T,l_3\}$. 
Because of the rotational symmetry along the beams' axis, the six-dimensional
observable space $\{\vec k,\vec l\}$ is in practice just five-dimensional.
One possible choice of variables is the set $(k+l)^2$, the
invariant mass of the 
lepton pair; $k_T$, $l_T$, the moduli of the transverse
momenta; and $\vec k_T\cdot \vec l_T$, or the familiar 
$\Delta\varphi=\arccos [\vec k_T\cdot \vec l_T/(k_T\,l_T)]$.
All four of these ``transverse" observables are invariant under longitudinal boosts along
the beams' axis. The remaining variable, for instance $k_3+l_3$, is not.

We shall derive two types of singularity variables: those which do --or do not--
depend only on transverse observables.  Transverse 
variables are preferable,
in that they are insensitive to the significant uncertainties associated
with the (longitudinal) parton distribution functions (pdfs). In practice the uncertainties are 
to a modest extent reintroduced via the angular coverage limitations
of an actual experiment, which are not invariant under longitudinal boosts.
The histograms of singularity variables that are not longitudinally invariant
do depend on the pdfs, but, particularly during a Higgs-hunting or initiatory epoch, this
is not a serious limitation.

\begin{figure}[htbp]
\begin{center}
\includegraphics[width=0.23\textwidth]{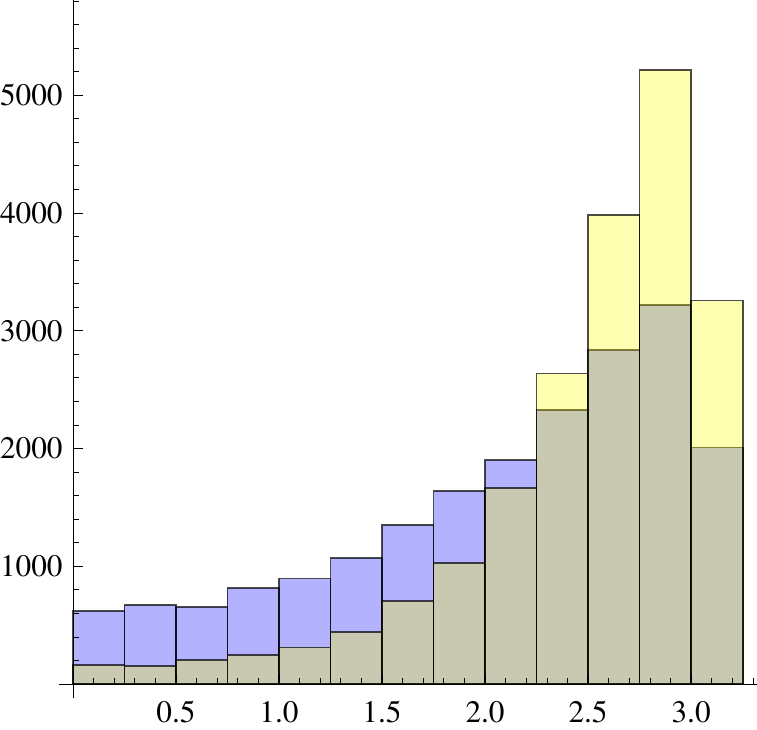}
\includegraphics[width=0.23\textwidth]{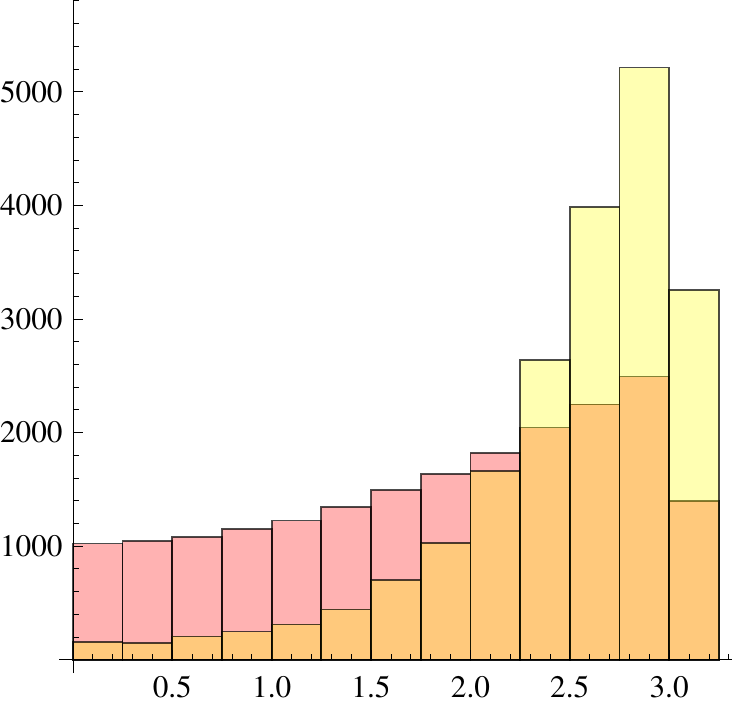}\\
\includegraphics[width=0.23\textwidth]{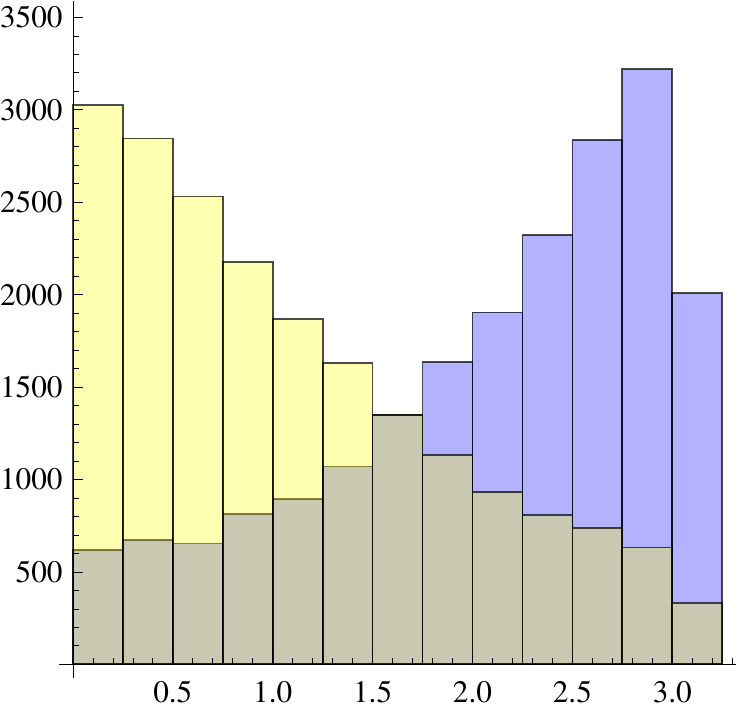}
\includegraphics[width=0.23\textwidth]{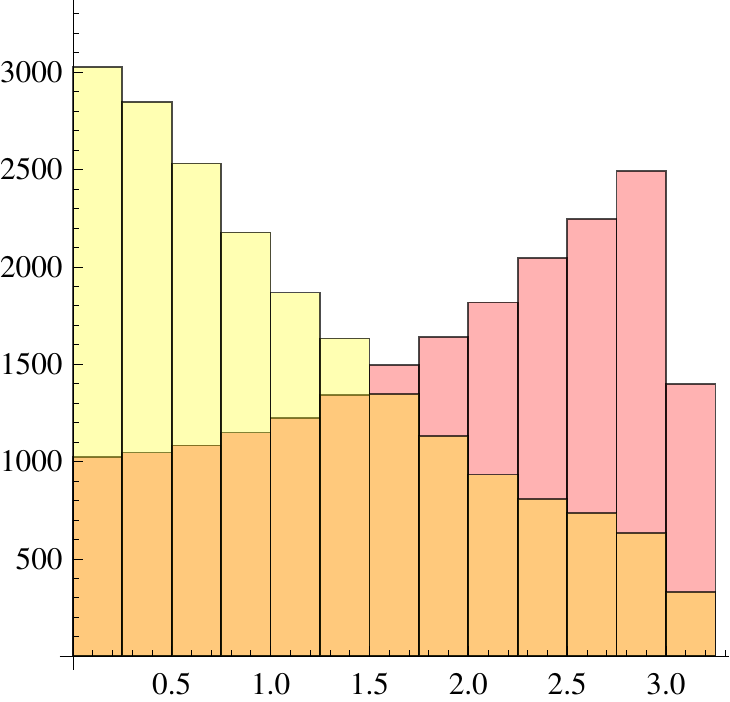}
\caption{Distributions of $\Delta\varphi$. Top row: $M_H=500$ GeV. Bottom row: $M_H=120$ GeV.
Left column: Comparison between the shape of the (yellow) signal distribution and that
of the (blue) $WW$
background. Right column: Comparison of signal with the (orange) $t\bar t$ background.
\label{fig:DeltaPhi}}
\end{center}
\end{figure}

In the problem at hand, the quintessential function of transverse variables
--in the sense of its ability to tell apart signal from backgrounds-- is $\Delta\varphi$,
the angle between the charged leptons in the transverse plane.
In Fig.~\ref{fig:DeltaPhi}, for comparison with our coming results, we recall this fact by showing the
(arbitrarily normalized) shapes of signal and background distributions
for two examples, with $M_H=500$ and 120 GeV. As is well known, the V-A nature of the weak
current and the specific spin zero nature of the $W$ pair in signal events, favours
collinear vs. anticollinear leptons. The effect weakens as $M_H$ increases and the 
leptons are boosted away from each other.

The simulations of Fig.~\ref{fig:DeltaPhi}, as well as all others in this note, were made
with use of the PYTHIA6 event generator \cite{SMS}. 
They are for the $H\to WW$, $W\to e\nu$, $W\to \mu\nu$ channel,
with leptons of transverse momentum  greater than 15 GeV, 
and satisfying the pseudorapidity cuts $\eta(e)<2.5$, $\eta(\mu)<2.1$.

Our goal is two-fold. First and foremost, to derive the complete set of phase-space 
singularity variables (functions analogous to $\Delta\varphi$) for the process at hand. 
Second, to illustrate with examples their potential phenomenological usage. 
At least at low $M_H$ values, $\Delta\varphi$ is more heavily dependent on dynamics
than on kinematics. The  singularity variables we shall derive
are the other way around. Individually, several of them are nearly  ``as good" as $\Delta\varphi$ 
in disentangling a signal
from the backgrounds. The ensemble of their distributions, particularly in conjunction 
with $\Delta\varphi$ itself, should be a powerful and relatively simple tool to search
for a Higgs boson, which, in the sense of signal kinematics, is guaranteed to be
optimal.

Since we investigate a plethora of singularity variables, this paper is long and detailed.
The reader mainly interested in results may be well advised to start reading it from the
end: \S \ref{Sec:Summary} and \S \ref{sec:conclusions}.

\section{Outline}

The simple example of single-$W$ production is used in \S \ref{1W} to clarify what
singularity conditions and singularity variables are. After posing the formal problem
in \S\ref{Formal} we proceed in \S\ref{sec:gluoncollider} to solve it
 in the center of mass (CM) reference system of the Higgs boson. There are two reasons
 for this. First, it is a necessary intermediate step in the theoretical derivation, in 
 \S \ref{sec:Back},
 of the general case with a boson which is not at rest. Second, the ``approximation"
 of a heavy particle made at rest by gluon-gluon or $q\bar q$ fusion
 in a hadron collider is not so bad, since the quark and
 gluon pdfs are fast falling functions of their fractional momenta.
 Our results will reflect this fact.
 
 We shall have to deal with the case $M_H<2\sp M_W$, in which at least one of the
 $W$s is off-shell; the relative probability for both of them having an invariant
 mass significantly different from $M_W$ is small. In \S\ref{SmallMH} we explain
 the simple way in which we treat this case.
 Further details of our data analysis not already discussed in the introduction
 are given in \S \ref{sec:details}.
 
We analize MC-generated data in \S \ref{CMdata} in the theoretical approximation
of a Higgs boson made at rest. To some extent, this section is a ``warm-up" for
the general results wherein we lift this approximation, discussed in
\S \ref{BoostedData}, to which the reader interested in the most powerful
results may prefer to jump.

 A summary of results is given in \S \ref{Sec:Summary}.
 Our data analysis is not as thorough as the theoretical one, it is only meant
 to illustrate our points. But it suffices to reach our conclusions, which,
 naturally,  are drawn in the last section. A very formal but important step
 in our theoretical analysis is relegated to the Appendix.

\section{Simple singularity variables}
\label{1W}

Our main result is the theoretical derivation of the phase space singularity variables 
and singularity conditions
for the process  $H\to W^+W^-$, $W^\pm\to \ell^\pm\nu$. To understand these concepts
 it is easiest to recall a simpler problem: the analogous one for single-$W$ production
at a hadron collider, followed by the same leptonic decay. In this case, the singularity condition
\cite{AA} is
 $\Sigma_T=0$, with:
\begin{eqnarray}
\!\!\!\!\!\!\!\!\!\!&&\Sigma_T(M,\vec l_{_T},\vec p_{_T})\equiv \nonumber\\
\!\!\!\!\!\!\!\!\!\!&&M^4-4\,M^2\,(\vec l_{_{T}}\cdot \vec p_{_{T}}+l_{_{T}}^2)
+4\,\left[(\vec l_{_{T}} \cdot \vec p_{_{_T}})^2-l_{_{T}}^2\,p_{_{T}}^2\right]
\label{calMeq2}
\end{eqnarray}
Of the four $M$-roots of
$\Sigma_T=0$, one is not unphysical:
\begin{equation}
{M_T}(\vec l_{_T},\vec p_{_T})
\!=\!+\sqrt{2\,\left[ |l_{_T}|\,|p+l|_{_T}+\vec l_{_T}\cdot (\vec l_{_T}+\vec p_{_T})\right]},
\label{mT2bis}
\end{equation}
which reduces to ${M_T}\! =\! 2\, |l_{_T}|$ for $\vec p_{_T}\!=\!0$.
The function in Eq.~(\ref{mT2bis}) is the habitual $M_{_T}^2$ originally
derived in \cite{mT2A,mT2B}.

The result of Eq.~(\ref{calMeq2}) is obtained by projecting the full phase space (which includes the
neutrino momentum) onto the observable phase space. The function $\Sigma_T$ is a
singularity variable which
--for a general non-singular event-- is a measure of its distance to the nearest singularity
at the singular $\Sigma_T=0$ border of the projected space. In Eq.~(\ref{calMeq2})
the mass of the $W$ appears in two ways: the physical $M_H$ is imprinted in the data
and also appears as an implicit ``trial" mass $M\to\cal M$ in the equation. In the $M_T$
singularity variable of Eq.~(\ref{mT2bis}) $M_H$ is only reflected in the observables.
In applying the phase-space singularity approach to our two-$W$ problem, we shall
encounter both types of singularity variables.

In the single-$W$ case one can refine
the result in the sense of finding the optimal singularity variable, that which would
result in the most precise measurement of $M_W$ \cite{AA}. In the two-$W$ case this is not
worthwhile, as there are  decay channels, such as 
$H\to\gamma\gamma$, $H\to ZZ$; $Z\to \ell^+ \ell^-$, for which the mass is reconstructible.

\section{The formal problem}
\label{Formal}
Back to the $H\to W^+W^-$, $W^\pm\to \ell^\pm\nu$ process, let
$y$ and $x$, respectively, be the four-momenta of the neutrinos
accompanying the charged leptons of four-momentum $k$ and $l$.
The full information relevant to the reconstruction of the boson's
mass for a signal event is embedded in the kinematical equations:

\begin{eqnarray}
&&E_1\Rrightarrow x^2=0\nonumber\\
&&E_2\Rrightarrow y^2=0\nonumber\\
&&E_3\Rrightarrow 2\,l\cdot x =M_W^2\nonumber\\
&&E_4\Rrightarrow 2\,k\cdot y =M_W^2\nonumber\\
&&E_5\Rrightarrow 2\, (l+x)\cdot (k+y)=M_H^2-2\,M_W^2\nonumber\\
&&E_6\Rrightarrow k_1+y_1+l_1+x_1+p_1=0\nonumber\\
&&E_7\Rrightarrow k_2+y_2+ l_2+x_2+p_2=0
\label{alleqs}
\end{eqnarray}
where we have made the approximation $l^2=k^2=0$ for the charged leptons
and --fleetingly in error for the $WW^*$ case-- set the masses of the two $W$s to
their central values.
There are 9 unknowns (2 neutrino four-momenta and $M_H$)
and only 7 equations. In spite of this, is there a systematic way 
to extract the kinematically most stringent information on $M_H$?
This is the problem to face.

Consider the 14D space of the components $\vec l,\vec k$ of the
three-momenta of the two (approximately massless)  charged leptons and the 
four-momenta $x,y$ of the two neutrinos. For a fixed $M_H$, the seven Equations (\ref{alleqs})
define a $14-7=7$D manifold, the {\it phase space}. 
This surface is to be projected onto the 6D hyper-plane of observable three-momenta.
The points in the full phase space that project onto the boundary of the 6D
space of observables are {\it singular:} at such points
one or more of the invisible directions are contained 
in the tangent plane to the full phase space \cite{Kim,AA} , and a tangent to a surface 
is singular in that it ``touches
it" at more than one single point. 

The equation, $\Sigma(\vec l,\vec k,M_W^2;M_H^2)=0$, describing the
boundary of the projected phase space is a {\it singularity condition}. 
A general event (i.e.~specific values of $\vec l$ and $\vec k$) is non-singular
and its corresponding value of $\Sigma$ is, once again,
 a measure of distance to the $\Sigma=0$ singularity.
The shape of the distribution of the values of the {\it singularity variable}
$\Sigma$  is sensitive to the unknown
mass $M_H$ in a manner that allows one to extract its true value,  be it physical
or Monte Carlo (MC) generated.

The formal modus operandi  to obtain singularity variables
is summarized in \cite{Kim} and discussed in detail in \cite{AA}. We recalled that
at a singularity one or more of the invisible directions are contained 
in the tangent plane to the full phase space.
The general condition for this to happen
is that, in the space $\{z\}=\{x,y\}$ of invisible directions, 
the row vectors of the Jacobian matrix
$J_{i j}\equiv\partial E_i/\partial z_j$ (with the row index $i$ running along the number of
equations and the column index
$j$ over the number of invisible coordinates) be linearly dependent. In other
words, at a singularity, the rank of $J_{i j}$ must be smaller than its rank at 
nonsingular points.

There are 7 equations and 8 invisible
directions in Eqs.~(\ref{alleqs}). The vanishing of the Jacobian $J_{ij}$ (a $7\times 8$ matrix) 
entails 8 conditions: the nullification of all $7\times 7$ minors. Two of these minors coincide,
up to their sign, with two others. Moreover the sums of two pairs of minors are of the forms
$D\,{\cal S}_0$, $D\,{\cal S}_3$, with
\begin{equation}
\begin{split}
&D\equiv \det(l,x,k,y)
\\
&{\cal S}_0\equiv k_0+l_0+x_0+y_0,
\\
&{\cal S}_3\equiv k_3+l_3+x_3+y_3.
\end{split}
\label{eq:Sigmas}
\end{equation}
Given that ${\cal S}_0>0$, one condition is:
\begin{equation}
\det(l,x,k,y)=0,
\label{eq:det0}
\end{equation}
that is, the coplanarity of the four lepton four-momenta, equivalent to
\begin{equation}
y=\alpha\, l+\beta\, x+(\gamma-1)\, k.
\end{equation}
Introducing this into the 8 original minors, it is easy to see that they all vanish provided
that
\begin{equation}
\gamma={{(\alpha-\beta)\,(l_3\,x_0-l_0\,x_3)}\over\alpha\,(k_3\,l_0-k_0\,l_3)+\beta\,(k_3\,x_0-k_0\,x_3)}
\end{equation}

The transposed Jacobian matrix, with
use of E6 and E7 of Eqs.~(\ref{alleqs}), is
\begin{equation}
J({\cal S}_0,{\cal S}_3)=
\left(
\begin{array}{ccccccc}
 x_0 & 0 & l_0 & 0
   &
   {\cal S}_0 & 0 & 0 \\
 - x_1 & 0 & -l_1 &
   0 & 0 & 1 & 0 \\
 - x_2 & 0 & -l_2 &
   0 & 0 & 0 & 1 \\
 - x_3 & 0 & -l_3 &
   0 &
   -{\cal S}_3 & 0 & 0 \\
 0 & y_0 & 0 & k_0
   &
   {\cal S}_0 & 0 & 0 \\
 0 & -y_1 & 0 &
   -k_1 & 0 & 1 & 0 \\
 0 & -y_2 & 0 &
   -k_2 & 0 & 0 & 1 \\
 0 & -y_3 & 0 &
   -k_3 &
   -{\cal S}_3 & 0 & 0
\end{array}
\right),
\label{eq:Jacobian}
\end{equation}
where the functional dependence of $J$ on
 ${\cal S}_{0,3}$ has been made explicit for later convenience.

\begin{figure}[htbp]
\begin{center}
\includegraphics[width=0.45\textwidth]{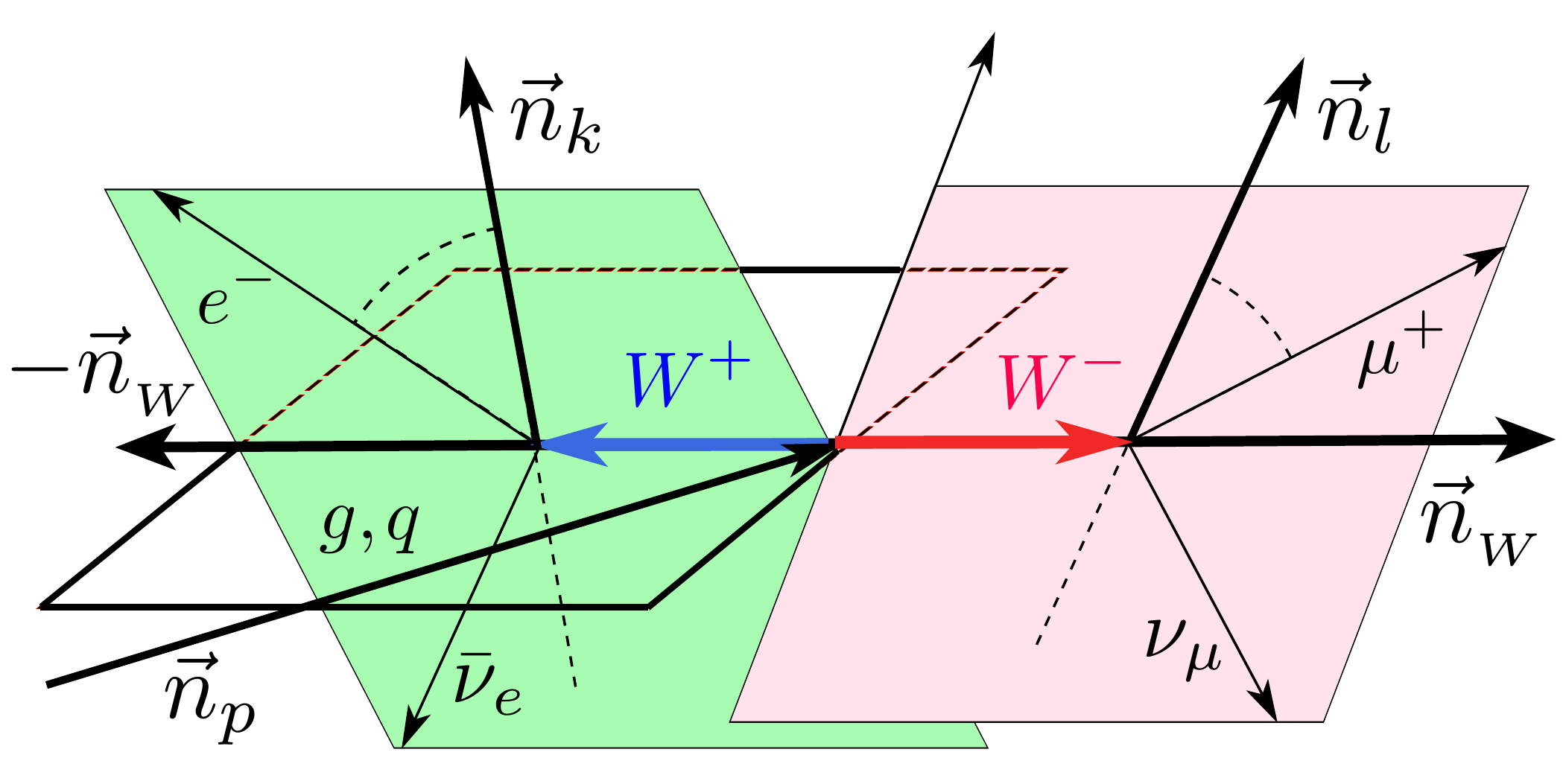}\\
\caption{Top:
The $H\to WW$, $W\to \ell\nu$ process in a Cabibbo-Maximovich parametrization
\cite{CM}. The vectors $\vec{n}_l$ and $\vec{n}_k$ are the directions of the charged
leptons (a $\mu^+$ and an $e^-$ in this illustration) in the respective rest systems 
of their parent $W$s. The overall $WW$ system, shown here at rest, is to be boosted
along the direction $\vec{n}_p$ of the gluon or $q\bar q$ pair that fuse to produce the
$W$ pair, resonantly (for the $H$ signal) or not (for the irreducible background).
\label{fig:Angles}}
\end{center}
\end{figure}

It turns out to be very useful to study the behaviour of the $7\times 7$ minors of $J$
under longitudinal Lorentz transformations, the boosts along the axis ``3" of the proton beams. 
To proceed recall that, for the reasons stated in the Introduction,
we are setting the hadron momenta $p_1=p_2=0$. Next,
parametrize an event in the usual
Cabibbo-Maximovich manner \cite{CM}, illustrated in our notation in Fig.~\ref{fig:Angles}. 
That is, consider the lepton momenta as
if both $W$ bosons were at rest, boost them by the antiparallel motion of
the $W$s in the $H$ rest system and finally boost the Higgs boson longitudinally
left or right along the beams' axis:
\begin{equation}
\begin{split}
& l = L(y_H,\vec{n}_p)\,L(y,\vec{n})\,( M_W/2)\{1,\vec{n}_l\},
\\
& x = L(y_H,\vec{n}_p)\,L(y,\vec{n})\,( M_W/2)\{1,-\vec{n}_l\},
\\
& k = L(y_H,\vec{n}_p)\,L(y,-\vec{n})\,( M_W/2)\{1,\vec{n}_k\},
\\
& y = L(y_H,\vec{n}_p)\,L(y,-\vec{n})\,( M_W/2)\{1,-\vec{n}_k\},
\end{split}
\label{eq:Lorentz}
\end{equation}
where $\vec{n}_k,\vec{n}_l,\vec{n}$ and $\vec{n}_p$ 
are unit vectors, $L(y,\vec{n})$ is
a Lorentz boost along $\vec{n}$ with velocity $\beta=\tanh(y)$,
$y={\rm arccosh}(\gamma)$, $\gamma=M_H/(2 \, M_W)$,  and analogously
for the longitudinal boost along $n_p$, of rapidity $y_H$. 

Label $m_j$, $j=1$ to 8, the $7\times 7$ minors of $J$ in Eq.~(\ref{eq:Jacobian}),
lacking the row $9-j$ of $J$.
Under a longitudinal 
boost $L(y_H,\vec{n}_3)$, they transform as $m_j\mapsto \bar m_j$, with:
\begin{equation}
\begin{split}
&\bar m_1=\gamma_H\,[m_1+\beta_H\, m_8+\beta_H\, {\cal S}_3\,D],
\\
&\bar m_i=m_i,\;\;\;i=2,\,3,\,6,\,7,
\\
& \bar m_4=\gamma_H\,[m_4+\beta_H\, m_5-\beta_H\, {\cal S}_0\,D],
\\
& \bar m_5=\gamma_H\,[m_5+\beta_H\, m_4+\beta_H\, {\cal S}_3\,D],
\\
&\bar m_8=\gamma_H\,[m_8+\beta_H\, m_1-\beta_H\, {\cal S}_0\,D],
\end{split}
\label{eq:LTmenoresbis}
\end{equation}
where we used the definitions in Eqs.~(\ref{eq:Sigmas}).

The conditions $m_j=0,\; \forall\, j$ imply that $D=0$, and consequently that
$\bar m_j=0,\; \forall\, j$. Thus, we reach a crucial point: the general singular configurations
can be obtained by boosts of the ones in the boson's rest system. 
This is one of the reasons why we pause to study  this latter simpler case.

\section{Lessons from a gluon collider}
\label{sec:gluoncollider}

A standard Higgs can be made in various ways, with top-mediated gluon fusion
being the dominant mechanism up to very high $M_H$ values. The gluonic pdfs,
as well as those of the other partons, are fast-falling functions of their fractional
momentum. This implies that Higgs bosons are made with a narrow distribution of rapidities, 
centered at $y_H=0$. The same is true for the backgrounds to the $H\to WW$ channel, 
e.g.~non-resonant pairs of relatively heavy objects, such as $W$-bosons, are also 
made with a moderate collective motion. Thus, the approximation
of a monochromatic ``gluon collider" (or $q\bar q$ collider) is a good starting point for 
our analysis.

\subsection{Derivation of the singularity conditions}
\label{seq:derivation}

In the $WW$ center of mass (CM) system an extra working condition 
is to be added to Eqs.~(\ref{alleqs}):
\begin{equation}
C_{\rm CM}\Rrightarrow {\cal S}_3\equiv l_3+x_3+k_3+y_3 = 0.
\label{eq:E8}
\end{equation}
and the Jacobian is now $J({\cal S}_0,0)$, with $J$ as in Eq.~(\ref{eq:Jacobian}).
Since ${\cal S}_0>0$ and the fifth column of $J({\cal S}_0,0)$
is proportional to ${\cal S}_0$, it suffices to
consider the vanishing of the eight $7\times 7$ minors of
$J(1,0)$,
of which only four, e.g.~$m_i$, $i=1,...4$, are independent modulo $D$.

In the CM system, let $E=M_H/2$ denote the $W$'s energy and $P$
the corresponding momentum modulus. The four-momenta of the individual $W$s
in the notation of Eq.~(\ref{eq:Lorentz}) are:
\begin{eqnarray}
&p_{_{Wl}}=\{E,+P\,{\vec n}\}\nonumber\\
&p_{_{Wk}}=\{E,-P\,{\vec n}\}
\label{eq:pWs}
\end{eqnarray}
and it is convenient to put the neutrino's momenta
in the form $x=p_{_{Wl}}-l$, $y=p_{_{Wk}}-k$.

The conditions $x^2=y^2=0$ now read
\begin{equation}
\begin{split}
& 2 \sp E\sp l_0 = M_W^2 + 2 P \,\vec{l}\cdot\vec{n}
\\
& 2 \sp E \sp k_0 = M_W^2 - 2 P \,\vec{k}\cdot\vec{n}.
\end{split}
\end{equation}

Stepping back to Eq.~(\ref{eq:Jacobian}) and
introducing the explicit lepton four-momenta in the minors of $J(1,0)$
and in $\det(l,x,k,y)$, the vanishing of the results requires, in particular, that 
$\det(\vec{l},\vec{k},\vec{n})=0$, that is, the 3D coplanarity
of  $\vec{l},\vec{k}$ and $\vec{n}$ and, consequently, of all four lepton three-momenta. 
We may write
\begin{equation}
\vec{n}=(a/l_0) \vec{l} + (b /k_0) \vec{k}.
\end{equation}

Gathering results and imposing $\vec{n}\cdot\vec{n}=1$, one may express $a,b,x$ and $y$
as functions of $l$ and $k$. Two families of CM critical configurations are obtained. They
differ by the sign of $\delta$ in
\begin{equation}
\delta^2=(4 k_0^2-2 M_H k_0+M_W^2)(4 l_0^2-2 M_H l_0+M_W^2)
\end{equation}
and satisfy:
\begin{eqnarray}
&& \vec{k}\cdot\vec{l}={-4 E^2 k_0 l_0 + 2 E (k_0 + l_0) M_W^2 - M_W^4 + M_W^2 \delta
\over 4\sp P^2},\nonumber
\\
&& a=\dfrac{l_0 \big[-2 E k_0 (\vec{k}\cdot\vec{l} + k_0 l_0) + (k_0^2 + 
\vec{k}\cdot\vec{l}) M_W^2\big]}{2 \big[(\vec{k}\cdot\vec{l})^2 - k_0^2 l_0^2\big]\sp P},\nonumber
\\
&& b=\dfrac{k_0 \big[2 E l_0 (\vec{k}\cdot\vec{l} + k_0 l_0) 
- (l_0^2 + \vec{k}\cdot\vec{l}) M_W^2\big]}{2 \big[(\vec{k}\cdot\vec{l})^2 - k_0^2 l_0^2\big]\sp P},
\label{eq:cc}
\end{eqnarray}

Substituting these expressions into $m_i$, $i=1,...4$
 one finds that $m_4$ vanishes automatically. The others independent minors acquire the form
\begin{equation}
\begin{split}
& m_{1}=(k_2 l_1 - k_1 l_2) \, {\cal N}/{\cal D},
\\
& m_{2}= (k_3 l_1 - k_1 l_3)\,  {\cal N}/{\cal D},
\\
& m_{3}=(k_3 l_2 - k_2 l_3) \, {\cal N}/{\cal D},
\end{split}
\label{eq:3minors}
\end{equation}
where $\cal N$ and $\cal D$ are lengthy functions of $k$ and $l$.

There are two alternative ways to satisfy $m_{i}=0$  $\forall i$. One of them is to
let all three parenthesis in Eqs.~(\ref{eq:3minors}) vanish simultaneously,
tantamount to imposing $\vec{k}\propto\vec{l}$, a specific case of the 
condition to be obtained anon from the second alternative: ${\cal N}=0$.
Eliminating the sign ambiguity of $\delta$ yields a first requirement for
an event to be singular, $C=0$, with
\begin{equation}
\begin{split}
&C =  \kappa\,C_1^2\,C_2^3\,C_3,\\
&\kappa\equiv -\sp 2\sp M_H^{10} M_W^4 \left(M_H^2-4 M_W^2\right)^3,
\\
&C_1 \equiv
k_0-l_0,
\\
   & 
C_2 \equiv   2 k_0 l_0 M_H-M_W^2 (k_0+l_0),
   \\
   &
C_3\equiv  4 M_W^2 (k_0 l_3+k_3 l_0)
   \big[2 M_H (k_0 k_3+l_0 l_3)+{}
   \\
   &
   2 (k_0-l_0) (k_0 l_3-k_3 l_0)- M_H^2(k_3+l_3)\big]+{}
   \\
   &
  M_W^4(k_3+l_3) \big[M_H (k_3+l_3)-2 (k_0-l_0)(k_3-l_3)\big]+{}
   \\
   &
   4 M_H(2 k_0-M_H) (2 l_0-M_H)(k_0 l_3+k_3 l_0)^2
   \label{eq:Cs}
\end{split}
\end{equation}

The non-trivial vanishing of $C_3$ implicitly presupposes 
${\rm det}(\vec{l},\vec{k},\vec{n})=0$. 
Up to non-vanishing overall factors, a second requirement
for an event to be singular is this coplanarity
condition, squared such as to eliminate the sign of $\delta$: $C_0=0$, with
\begin{equation}
\begin{split}
\\
&C_0=2\sp M_W^2
(l\!\cdot\! k-2k_0l_0)\big[2\sp l\cdot k-M_H(k_0+l_0)\big]
\\
& - M_H^2(l\!\cdot\! k-2k_0l_0)^2+
M_W^4\big[2\sp l\!\cdot\! k-(k_0+l_0)^2\big],
\end{split}
\label{eq:det}
\end{equation}
where $l\!\cdot\! k$ has its customary Minkowskian meaning.

\subsection{Questions of nomenclature}

For a singular event the values of $C$ in Eq.~(\ref{eq:Cs}) and
$C_0$ in Eq.~(\ref{eq:det}) must both vanish. Given the form of $C$, there are three 
nontrivial ways for
this to happen: $C_i\!=\!C_0\!=\!0$, $i=1$ to 3, which we call {\it complete singularity conditions}.
Of these, only $C_3\!=\!C_0\!=\!0$ guarantees that all minors of the Jacobian vanish.
The other two conditions, $C_i\!=\!C_0\!=\!0$, $i=1$ to 2, are {\it mock} singularity conditions,
in a sense occasionally used in mathematics, that is,
they do not satisfy all wanted conditions, but are useful for one's purposes.
As it turns out, even the four {\it partial singularity conditions} $C_j\!=\!0$, $j=0$ to 3, are
of interest. 

We choose $C_0$ as the example to make our next linguistic points.
Consider a real or MC-generated event due to the production and decay 
of a Higgs boson.  Its corresponding value of the $C_0$ function in Eq.~(\ref{eq:det})
--a (partial) measure of distance to the $C_0\!=\!0$ singularity--
depends on the Higgs boson mass in two distinct senses.
The first is that $k_0$, $l_0$ and $l\!\cdot k\!$ are contingent on this ``input" mass.
The second is the explicit $M_H$ in the expression of $C_0$, which is
a variable that one may --naturally-- vary at will. To emphasize this
point, we  label this analyst's mass calligraphically:
$M_H\to {\cal M}$.

It is convenient to rescale and rewrite $C_0$ as:
\begin{eqnarray}
&&C_0=-\,M_E^4\; \Sigma_0({\cal M})/4
\nonumber\\
&&\Sigma_0({\cal M})\equiv ({\cal M}-M_+)({\cal M}-M_-)
\nonumber\\
&&M_\pm=
\frac{2 M_W}{M_E^2} \left[M_W
   ({k_0}+{l_0})\pm M_M \sqrt{
   M_W^2-M_E^2}\right]
\nonumber\\
&&M^2_{M,E}\equiv 2\sp (k_0\sp l_0 \mp \vec{k} \,\vec{l}\;),
\label{eq:Sigma4}
\end{eqnarray}
where $M_\pm$ are the non-zero roots of $C_0=0$ and 
$M^2_{M,E}$ are the Minkowski and ``Euclidean" masses of the
(approximately massless)
charged lepton pair. Notice that $\Sigma_0$ depends on the implicit
variable $\cal M$, while its roots, $M_\pm$ do not. That is why we
refer to $\Sigma_0$ and $M_\pm$ with different symbols, even though their distributions
are in all cases diagnostics of the value of the real or simulated 
Higgs boson mass  (we reserve the nomenclature
``$M$'' for all singularity variables of the later kind). Implicit masses become
theoretically inevitable in cases for which, unlike for $\Sigma_0$,
the ${\cal M}$ roots cannot be made explicit.

Functions of an implicit mass $\cal M$, such as $\Sigma_0$,
 are also singularity  variables.
They vanish at singular points of phase space, {\it iff} the correct choice
${\cal M}=M_H$ has been made, with $M_H$ the physical or Monte Carlo ``truth".

\subsection{Partial and complete singularity conditions and variables}

The singularity condition $C=0$ of Eq.~(\ref{eq:Cs}) can be satisfied
in various ways. Two of them ($M_H=0$ and $M_H=2\,M_W$) are of little
practical relevance. Two others correspond to the na\"ive-looking observables
\begin{eqnarray}
M_1&=&\vert k_0-l_0\vert\label{eq:C1}\\
M_2&=& M_W^2\,{k_0+l_0 \over 2\,k_0\,l_0}
\label{eq:C2}
\end{eqnarray}

The remaining possibility is $C_3=0$ in Eq.~(\ref{eq:Cs}),
a cubic  polynomial in the Higgs boson mass.
In analogy with Eq.~(\ref{eq:Sigma4}) for $C_0$, we 
introduce its roots:
\begin{eqnarray}
&&C_3=F\;\Sigma_3({\cal M})
\nonumber\\
&&\Sigma_3({\cal M})=({\cal M}-\tilde M_1)({\cal M}-\tilde M_2)({\cal M}-\tilde M_3)
\nonumber\\
&&F\equiv 4\,(k_3\,l_0+k_0\,l_3)^2
\label{eq:Sigma3}
\end{eqnarray}
where the explicit forms of $\tilde M_i$ are lengthy.

It is not useless to rewrite Eqs.(\ref{eq:Sigma4},\ref{eq:C2},\ref{eq:Sigma3}) in the
form:
\begin{eqnarray}
\tilde\Sigma_0&\equiv&  {\rm Sign}(\Sigma_0)\,\vert \Sigma_0\vert^{1/2}\nonumber\\
\Sigma_1&\equiv& M_1\nonumber\\
\Sigma_2&\equiv& {\cal M}-M_2\nonumber\\
\tilde\Sigma_3&\equiv&{\rm Sign}(\Sigma_3)\,\vert \Sigma_3\vert^{1/3}
\label{eq:Sigma123}
\end{eqnarray}
This is because to
construct true singularity variables that reflect a complete set
of singularity conditions we must introduce a measure of the distance
between a data point (given values of $\vec k$ and $\vec l\,$) and one
of the three center-of-mass singular manifolds: the points $\{0,0\}$ of the 
planes  $\{C_i =0\}\cap\{C_0=0\}$, $i=1$ to 3. With the help of 
Eqs.~(\ref{eq:Sigma4},\ref{eq:Sigma123})
we define the following quantities with unit mass dimension: 
\begin{eqnarray}
D_1&=&\left(\tilde\Sigma_0^2+\Sigma_1^2\right)^{1/2}
\nonumber\\
D_2&=&\left(\tilde\Sigma_0^2+\Sigma_2^2\right)^{1/2}
\nonumber\\
D_3&=&\left(\tilde\Sigma_0^2+\tilde\Sigma_3^{2}\right)^{1/2}
\label{eq:Ds}
\end{eqnarray}
The functions $D_i({\cal M})$ are the full set of complete center-of-mass
singularity variables for the case at hand.

\subsection{From Algebra to Geometry}
\begin{figure}[htbp]
\begin{center}
\includegraphics[width=0.34\textwidth]{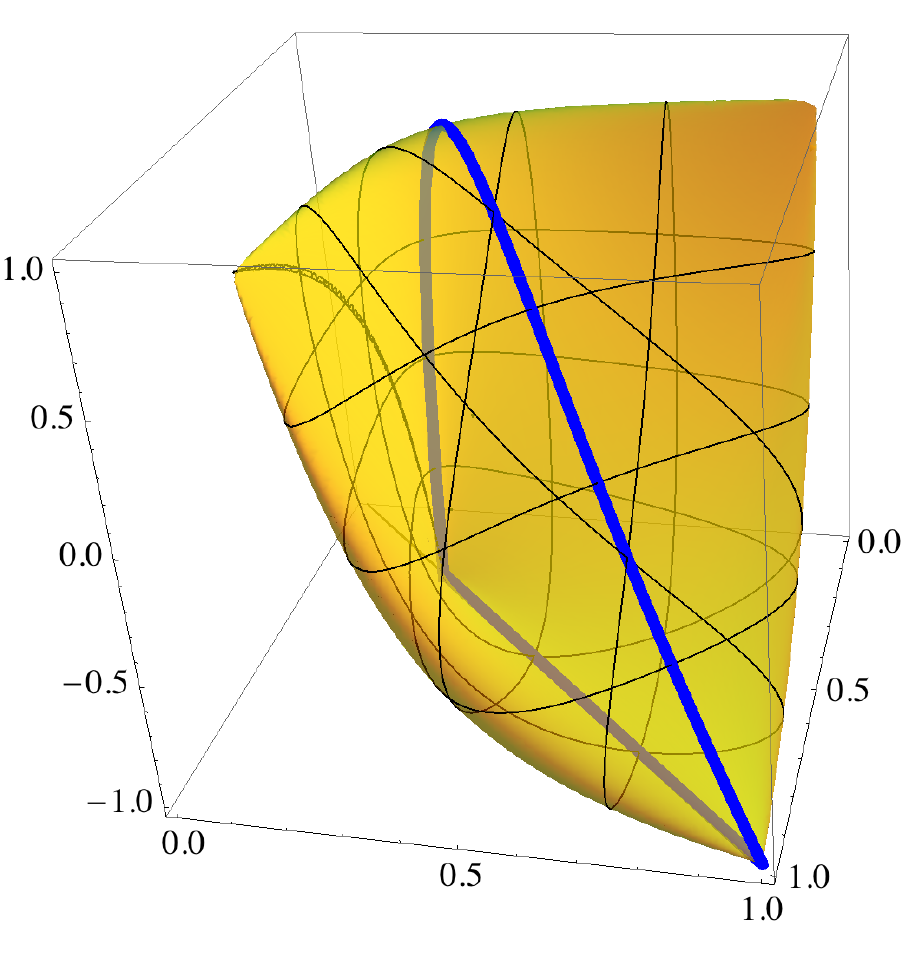}\\
\includegraphics[width=0.37\textwidth]{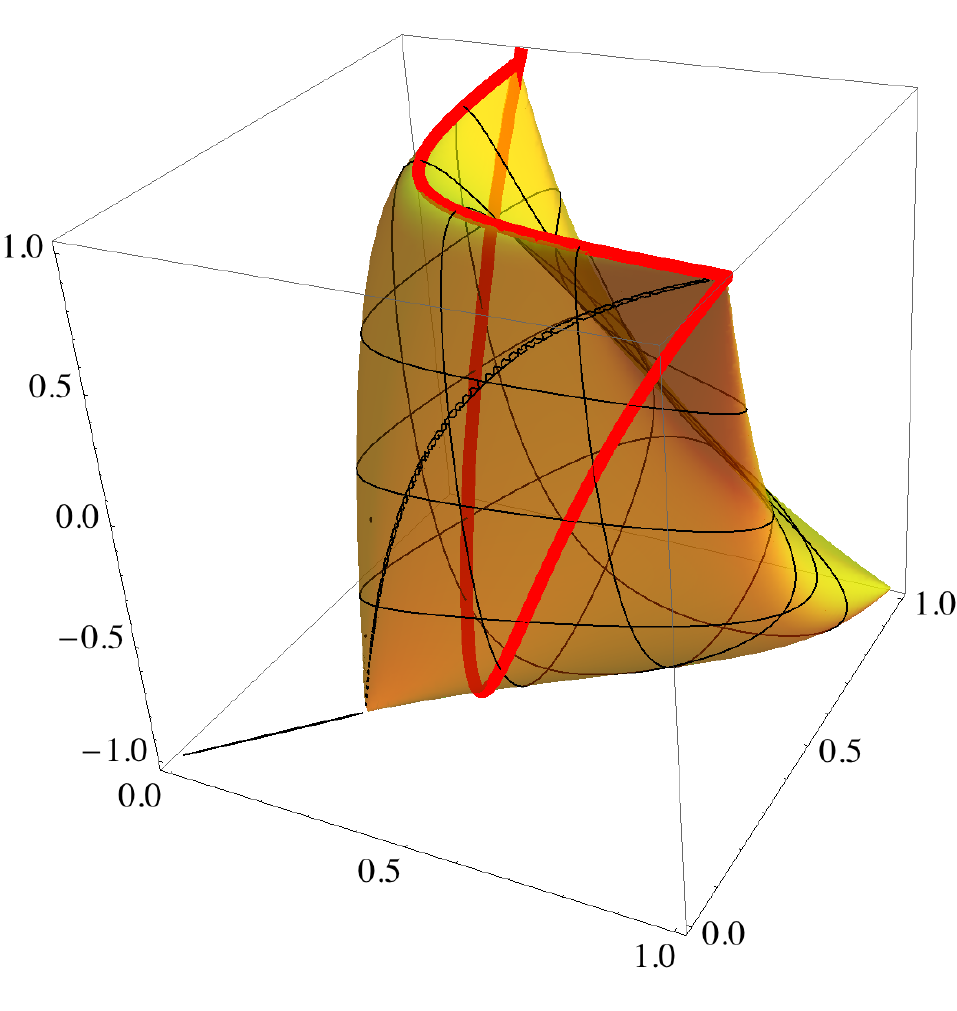}\\
\includegraphics[width=0.34\textwidth]{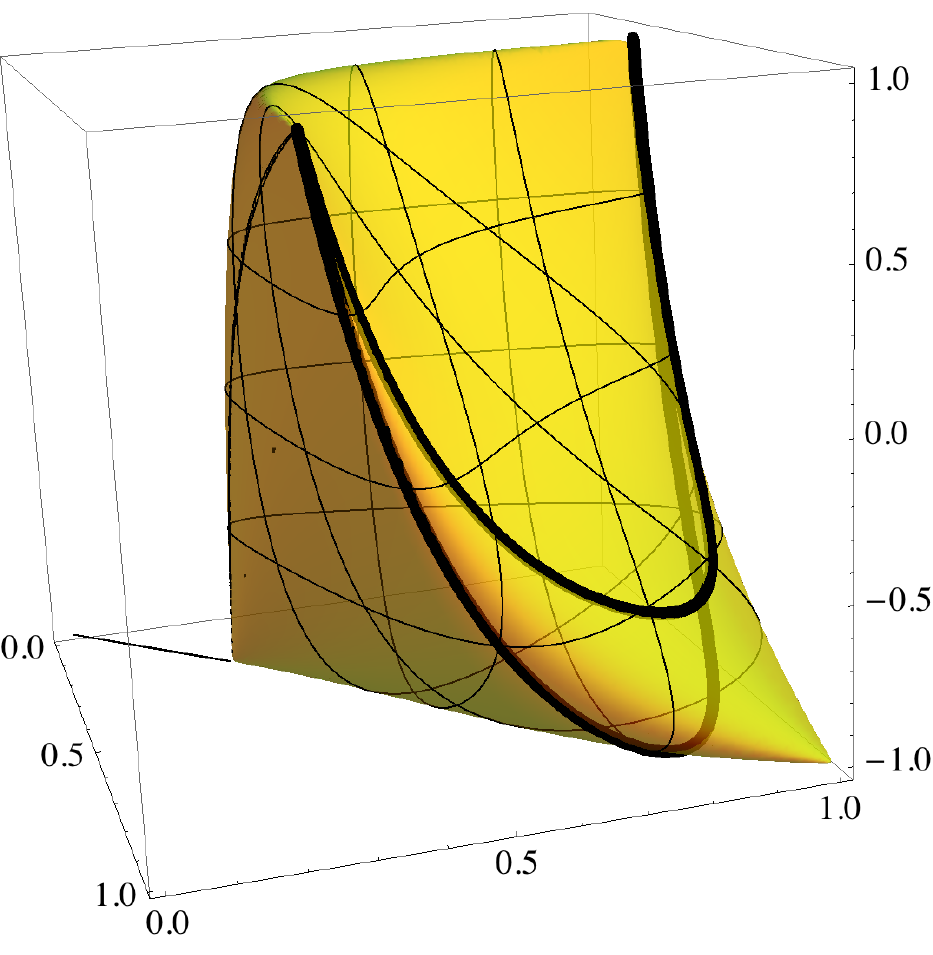}\\
\caption{Visualizing the singular CM phase space. The closed surface in all figures is 
$C_0=0$, see Eq.~(\ref{eq:det}).
The thick lines in the top (middle) figure are the singularity 
conditions $C_0=C_1=0$ ($C_0=C_2=0$), see Eqs~(\ref{eq:Cs}).
The thickest black line in the bottom figure 
is the singular phase space 
$C_0=C_3=0$ for the particular choice $k_3+l_3=0$.
The entire illustration is for $M_W=1,M_H=2.5$.
The horizontal axes are $k_T$ and $l_T$, the vertical one is 
$c_\varphi\equiv  \vec k_T\cdot \vec l_T/(k_T\,l_T)$.
  \label{fig:AbsPhaseCorr}}
\end{center}
\end{figure}

An advantage of the approximation in which Higgs bosons would be produced at
rest is that the locus of the singular points in the observable $\{\vec k,\vec l\}$ 
phase space can be visualized. Let 
$c_\varphi \equiv \cos \Delta\varphi= \vec k_T\cdot \vec l_T/(k_T\,l_T)$.
The singular phase space is shown in Fig.~\ref{fig:AbsPhaseCorr} in the
variables $\{k_0,l_0,c_\varphi \}$, in an example wherein we have chosen $M_H=2.5$
in $M_W=1$ units. The closed surface in the three subfigures is 
the coplanarity condition $C_0=0$, see Eq.~(\ref{eq:det}).
The thick lines in the top and middle figure correspond to the singularity 
conditions $C_0=C_1=0$ and $C_0=C_2=0$, see Eqs~(\ref{eq:Cs}).
The last figure partly describes the singular phase space 
$C_0=C_3=0$ for the choice $k_3+l_3=0$; the complete space would
be the direct product of this latter line in $\{k_3,l_3\}$ space
 with the thick line in the figure.

The $C_0=C_3=0$ singularity condition, as one varies $k_3+l_3$, covers
all of the $C_0=0$ coloured surface of Fig.~\ref{fig:AbsPhaseCorr}.
This reflects the fact that the other two conditions are
mock, and of zero measure relative to $C_0=C_3=0$.

\section{Back to a hadron Collider}
\label{sec:Back}

The derivation of a singularity variable for the more realistic case of an $H$ boson
of rapidity $y_H\neq 0$ is akin to that of the  $y_H= 0$ case, requiring only one
extra step. Naturally, this is to start by applying the Lorentz boost $L(y_H,\vec{n}_p)$ to
the $W$ momenta of Eq.~(\ref{eq:pWs}), to obtain:
\begin{eqnarray}
&p_{_{Wl}}\!=\!
\{c\mspace{1mu}E+s\mspace{1mu} P\mspace{1mu} n_3,+P \mspace{1mu}n_1,
+P \mspace{1mu} n_2,c\mspace{1mu}P\mspace{1mu} n_3+s\mspace{1mu}E\}
\nonumber\\
&p_{_{Wk}}\!=\!
\{c\mspace{1mu}E-s\mspace{1mu} P\mspace{1mu} n_3,-P \mspace{1mu}n_1,
-P \mspace{1mu} n_2,c\mspace{1mu}P\mspace{1mu} n_3-s\mspace{1mu}E\}
\label{eq:pWsbis}
\end{eqnarray}
where $c\equiv \cosh(y_H)$ and $s\equiv\sinh(y_H)$.  
Following precisely the same steps as in \S\ref{seq:derivation},
one concludes that the partial singularity conditions are $C'_i=0$,
$i\!=\!1$ to 4, with
\begin{eqnarray}
C_1'&=& 2\sp\xi\sp C_1(l_0',l_3',k_0',k_3') \nonumber\\
&=&k_0+k_3-l_0-l_3+
   \left(k_0-k_3-l_0+l_3\right)\xi^2\nonumber\\
C_2'&=& 2\sp\xi^2\sp C_2(l_0',l_3',k_0',k_3')\nonumber\\
&=&\left(k_0+k_3\right)
   \left(l_0+l_3\right) M_H\; -\, ...\nonumber\\
   &&+ \left(k_0-k_3\right)
   \left(l_0-l_3\right)
   M_H\xi^4\nonumber\\
C_3'&=&4\sp\xi^6\sp C_3(l_0',l_3',k_0',k_3'),\nonumber\\
&=&4
   \left(k_0+k_3\right){}^3
   \left(l_0+l_3\right){}^3 M_H\; -\, ...\nonumber\\
&&+4 \left(k_0-k_3\right){}^3
   \left(l_0-l_3\right){}^3 M_H\sp \xi^{12}\nonumber\\
C_0'&=&4\sp \xi^4\sp C_0(l_0',l_3',k_0',k_3')\nonumber\\
&=&-\left(k_0+k_3\right){}^2
   \left(l_0+l_3\right){}^2 M_H^2\; +\, ...\nonumber\\
&&-
   \left(k_0-k_3\right){}^2
   \left(l_0-l_3\right){}^2
   M_H^2\sp \xi ^8
\label{eq:CPrimes}
\end{eqnarray}
where
\begin{eqnarray}
\xi&=& c+s=e^{y_H}\label{eq:xi}\\
l_0'&=&c\mspace{1mu} l_0 - s\mspace{1mu} l_3,
\;\;\;\;\;
l_3'= c\mspace{1mu} l_3 - s\sp l_0,\nonumber\\
k_0'&=&c\mspace{1mu} k_0 - s\mspace{1mu} k_3,
\;\;\;
k_3'= c\mspace{1mu} k_3 - s\mspace{1mu} k_0\nonumber
\end{eqnarray}
In $C'_{2,3,0}$, whose expressions in terms of unprimed momenta 
 are easily obtained and lengthy, we have only given the first and last term
in their expansion in $\xi$, which are sufficient to specify their mass dimension and their
grade as polynomials in $\xi$, two numbers that we shall need.

To obtain  longitudinally boost-invariant results analogous to
the ones in Eqs.~(\ref{eq:Ds}) one must eliminate the unknown boost parameter
$\xi$ between the pairs of polynomials $\{C'_j(\xi),C'_0(\xi)\}$, $j\!=\!1$ to 3.
The first and simplest of these results, for $j=1$, is the singularity condition
$\Delta_1=0$, with
\begin{equation}
\begin{split}
\Delta_1\propto {} &M^2\;\Sigma,
\\
M ={} &A-B\,l\cdot k,
\\
\Sigma={}&\big[4A^2 E^4+B^2(M_W^4-2 P^2\,l\cdot k)^2\\
&-
4A\,B\,E^2(M_W^4+2 P^2\,l\cdot k)\big],\\
 A\equiv {}& 2\left[(\vec l_T\cdot \vec k_T+l\cdot k)^2-k_T^2\,l_T^2\right],
\\
 B\equiv {} & 2\,(\vec l_T\cdot \vec k_T+l\cdot k)-k_T^2-l_T^2,
\end{split}
\label{eq:Delta1}
\end{equation}
where $E={\cal M}/2$ is the energy of a $W$ in  the rest system of
a Higgs boson of trial mass ${\cal M}$
and $P$ is the corresponding momentum.  
We have followed our convention to label $M$ the singularity variables
that do not depend on ${\cal M}$,  and $\Sigma$ (and now $\Delta$)
those which do. Notice that $\Delta_1$, by construction and demonstration, 
is a function of longitudinally boost-invariant observables.

The derivation of analytical results for the remaining polynomial pairs is not
as simple as it was for $\{C'_0,C'_1\}$,  $C'_1$ being merely quadratic
in $\xi$. The expressions for 
$C'_{2,3,0}$ are polynomials in $\xi$ of degrees 4, 12 and 8, respectively.
The condition for two polynomials 
$\sum_{i=0}^n a_i\sp \xi^i$ and $\sum_{j=0}^m b_j\sp \xi^j$
to vanish simultaneously (to have common roots)
is called their {\it resultant}, and is a sum of products of powers of $a_i$ and $b_j$.
The resultant of $C'_1=0$ and $C'_0=0$ is the condition
${\rm Res}\{C'_1,C'_0\}\equiv\Delta_1=0$, see Eq.~(\ref{eq:Delta1}). 
The number of terms of a resultant grows very rapidly with $m\times n$, it is 
95 for $(m,n)=(2,8)$, 4970 for ${\rm Res}\{C'_2,C'_0\}$, for which $(m,n)=(4,8)$. For
this case, after considerable simplifications, the singularity condition is
$\Delta_2=0$, with
\begin{widetext}
\begin{equation}
\begin{split}
& {\rm Res}\{C'_2,C'_0\}\equiv\Delta_2 \propto E^4 \,k_T^6\, l_T^6\,(l\!\cdot\!  k)^4 
 (M_W^{16} (16\sp E^4 k_T^2 l_T^2 + (-4\sp (\vec l_T\!\cdot\!  \vec k_T)^2 +2\sp \vec l_T\!\cdot\!  \vec k_T M_W^2 + l_T^2 M_W^2 + k_T^2 (4 \sp l_T^2 + M_W^2))^2 {}
\\
& - 8\sp E^2 (2\sp k_T^4 l_T^2 + \vec l_T\!\cdot\!  \vec k_T l_T^2 (-2\sp \vec l_T\!\cdot\!  \vec k_T + M_W^2) +
         k_T^2 (-2\sp (\vec l_T\!\cdot\!  \vec k_T)^2 + \vec l_T\!\cdot\!  \vec k_T M_W^2 + 2\sp l_T^2 (l_T^2 + M_W^2)))) 
\\
&  +
   4\sp l \!\cdot\!  k \sp M_W^{12} (-32\sp E^6 k_T^2 l_T^2 +
      M_W^4 (8\sp (\vec l_T\!\cdot\!  \vec k_T)^3 + 2\sp \vec l_T\!\cdot\!  \vec k_T M_W^4 + l_T^2 M_W^2 (l_T^2 + M_W^2)
\\
& +
         k_T^4 (4 \sp l_T^2 + M_W^2) - 4\sp (\vec l_T\!\cdot\!  \vec k_T)^2 (l_T^2 + 2\sp M_W^2) +
         k_T^2 (-4\sp (\vec l_T\!\cdot\!  \vec k_T)^2 - 8\sp \vec l_T\!\cdot\!  \vec k_T l_T^2 + 4\sp l_T^4 + 6\sp l_T^2 M_W^2 + M_W^4))      
 \\
 &
 -
      E^2 M_W^2 (-8\sp (\vec l_T\!\cdot\!  \vec k_T)^3 + 8\sp \vec l_T\!\cdot\!  \vec k_T l_T^2 M_W^2 + 4\sp  (\vec l_T\!\cdot\!  \vec k_T)^2 (-5\sp l_T^2 + M_W^2) +
          k_T^4 (20\sp l_T^2 + M_W^2) + l_T^2 M_W^2 (l_T^2 + 2\sp M_W^2) 
\\
&          
          +
         2\sp k_T^2 (-10\sp (\vec l_T\!\cdot\!  \vec k_T)^2 + 10\sp l_T^4 + 13 l_T^2 M_W^2 + M_W^4 +
            4\sp \vec l_T\!\cdot\!  \vec k_T (l_T^2 + M_W^2)))     
\\
&  +
      4\sp E^4 (4\sp k_T^4 l_T^2 + \vec l_T\!\cdot\!  \vec k_T l_T^2 (-4\sp \vec l_T\!\cdot\!  \vec k_T + 3\sp M_W^2) +
         k_T^2 (-4\sp (\vec l_T\!\cdot\!  \vec k_T)^2 + 3\sp \vec l_T\!\cdot\!  \vec k_T M_W^2 + 2\sp l_T^2 (2\sp l_T^2 + 7\sp M_W^2))))          
\\
&  +
   16\sp E^4 (l \!\cdot\!  k)^4 (M_W^6 + 4\sp k_T^2 P^4) (M_W^6 + 4\sp l_T^2 P^4) +
   4\sp (l \!\cdot\!  k)^2 M_W^8 (-4\sp \vec l_T\!\cdot\!  \vec k_T M_W^2 (2\sp E^2 + M_W^2) 
\\
&   \times (3\sp E^4 (k_T^2 + l_T^2) -
         4\sp E^2 (k_T^2 + l_T^2) M_W^2 + M_W^4 (k_T^2 + l_T^2 + M_W^2))         
\\
&  +
      4 (\vec l_T\!\cdot\!  \vec k_T)^2 (4\sp E^6 (k_T^2 + l_T^2) + M_W^8 +
         E^4 M_W^2 (-8\sp k_T^2 - 8\sp l_T^2 + M_W^2) +
         4\sp E^2 M_W^4 (k_T^2 + l_T^2 + M_W^2)) 
\\
&  +
      M_W^4 (M_W^4 (l_T^2 + M_W^2)^2  - 2\sp E^2 l_T^2 M_W^2 (l_T^2 + 6\sp M_W^2) +
         E^4 l_T^2 (l_T^2 + 12\sp M_W^2)) +
      2\sp k_T^2 (48\sp E^8 l_T^2 + M_W^8 (l_T^2 + M_W^2) 
\\
& -
         8 E^6 l_T^2 (l_T^2 + 15\sp M_W^2) +
         E^4 (16\sp l_T^4 M_W^2 + 97\sp l_T^2 M_W^4 + 6\sp M_W^6) -
         2\sp E^2 (4\sp l_T^4 M_W^4 + 15\sp l_T^2 M_W^6 + 3\sp M_W^8)) 
\\         
 &        +
      k_T^4 (-16\sp E^2 l_T^2 + M_W^4) P^4)   -
   16\sp E^2 (l \!\cdot\!  k)^3 M_W^4 (M_W^{12} + 3\sp (k_T^2 + l_T^2) M_W^8 P^2    
\\
&+
      6\sp (k_T^2 + l_T^2) M_W^6 P^4 + 24\sp k_T^2 l_T^2 M_W^2 P^6 +
      32\sp k_T^2 l_T^2 P^8 -
      2\sp \vec l_T\!\cdot\!  \vec k_T M_W^2 (E^2 + M_W^2) (M_W^6 + 2\sp (k_T^2 + l_T^2) P^4)))   
\end{split}
\label{eq:Delta2}
\end{equation}    
\end{widetext}  
Notice that $\Delta_2$, as was the case for $\Delta_1$, 
is a function of only longitudinally boost-invariant observables.

What is the number of terms in
 ${\rm Res}\{C'_3,C'_0\}$, for which the degrees of the polynomials
in $\xi$ are $(m,n)=(12,8)$?
For $m,n$ larger than a small integer
the resultant soon becomes obdurately complex. Not even the number
of addends in the (monomial) products of its formal coefficients is known. Only
upper bounds to that number are, to our knowledge, published. The
tightest one is  \cite{Ka}:
$$
S(m,n)=F(m,n,\lfloor m\,n/2\rfloor)\binom{m+n}{n},
$$
where, for integer $a,b,c$, $F$ satisfies the recurrence
$$
F(a,b,c)=\sum_{j=0}^b F(a-1,j,c-j),
$$
with
$$
F(1,b,c)=
\begin{cases}
1 & \text{if $0\leq c\leq b$}
\\
0 & \text{otherwise}
\end{cases}
$$

For the polynomial pair $\{C'_2,C'_0\}$, $(m,n)=(4,8)$ and $S(4,8) = 16335$
(an overestimate by a factor $\sim\! 3$),
while for $\{C'_3,C'_0\}$, $(m,n)=(12,8)$ and $S(12,8) = 477\,174\,360$.
This last upper limit is the best current
estimate (by expert mathematicians) of the number of terms in the expression
for the remaining singularity variable we are after, in terms of products 
of powers of the $m+n+2=22$ coefficients of $\xi$ in the polynomial pairs,
each of which is a complicated function of $M_W,{\cal M},\vec{k}$ and $\vec{l}$.

We shall not be discouraged by the mathematical hardship of constructing
explicit algebraic resultants. In analogy with $\Delta_1$ in Eq.~(\ref{eq:Delta1}) 
and given the complexity of $\Delta_2$ in Eq.~(\ref{eq:Delta2}), we
shall simply define:
\begin{eqnarray}
\Delta_2 &\equiv& {\rm Res}\{C'_2(\xi),C'_0(\xi)\}
\nonumber\\
\Delta_3 &\equiv& {\rm Res}\{C'_3(\xi),C'_0(\xi)\}
\label{eq:Res23and0}
\end{eqnarray}
and find, event by event, the resultant numerically. The coefficients of
the powers of $\xi$ in $C'_{2,3,0}$ being --for a given event-- numbers as opposed
to symbols, this is doable and --for the computer-- trivial.

The formal proof that the resultants in Eqs.~(\ref{eq:Res23and0})
ought to be boost-invariant is given in the Appendix.

\section{Dealing with the $M_H<2 M_W$ case}
\label{SmallMH}

In an $H\to WW^*$ process followed by leptonic decays of both $W$s, 
there is no way to assign a mass, $M_*$, to the $W$ which is putatively
off-shell, even for a fixed $M_H$. Moreover, there is no
deterministic way to decide which $W$ was approximately on-shell.
Finally,
except close to the $M_H=2M_W$ threshold, the theoretical distribution 
of off-shell masses, $d\Gamma/dM_*$, is very wide. To confront this situation
we choose to analize this case by assigning to {\it both} $W$s an adequately 
averaged squared mass:
\begin{equation}
\langle M^2 \rangle={M_W^2+\langle M_*^2\rangle \over 2}
\label{eq:avMass2}
\end{equation}
Since the non-observation of two neutrinos results in wide distributions for all
observables, there is very little difference between using this prescription and
other sensible ones, such as substituting the average $M_*^2$ in Eq.~(\ref{eq:avMass2})
by its most probable value.

\begin{figure}[htbp]
\begin{center}
\includegraphics[width=0.45\textwidth]{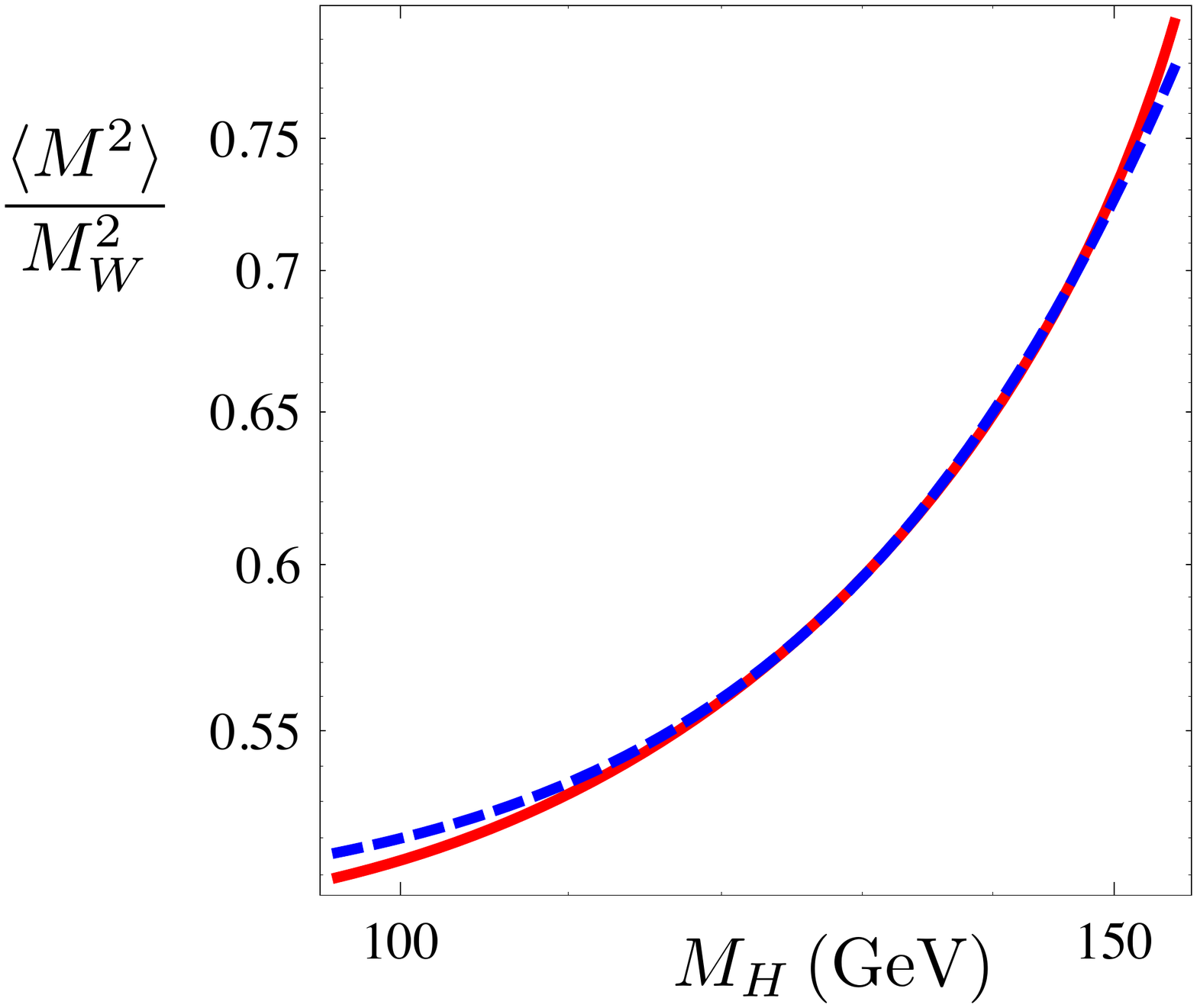}
\vspace{-4.5cm}
\caption{
The average squared mass of Eq.~(\ref{eq:avMass2}), 
as a function of $M_H$, in the process $H\to WW^*$, for $M_H<1.93\sp M_W$.
The continuous line is the leading-order calculation in the standard model.
The dashed line is the approximation of Eq.~(\ref{eq:approxM2}).
\label{fig:avMass2}}
\end{center}
\end{figure}

For $M_H>M_W$, up to a few $W$ widths, $\Gamma_H$, below the two-$W$ threshold, 
and for the leading order standard-model matrix element
for the $H\to WW^*$ decay, the distribution of $W^*$ masses, in the 
excellent approximation of neglecting $\Gamma_W^2/M_W^2$, is:
\begin{eqnarray}
&&{d\Gamma\over dM_*}\propto{\sqrt{M_W^4-2 M_W^2
   \left(M_H^2+M_*^2\right)+\left(M_H^2-M_*^2\right)^2}\over (M_*^2-M_W^2)^2}\nonumber\\
&&\times\left[M_W^4-2 M_W^2 \left(M_H^2-5
   M_*^2\right)+\left(M_H^2-M_*^2\right)^2\right] 
  \end{eqnarray} 
The corresponding $\langle M_*^2 \rangle$ distribution is shown in Fig.~(\ref{fig:avMass2})
for the relevant range of $M_H$ values. In this range and to a good approximation
\begin{equation}
\langle M^2(M_H) \rangle = {M_W^2\over 2}\,
\left[1+ \left({M_H\over 1.02456\,M_W}\right)^6\right],
\label{eq:approxM2}
\end{equation}
also shown in the figure as the dashed line.

\section{Details of our data analysis}
\label{sec:details}

We have derived singularity variables only for the
signal process, not for its backgrounds, and we use the signal singularity
variables to compare the distributions of MC-generated signals and backgrounds.

We present results only for the $H\to WW$, $W\to e\nu$, $W\to \mu\nu$ channel
and its non-resonant $WW$ and $t\bar t$ backgrounds,
with leptons of transverse momentum  greater than 15 GeV, 
and satisfying the pseudorapidity cuts $\eta(e)<2.5$, $\eta(\mu)<2.1$ \cite{SMS}.

Given the delicacies of measuring or simulating (at a ``reconstruction level")
the transverse momentum
of hadrons, $p_T$, we have not boosted each event to the approximate frame wherein
the putative Higgs boson is transversally at rest.
Our MC-simulations are for ``generator level" events and do not have a
$p_T=0$ requirement. No doubt this makes our results
look somewhat weaker than they might otherwise be.

The ratios of signal to background yields are fast-varying functions of $M_H$.
The selections made by experimentalists on the way to focus on signal events
are many and are also mass-dependent. These are reasons why we shall limit
ourselves to illustrating only the different shapes (and not the absolute scales)
of the signal and background histograms of various singular variables.

In discussing singularity variables such as $\Sigma_0({\cal M})$ of
Eq.~(\ref{eq:Sigma4}) or $\Delta_1({\cal M})$ of Eq.~(\ref{eq:Delta1}),
it is informative to do it not only for the correct ``guess"
${\cal M}=M_H$, but also for incorrect ones.
Naturally, the  histograms for a fixed $M_H$ and various $\cal M$
contain precisely the same statistical information. 
An experimentalist using an observable such as $\Sigma_0$ or $\Delta_1$ would 
deal with data (with $M_H$ not known a priori) armed with a plethora of 
``diagonal" MC templates with ${\cal M}=M_H$, with which to compare the
observed distributions.

\section{Data analysis in the CM approximation}
\label{CMdata}

In this section we sketch a numerical analysis of the
partial and complete ``Higgs at rest" singularity variables
derived in \S \ref{sec:gluoncollider}.  Recall that these theoretically-obtained expressions
ignore both the longitudinal and transverse momentum of the Higgs-boson
signal to be analized. 

\subsection{Partial singularity conditions}

We  start this part of the discussion with the singularity 
variable $\Sigma_0$, a measure of distance of an event to the partial
singularity condition of coplanarity: $\Sigma_0=0$.
We chose $M_H=500$ and 120 GeV as examples of the ``true" mass of the events
in this first illustration.

\begin{figure}[htbp]
\begin{center}
\includegraphics[width=0.22\textwidth]{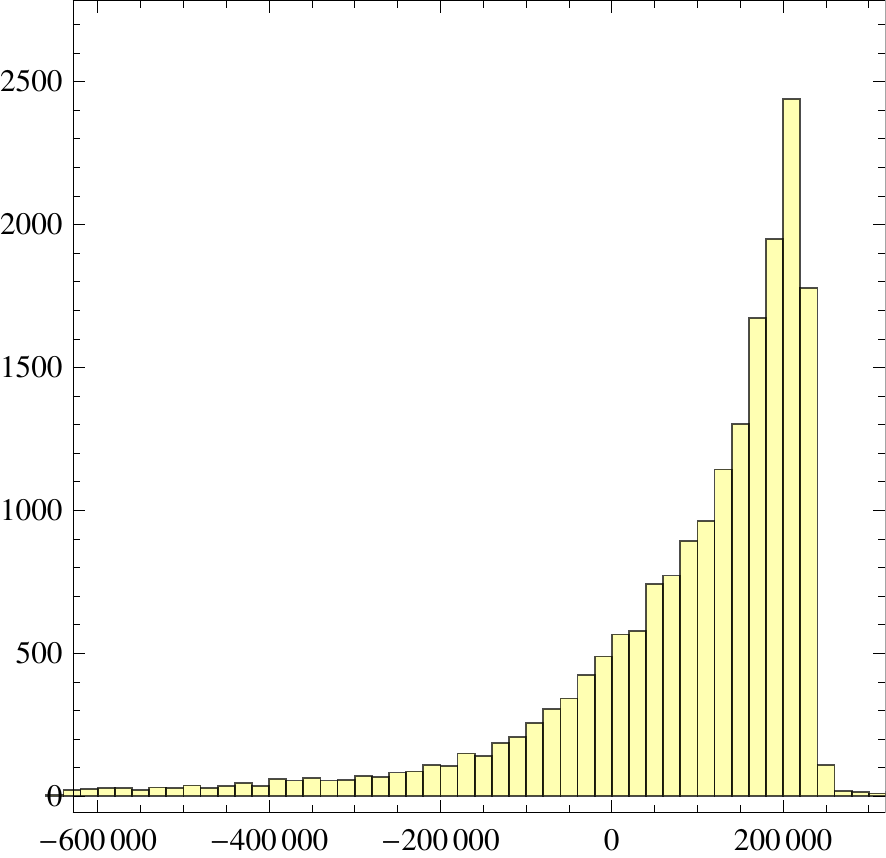}
\includegraphics[width=0.23\textwidth]{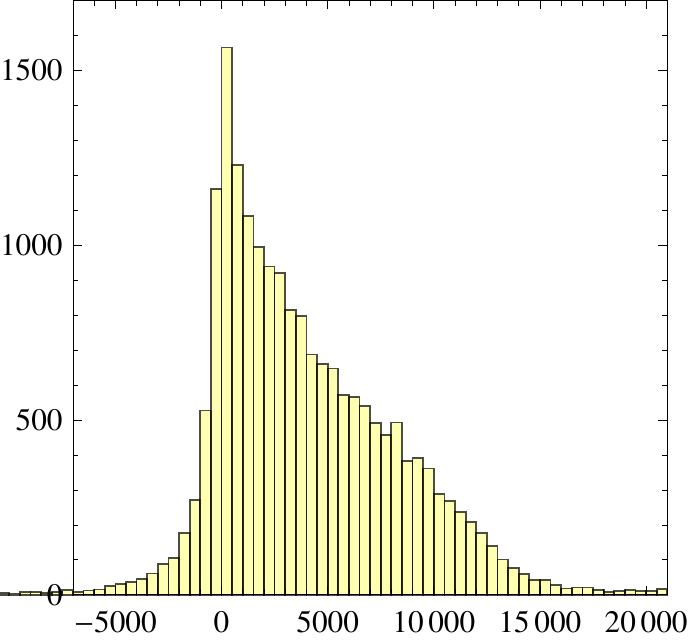} \\
\includegraphics[width=0.22\textwidth]{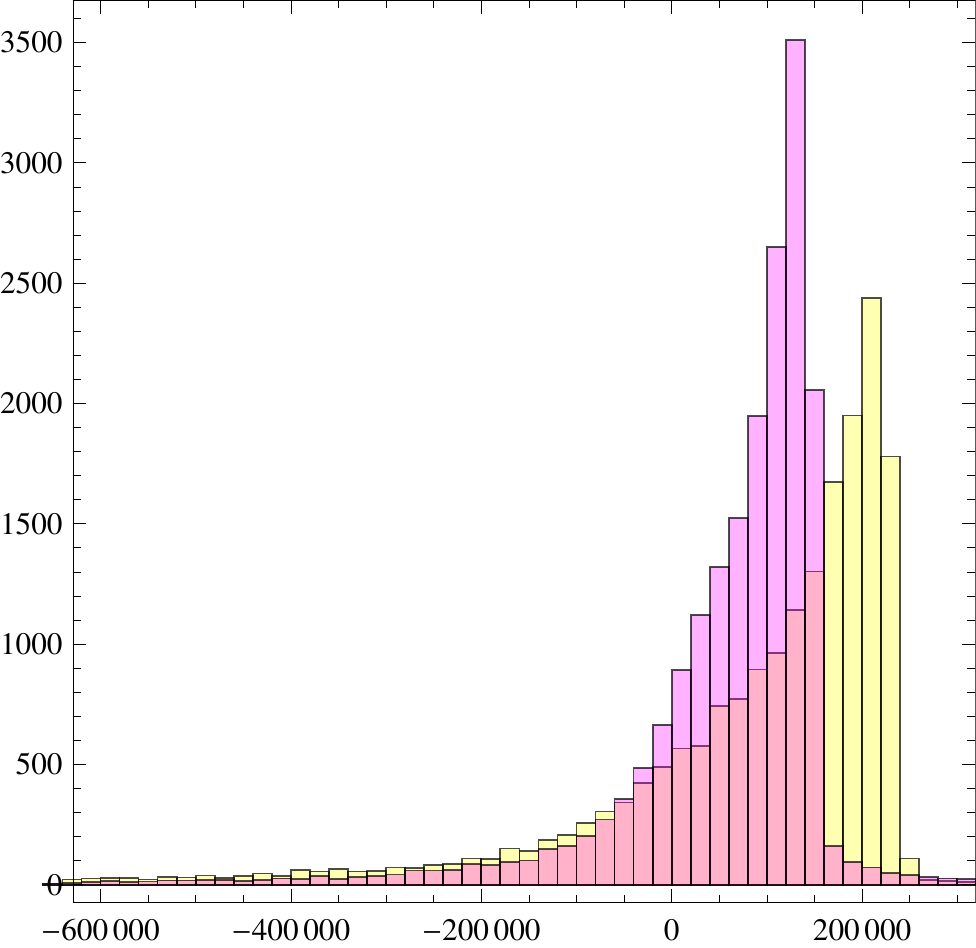}
\includegraphics[width=0.23\textwidth]{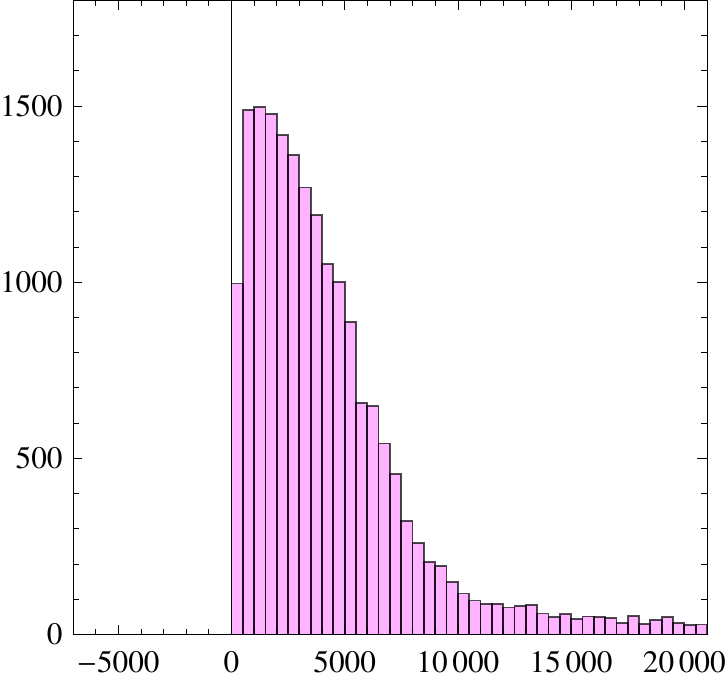} \\
\includegraphics[width=0.22\textwidth]{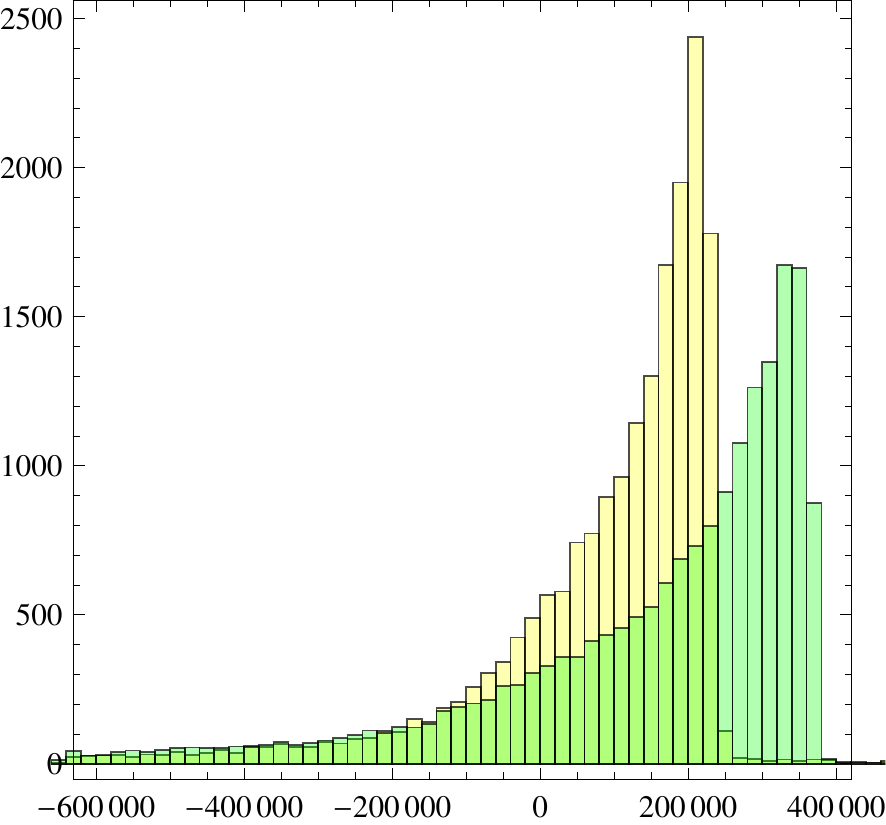}
\includegraphics[width=0.23\textwidth]{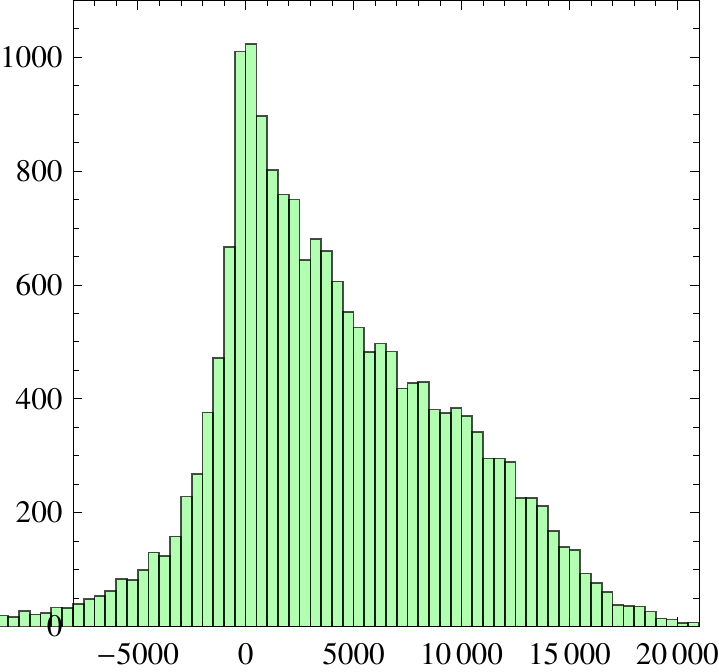}
\caption{The distribution of $\Sigma_0$ values. The horizontal scales are
in units of GeV$^2$. The left column in for $M_H=500$ GeV. Its top figure is
for ${\cal{M}}=M_H$. In its middle (lower) ones the result is compared with
that of the incorrect ${\cal{M}}=(4/5) M_H$ (${\cal{M}}=(5/4) M_H$).
The right panels are for $M_H=140$ GeV.
The top one is for the correct ${\cal{M}}=M_H$. The middle panel
is for ${\cal{M}}=120$ GeV, the lower one for ${\cal{M}}=160$ GeV. 
In the figures the correct-guess histogram is yellow.
  \label{fig:C4Determinant}}
\end{center}
\end{figure}

The distribution of values of $\Sigma_0$ for 20000 signal events generated with
$M_H=500$ GeV is shown in the left panels of Fig.~\ref{fig:C4Determinant}.
The top left panel is for the correct assumption
${\cal{M}}=M_H$, the two other left panels show comparisons
with the incorrect assumptions ${\cal M}=4 \sp M_H/5$ (middle) and ${\cal M}=5\sp M_H/4$
(bottom).
The right panels of Fig.~\ref{fig:C4Determinant} show results for $M_H\!=\!140$ GeV.
The top panel is for the correct assumption ${\cal{M}}\!=\!M_H$. The middle panel
is for ${\cal{M}}\!=\!120$ GeV and the lower one for ${\cal{M}}\!=\!160$ GeV.
At $M_H\!=\!500$ GeV the distribution of $\Sigma_0$ is very sensitive to the
boson's mass, as exemplified in Fig.~\ref{fig:C4Determinant} by the sensitivity
to ${\cal M}$. At $M_H\!=\!140$ GeV this is less so.

\begin{figure}[htbp]
\begin{center}
\includegraphics[width=0.23\textwidth]{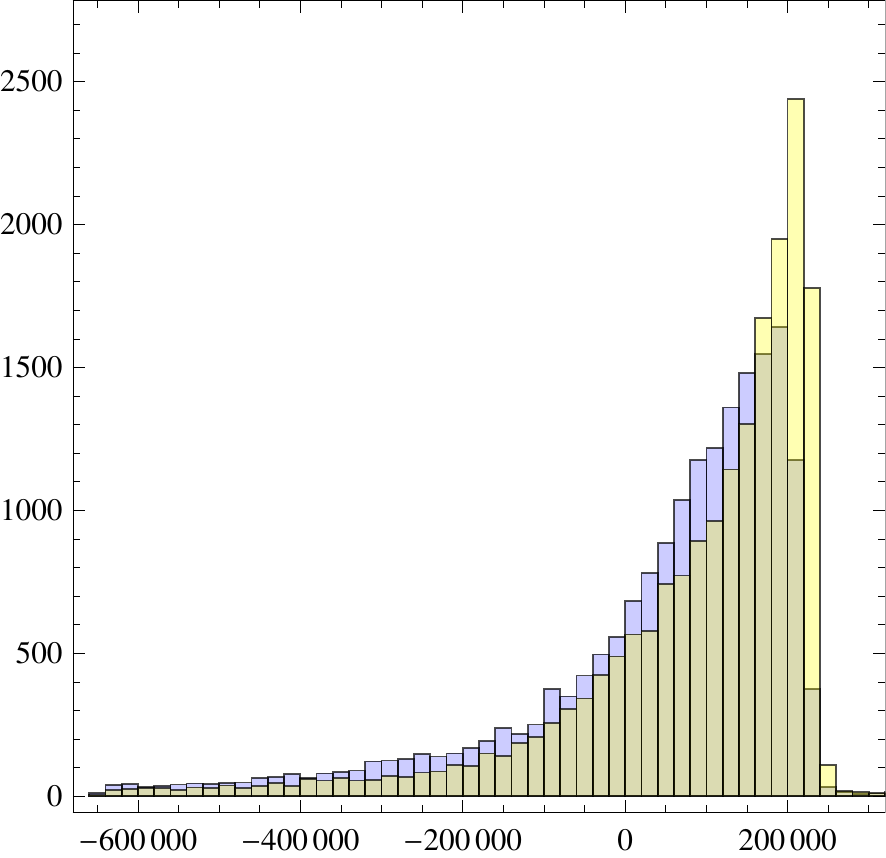}
\includegraphics[width=0.24\textwidth]{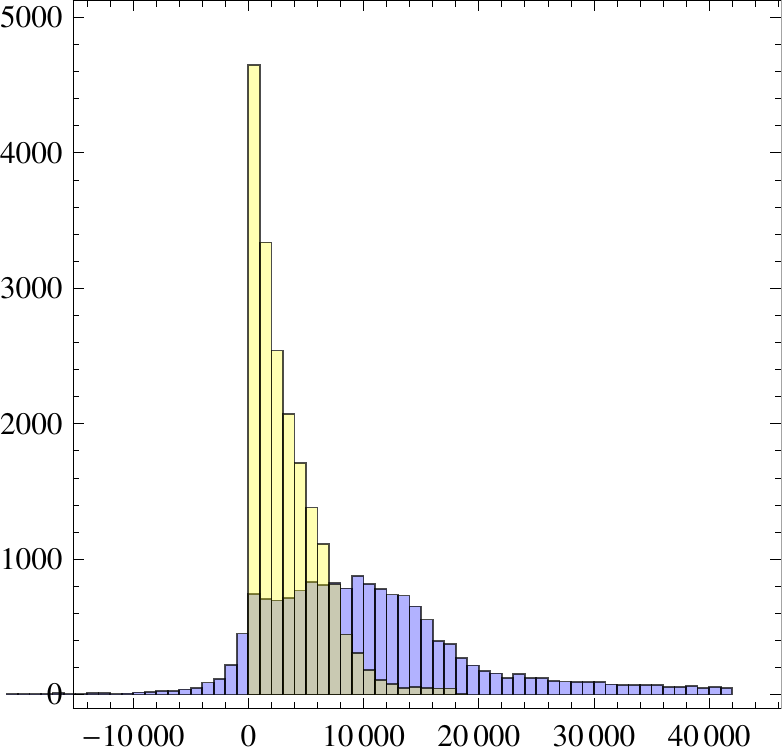}\\
\vspace{.3cm}
\includegraphics[width=0.23\textwidth]{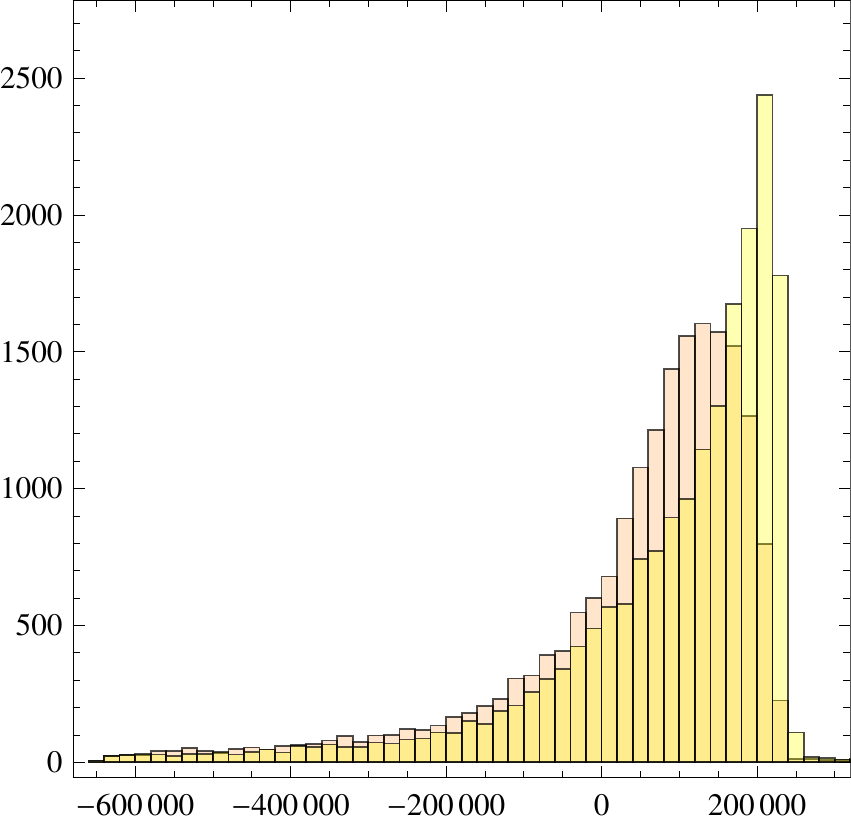}
\includegraphics[width=0.24\textwidth]{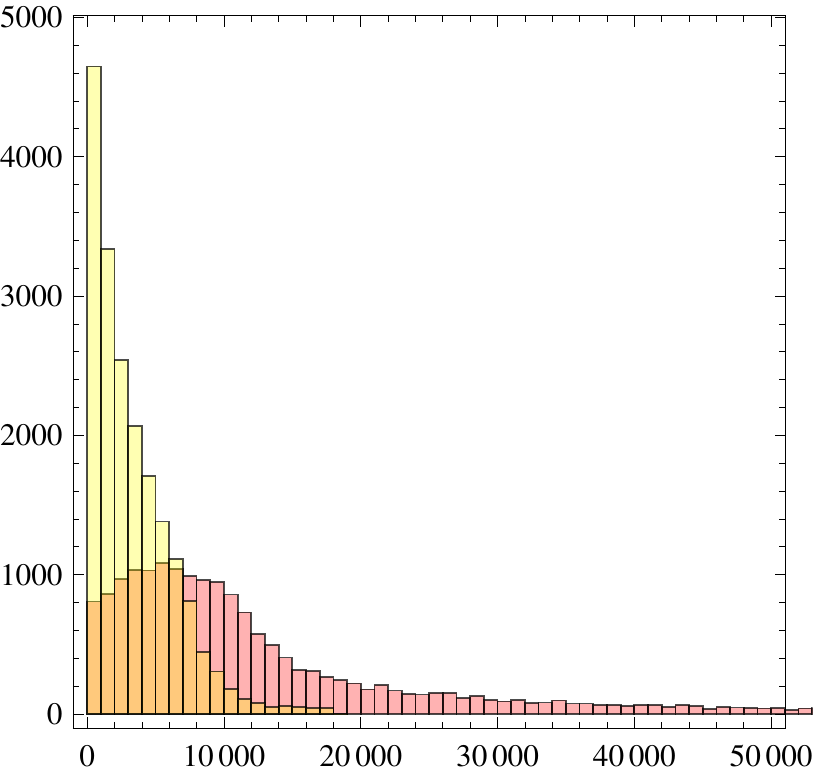}\\
\caption{
Shapes of the $\Sigma_0$ distributions for the signal --the yellow tallest histogram--
and the $WW$ and $t\bar{t}$ backgrounds (upper and lower rows). The horizontal scales are
in GeV$^2$ units.
The left (right) column is for $M_H=500$ (120) GeV.
  \label{fig:C4DeterminantCompare}}
\end{center}
\end{figure}

The ability of the $\Sigma_0$ distribution to sieve apart signal and background shapes
is illustrated in Fig.~\ref{fig:C4DeterminantCompare}. Its left (right) columns are
for $M_H=500$ (120) GeV, both with ${\cal M}$ set to its corresponding correct
value. The top (bottom) lines refer to the $WW$ and $t\bar{t}$ backgrounds.
In all cases we have simulated equally many signal and background events, so that
the figure reflects the shape of the distributions, not their relative weights. 
At $M_H=120$ GeV the shape of signal and backgrounds are very  different,
while at $M_H=500$ GeV this is less so.

The conclusions on the ability to distinguish signal and backgrounds
or different Higgs masses are, as we saw,  very mass dependent. The rest
of the questions to be discussed in this chapter are quite insensitive to $M_H$.
We shall study them only for the $M_H=120$ GeV example.

\begin{figure}[htbp]
\begin{center}
\includegraphics[width=0.50\textwidth]{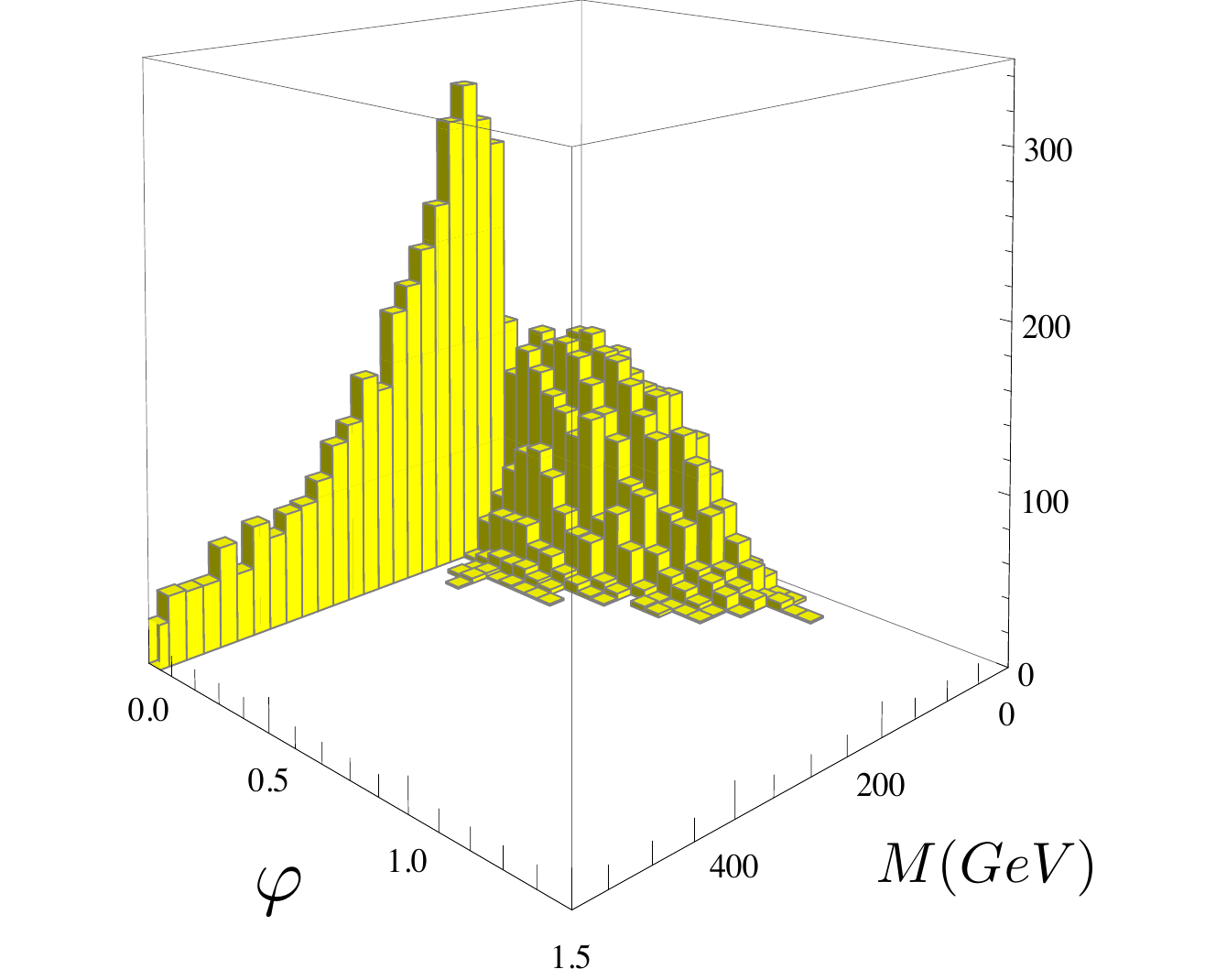}\\
\includegraphics[width=0.69\textwidth]{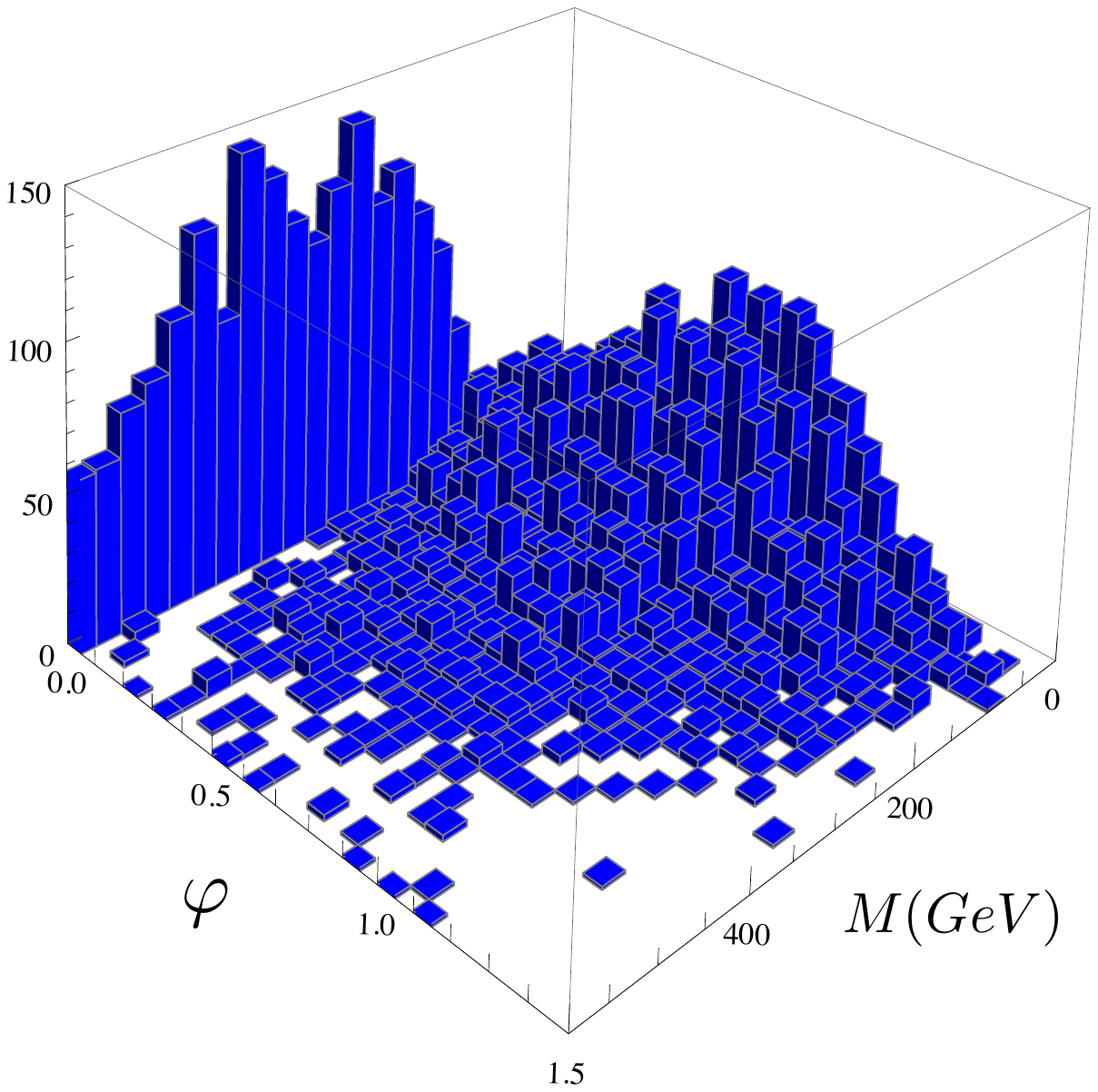}\\
\vspace{-10cm}
\includegraphics[width=0.43\textwidth]{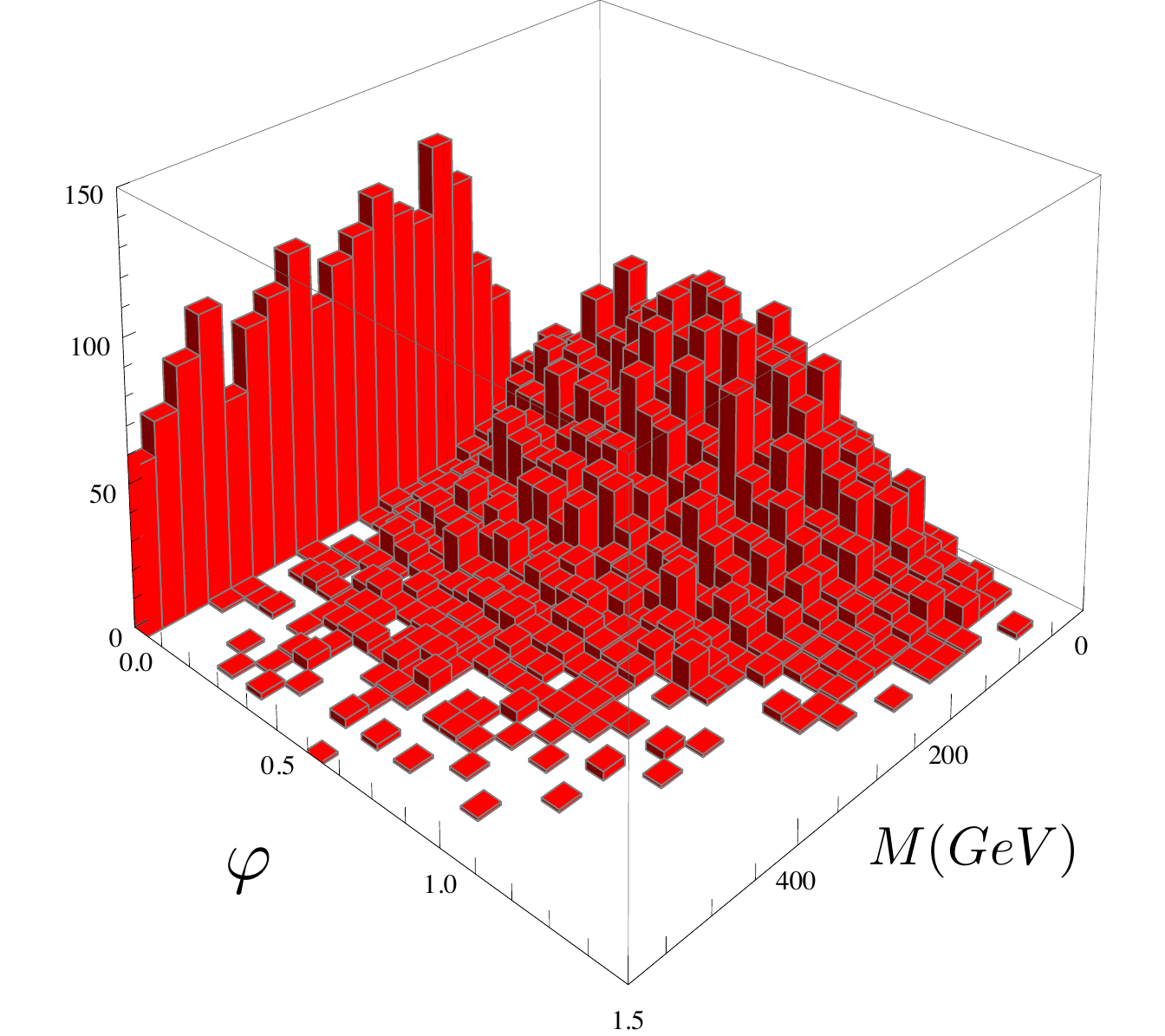}\\
\caption{Correlation between absolute value $M$ and the phase
$\varphi$ of the roots of $\Sigma_0$. Top: $M_H=120$ GeV signal.
Middle:  $WW$ background. Bottom: $t \bar t$ background.
The real roots gather along the $\varphi=0$ axis.
  \label{fig:AbsPhaseCorr2}}
\end{center}
\end{figure}

The quantity $\Sigma_0$ of Eq.~(\ref{eq:Sigma4}) is real, but its
roots, $M_\pm$, need not be. For input MC data corresponding to
$M_H=120$ GeV, about 14\% of the roots are a real pair, the rest being
two conjugate complex numbers. The conclusion that the complex roots
are useless would be most premature. A feature of these roots to be studied
ab initio is the correlation between their absolute value and phase.
This is done for the $M_H=120$ GeV signal and the $WW$ and $t\bar t$
backgrounds in Fig.~\ref{fig:AbsPhaseCorr2}, where the mass axis is
the absolute value of the roots (shown once for each complex root
and for its two values for each real pair). The $\varphi$ axis is the
phase of the roots having a non-negative imaginary part. We see
that the $\{\varphi,\vert M \vert  \}$ correlation is weak and the distributions are
significantly different for signal and background. 

\begin{figure}[htbp]
\begin{center}
\includegraphics[width=0.23\textwidth]{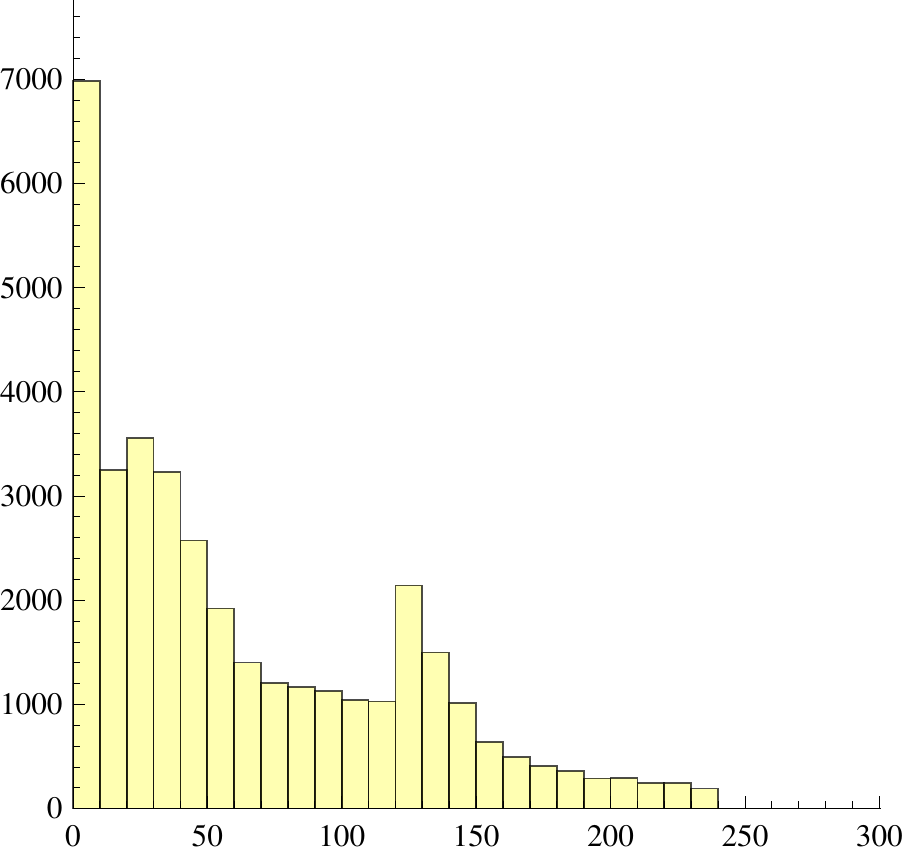}
\includegraphics[width=0.23\textwidth]{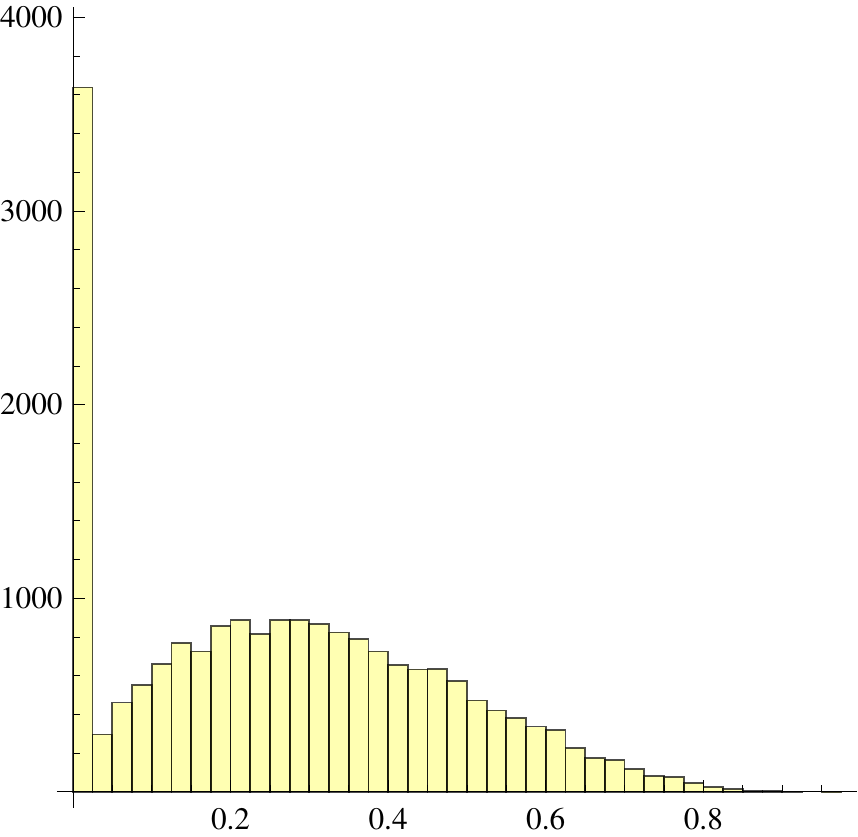}\\
\includegraphics[width=0.23\textwidth]{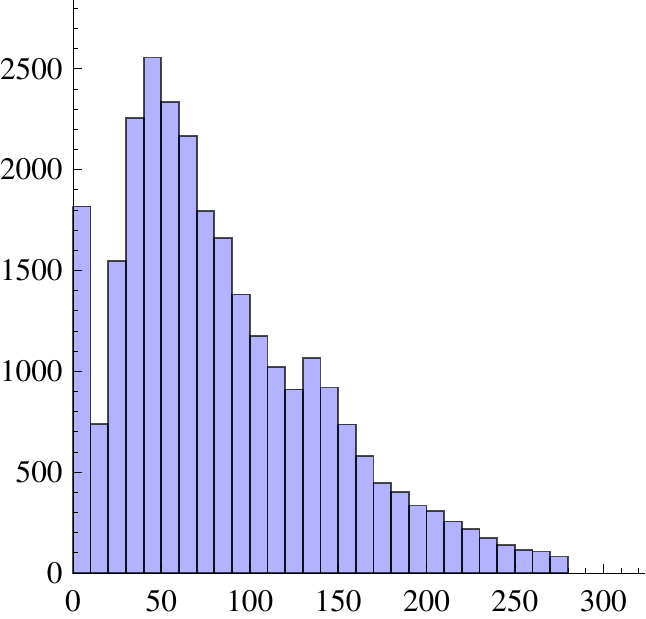}
\includegraphics[width=0.23\textwidth]{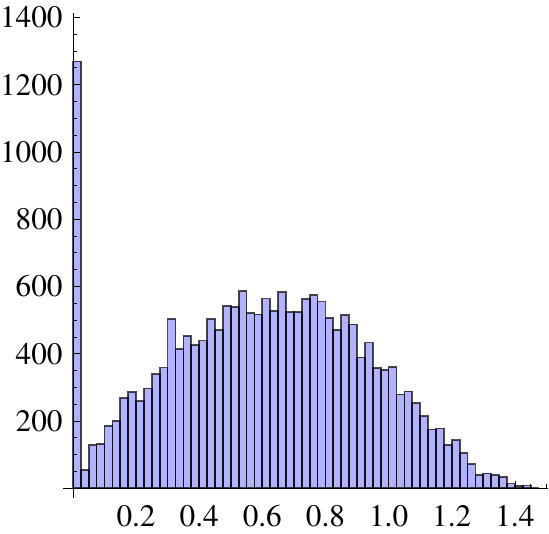}\\
\includegraphics[width=0.23\textwidth]{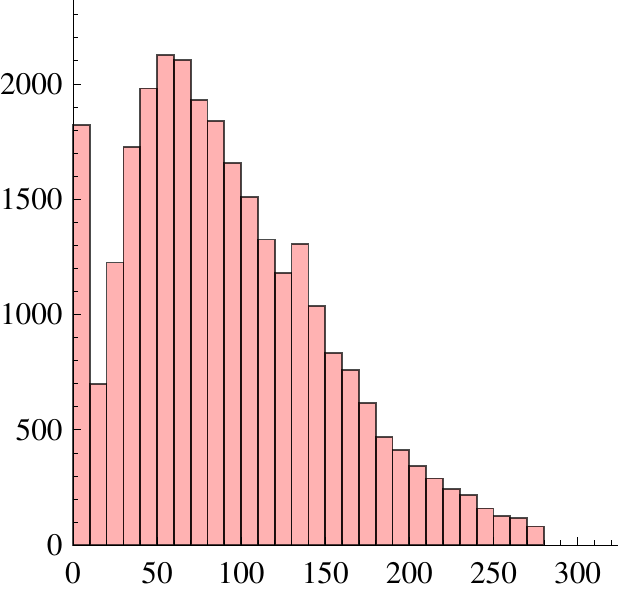}
\includegraphics[width=0.23\textwidth]{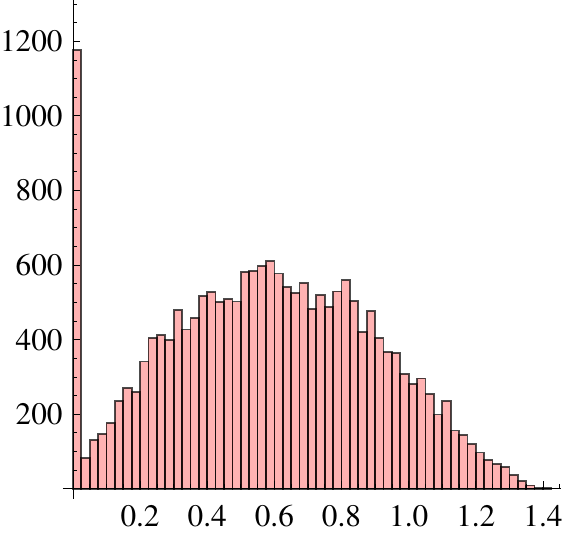}
\caption{Shapes of the distributions of $\vert M\vert$ (left column)
and $\varphi$ (right column) of the roots of $\Sigma_0$. The top
row is the $M_H=120$ GeV signal. The $WW$ and $t \bar t$
backgrounds are shown in the middle and lower rows.
$\vert M\vert$  is in GeV units.
  \label{fig:RootsofC4}}
\end{center}
\end{figure}

The distribution of absolute values and phases of the roots of 
$\Sigma_0$, that is the projections of the results of Fig.~\ref{fig:AbsPhaseCorr}
onto the $\vert M\vert$ and $\varphi$ axis, are shown in Fig.~\ref{fig:RootsofC4}.
The results for signal and $WW$ and $t\bar t$ backgrounds are significantly different.
Notice in particular how the signal has a much higher fraction than the background
of events with $\vert M\vert $ and $\varphi$ close to zero.

\begin{figure}[htbp]
\begin{center}
\includegraphics[width=0.23\textwidth]{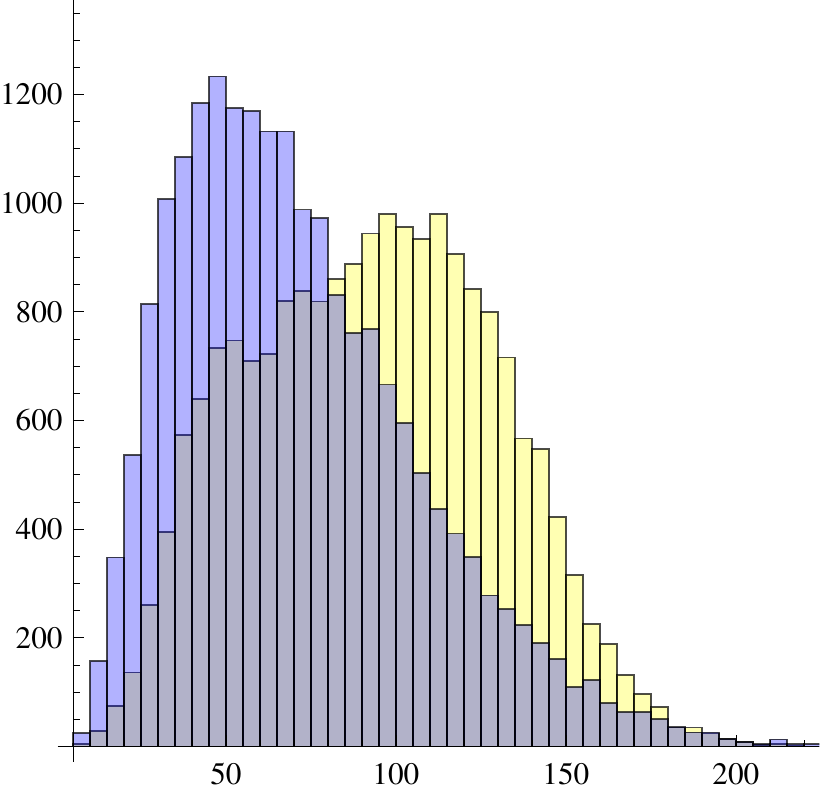}
\includegraphics[width=0.23\textwidth]{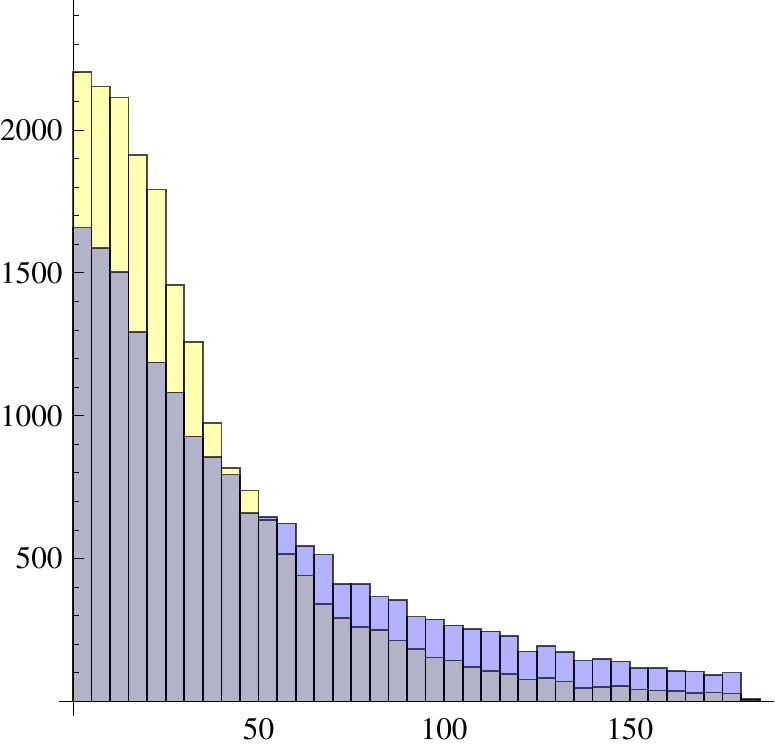}\\
\vspace{.3cm}
\includegraphics[width=0.23\textwidth]{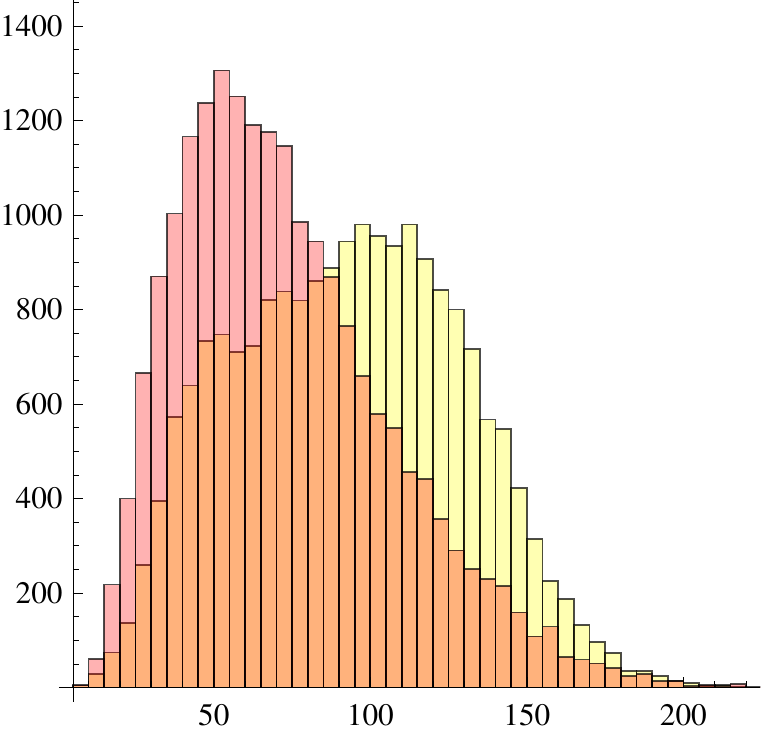}
\includegraphics[width=0.23\textwidth]{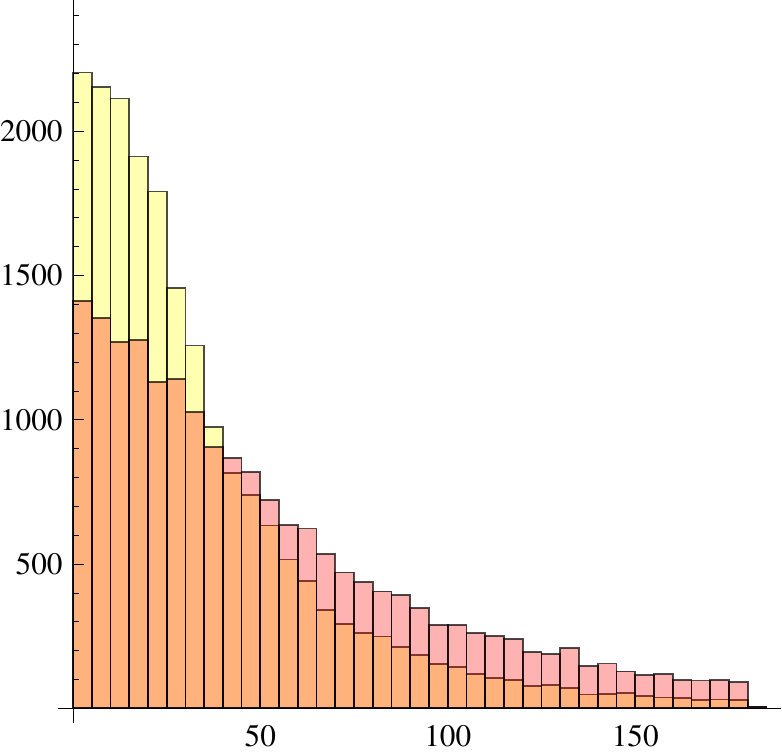}\\
\caption{Comparison of the shapes of the distributions
of $M_2$ (left column) and $M_1$ (right column) for
an $M_H=120$ GeV signal (yellow) and the $WW$ (top row)
and $t\bar t$ (lower row) backgrounds. See Eqs.~(\ref{eq:C2},\ref{eq:C1}).
Abscissae in GeV units.
  \label{fig:C2&C1}}
\end{center}
\end{figure}

The variables $M_{1,2}$ of Eqs.~(\ref{eq:C1},\ref{eq:C2}) are akin to $M_\pm$ in that
they do not refer to an ansatz mass ${\cal M}$.
In spite of their naivet\'e, these observables, particularly $M_2$, are quite 
good at telling signal from backgrounds. Their shapes for an $M_H=120$ GeV
signal and the $WW$ and $t\bar t$ backgrounds are shown in  Fig.~\ref{fig:C2&C1}.

\begin{figure}[htbp]
\begin{center}
\includegraphics[width=0.23\textwidth]{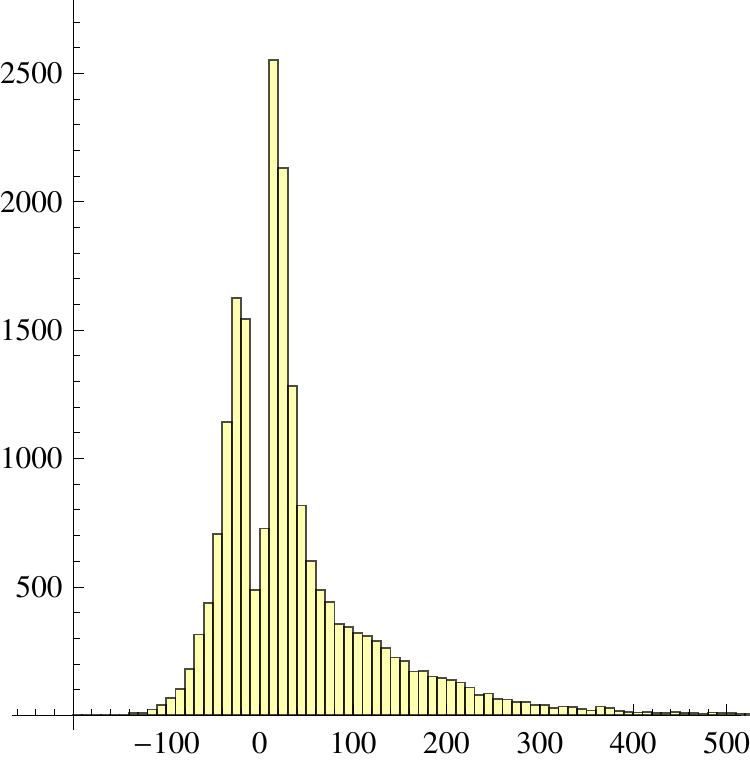}
\includegraphics[width=0.23\textwidth]{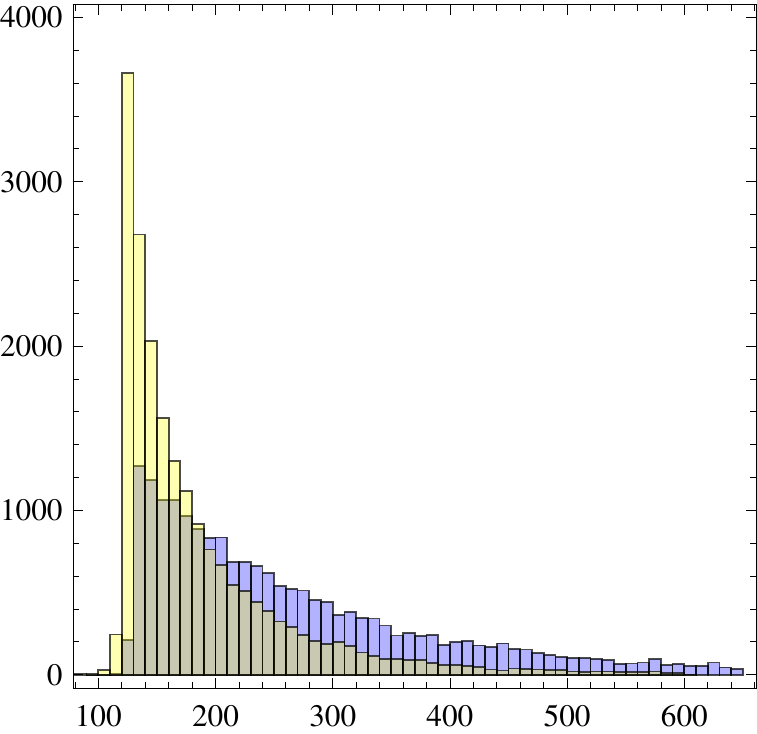}\\
\vspace{.3cm}
\includegraphics[width=0.23\textwidth]{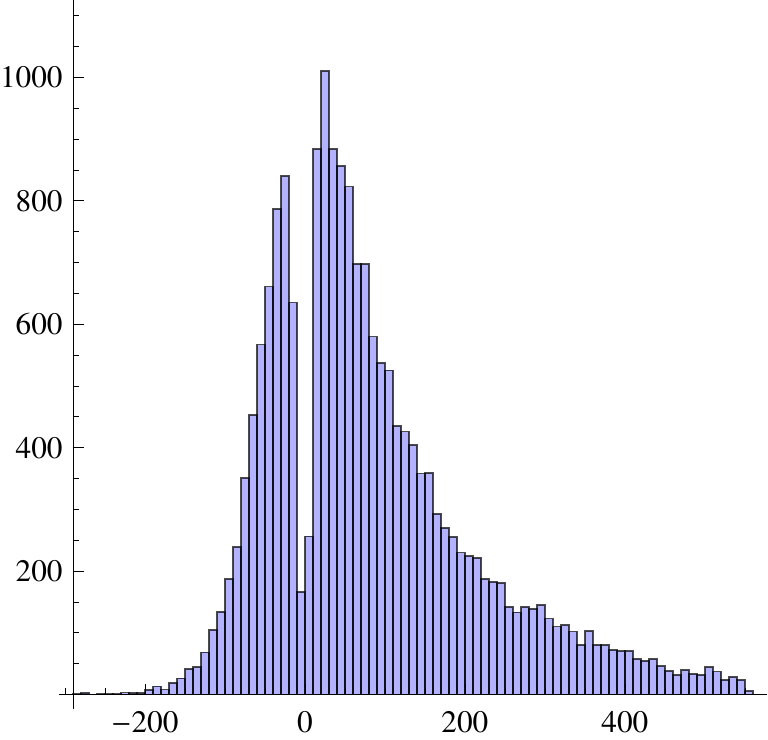}
\includegraphics[width=0.23\textwidth]{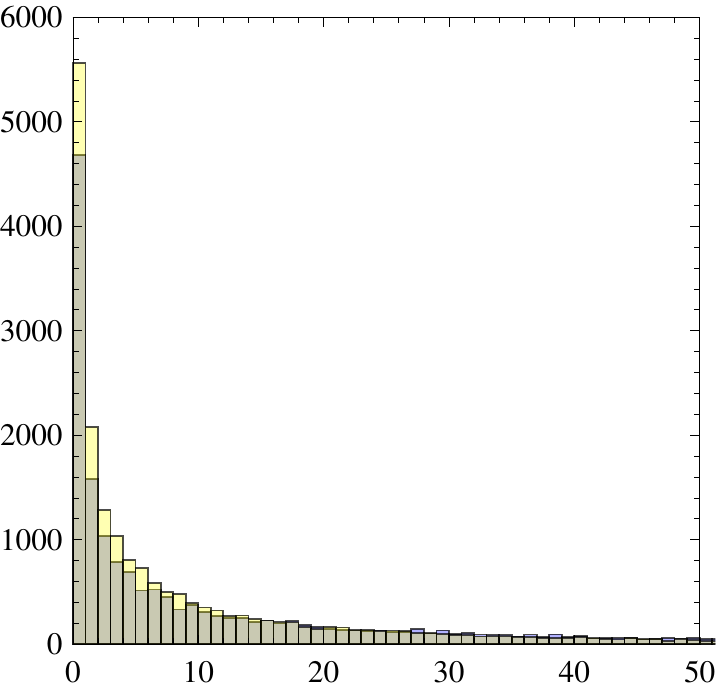}\\
\vspace{.3cm}
\includegraphics[width=0.23\textwidth]{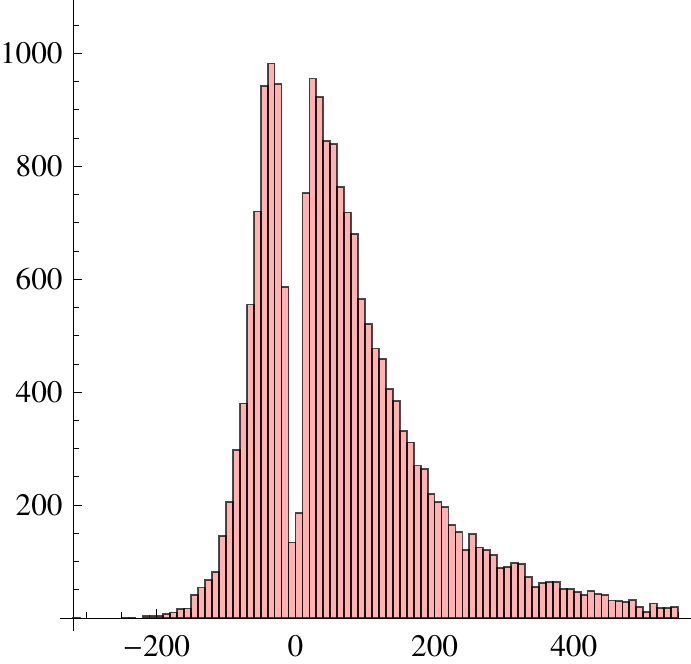}
\includegraphics[width=0.23\textwidth]{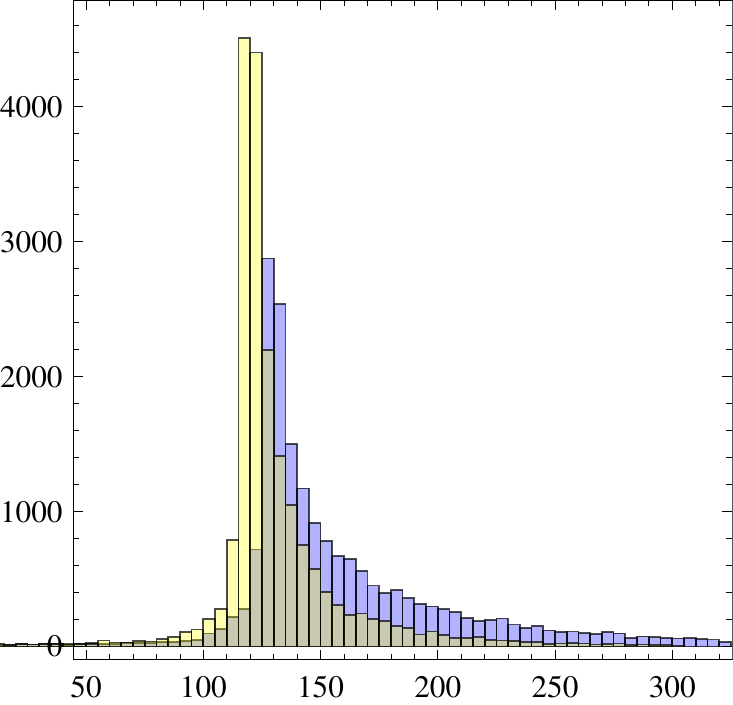}\\
\caption{Left: distributions of the variable $\tilde\Sigma_3$ of
Eq.~(\ref{eq:Sigma123}); top is the $M_H=120$ GeV signal, middle
is the $WW$ background, bottom is the $t\bar t$ one. Right:
distributions of the three roots of $\Sigma_3({\cal M})=0$,
for the quoted signal (tall and yellow) compared with the
$WW$ background. Abscissae in GeV units.
  \label{fig:C3}}
\end{center}
\end{figure}

Because the variable $\Sigma_3$ of Eq.~(\ref{eq:Sigma3})
has mass dimension 3, it is convenient to plot its sign-recalling cubic root
of Eq.~(\ref{eq:Sigma123}).
This we do in the left column of Fig.~\ref{fig:C3} for an $M_H=120$ GeV 
signal and the $WW$
and $t\bar t$ backgrounds.
The signal and the illustrated backgrounds are seen to result in
distributions with similar looks but significantly different details.
In the right column of Fig.~\ref{fig:C3} we show the three roots
of the cubic equation $\Sigma_3({\cal M})=0$, see Eq.~(\ref{eq:Sigma3}). The taller (yellow)
histograms are the $M_H=120$ GeV signal, they are compared
with those of the $WW$ background (the $t\bar t$ distributions, not
shown, differ a bit more than the $WW$ ones from the signal distributions).
The three roots $\tilde M_i$ of $\Sigma_3$, unlike the roots $M_\pm$ 
of $\Sigma_0$, are not so useful in telling signal from backgrounds.

\subsection{Correlations between partial singularity variables}

\begin{figure}[htbp]
\begin{center}
\includegraphics[width=0.23\textwidth]{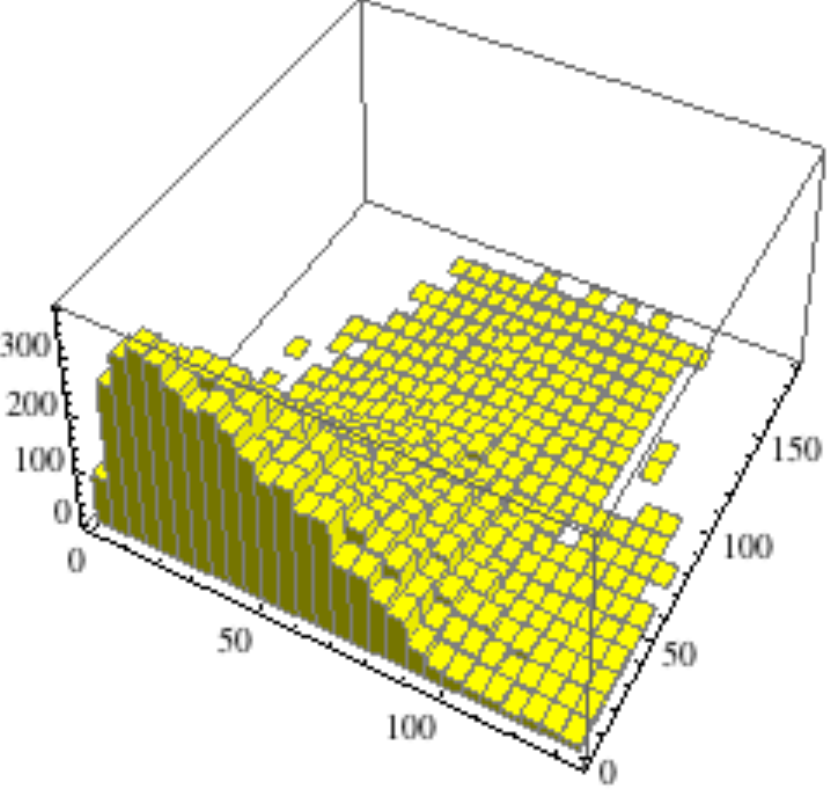}
\includegraphics[width=0.23\textwidth]{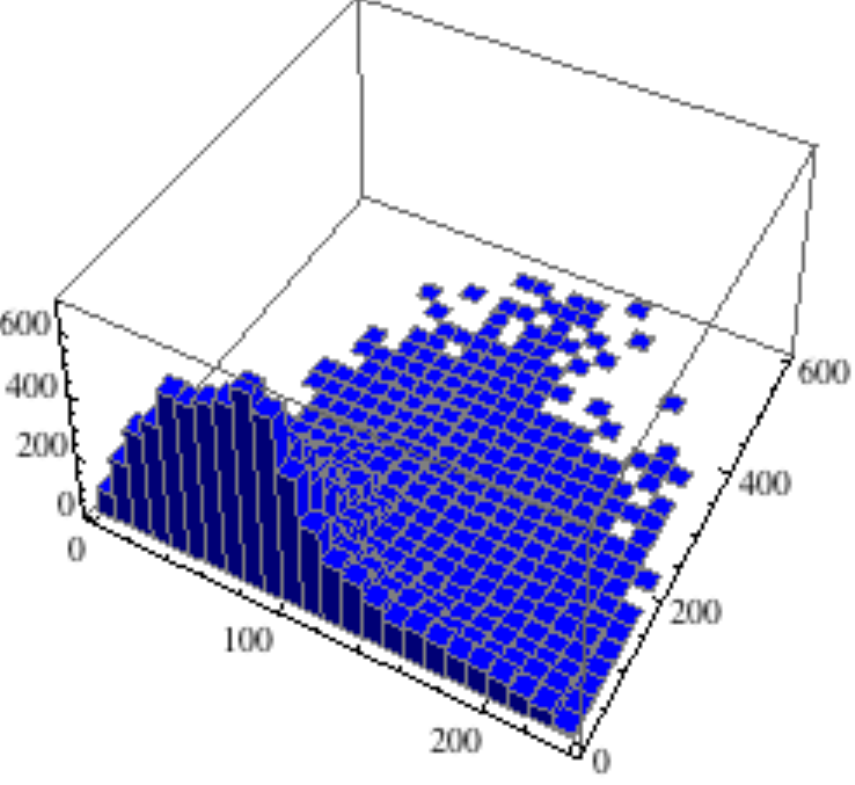}\\
\vspace{.3cm}
\includegraphics[width=0.23\textwidth]{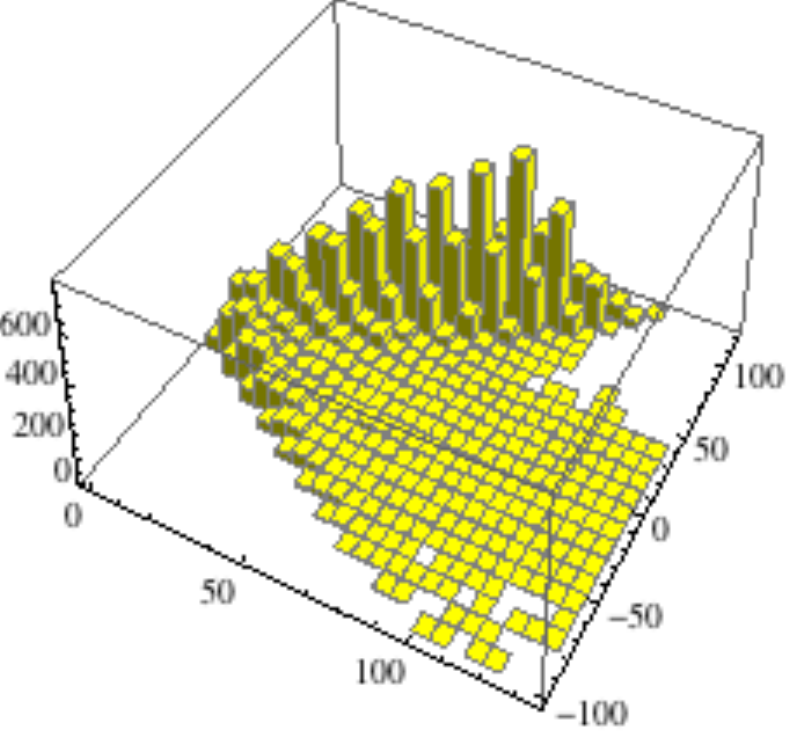}
\includegraphics[width=0.23\textwidth]{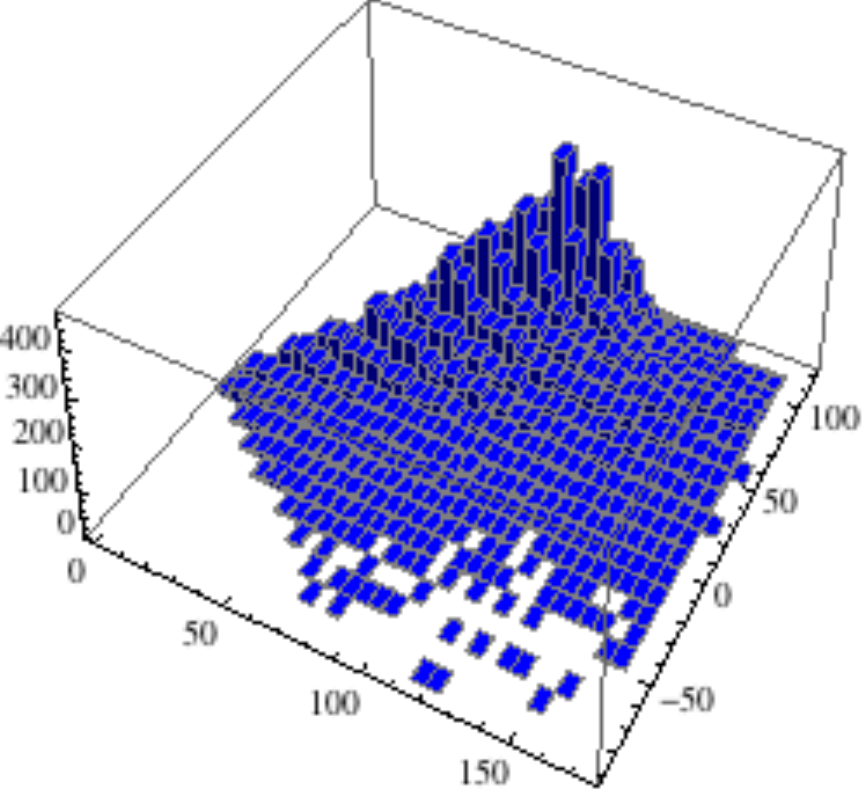}\\
\vspace{.2cm}
\includegraphics[width=0.23\textwidth]{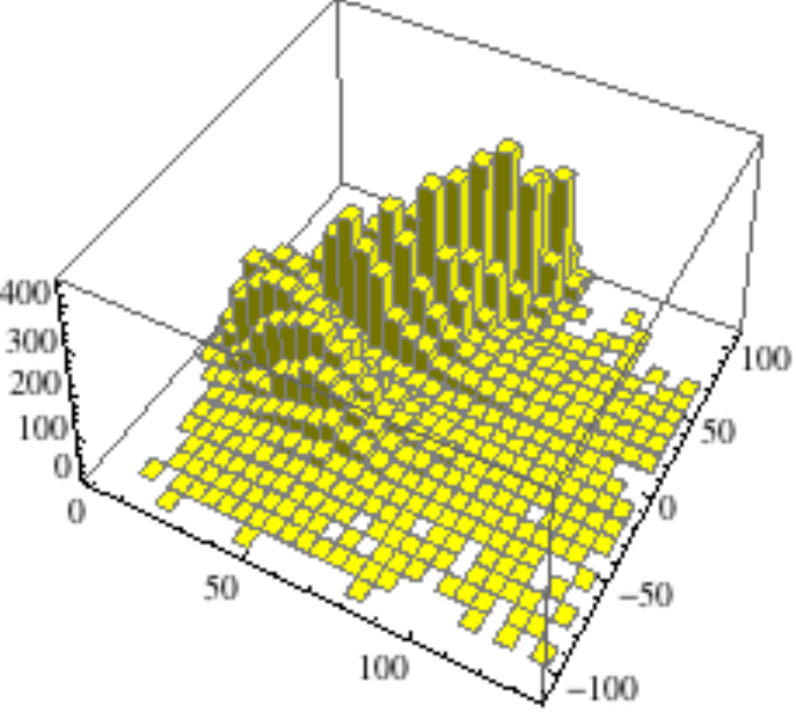}
\includegraphics[width=0.23\textwidth]{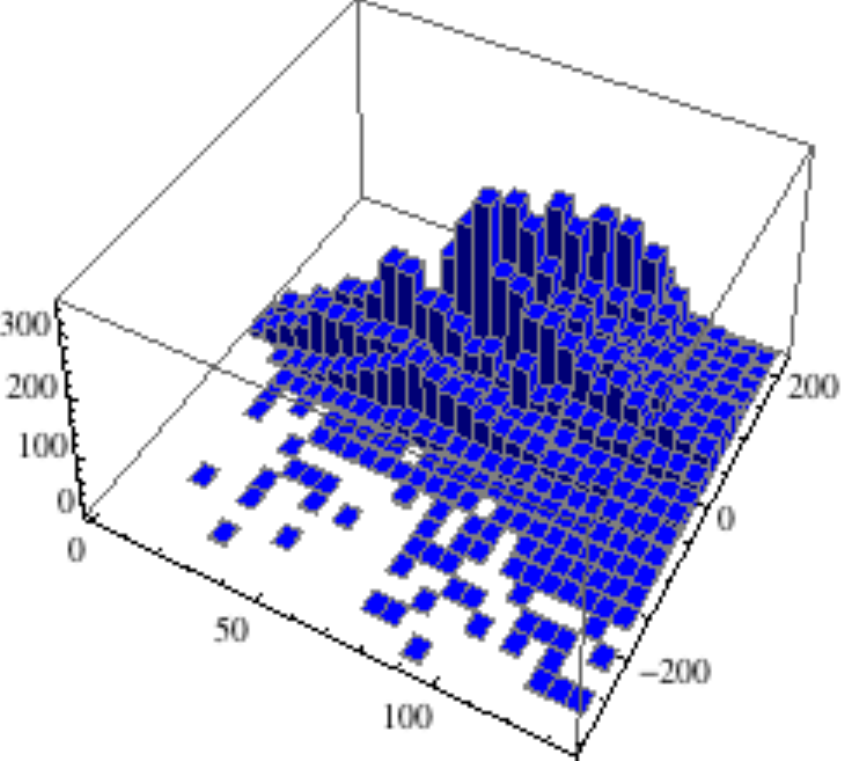}\\
\caption{Top left: Histogram of the values of $\tilde\Sigma_0$ and 
$\Sigma_1$, for a
signal with $M_H=120$ GeV. Top right: for the $WW$ background.
Middle row: the same as the first row, for $\{\tilde\Sigma_0,\Sigma_2\}$.
Third row:  $\{\tilde\Sigma_0,\Sigma_3\}$. All plots are made
for ${\cal M}=120$ GeV and all horizontal scales (which do not
have the same extent in all figures) are in GeV units.
The relevant definitions are in
 Eqs.~(\ref{eq:Sigma4}, \ref{eq:Sigma3}, \ref{eq:Sigma123}).
  \label{fig:Scatters}}
  \end{center}
\end{figure}

A question of practical interest is the extent to which the
$C_0$ and $C_{1,2,3}$ distributions of Eqs.~(\ref{eq:Sigma4},\ref{eq:Cs})
are correlated, for a putative
signal, and for the backgrounds. It can be answered, pictorially,
by contemplating 2D histograms in the three $\{C_0,C_i\}$ planes.
In the $\{C_0,C_3\}$ case, for which the functions depend
on  ${\cal M}$, we choose to plot the results
in the $\{\tilde\Sigma_0,\tilde\Sigma_3\}$ plane, see Eqs.~(\ref{eq:Sigma4},\ref{eq:Sigma3}).
The singularity, for the correct assignment ${\cal M}=M_H$, is at the origin of the plane.
For this to be the case in the two other pairs, we plot results
for  $\{\tilde\Sigma_0,\Sigma_1\}$ and  $\{\tilde\Sigma_0,\Sigma_2\}$, 
see Eq.~(\ref{eq:Sigma123}). All this we do in Fig.~(\ref{fig:Scatters}).

Two conclusions are to be extracted from the quoted figure,
after noticing that its horizontal scales for signal and 
backgrounds are not always the same. 
The signal variable pairs are quite correlated, with the exception of 
$\{\Sigma_0,\Sigma_1\}$.
The $WW$ background is less correlated and its distribution is
significantly different from that of the signal, for all variable pairs.
These statements are more so for the $t\bar t$ background,
which we have not shown.

\subsection{Complete CM singularity conditions}

\begin{figure}[htbp]
\begin{center}
\includegraphics[width=0.23\textwidth]{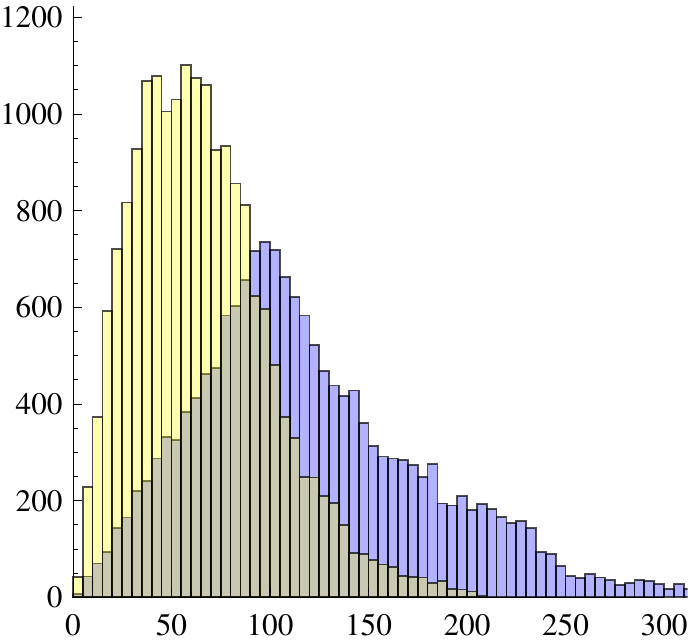}
\includegraphics[width=0.23\textwidth]{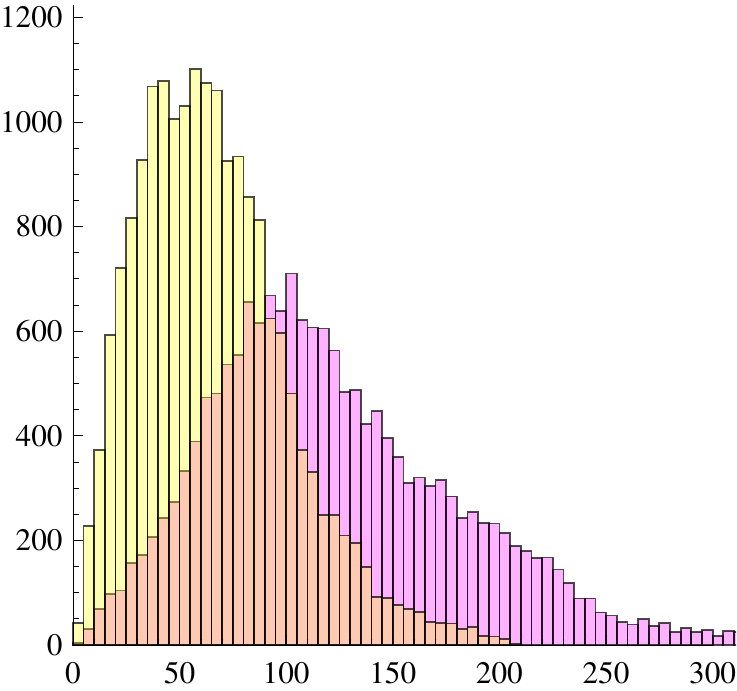}\\
\vspace{.3cm}
\includegraphics[width=0.23\textwidth]{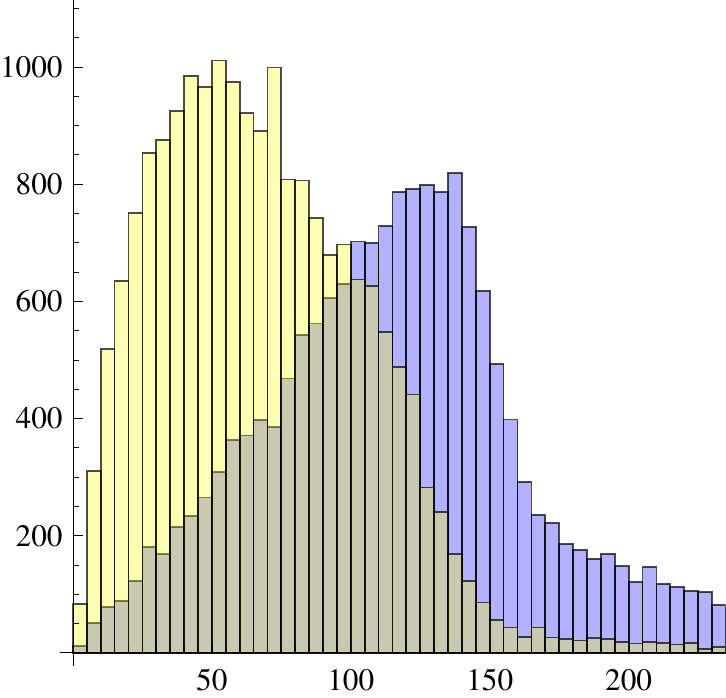}
\includegraphics[width=0.23\textwidth]{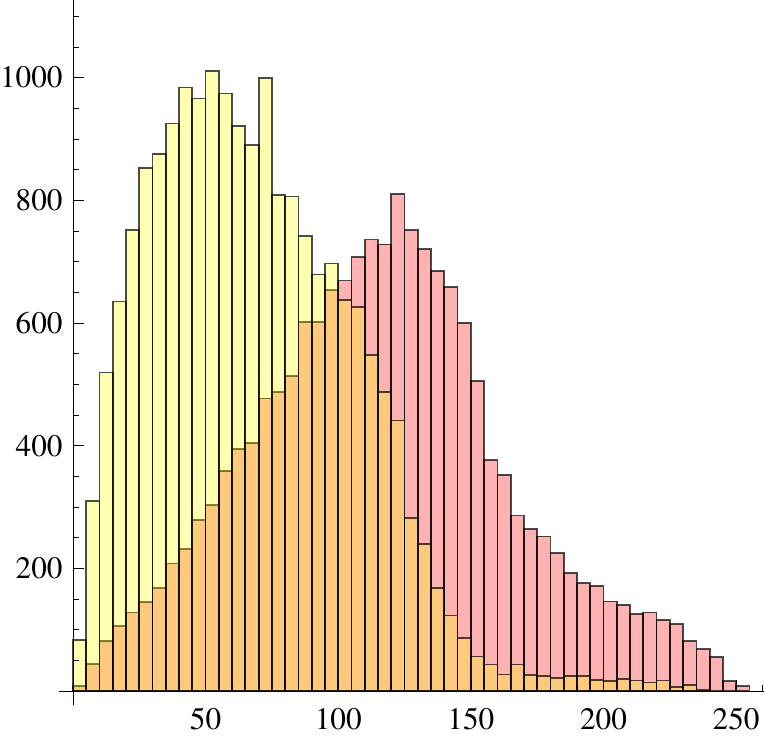}\\
\vspace{.2cm}
\includegraphics[width=0.23\textwidth]{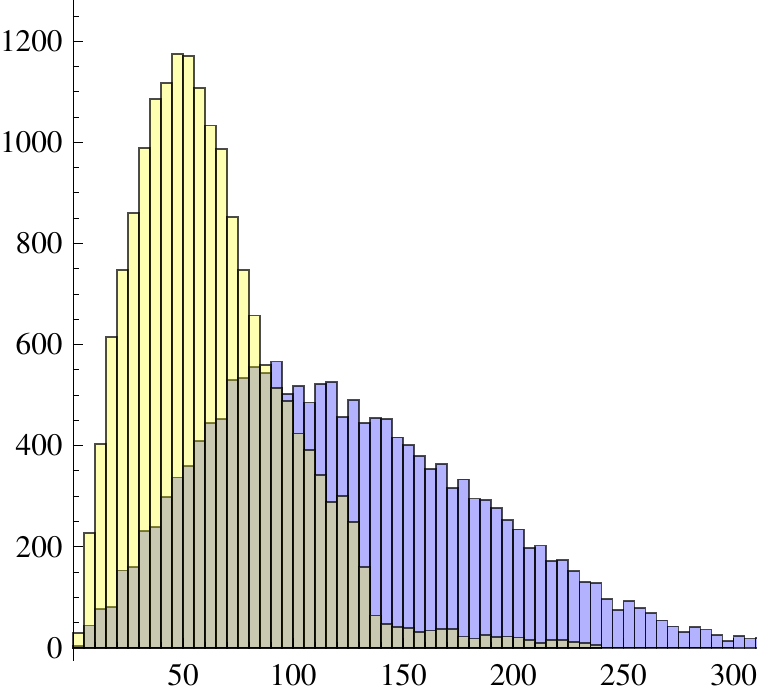}
\includegraphics[width=0.23\textwidth]{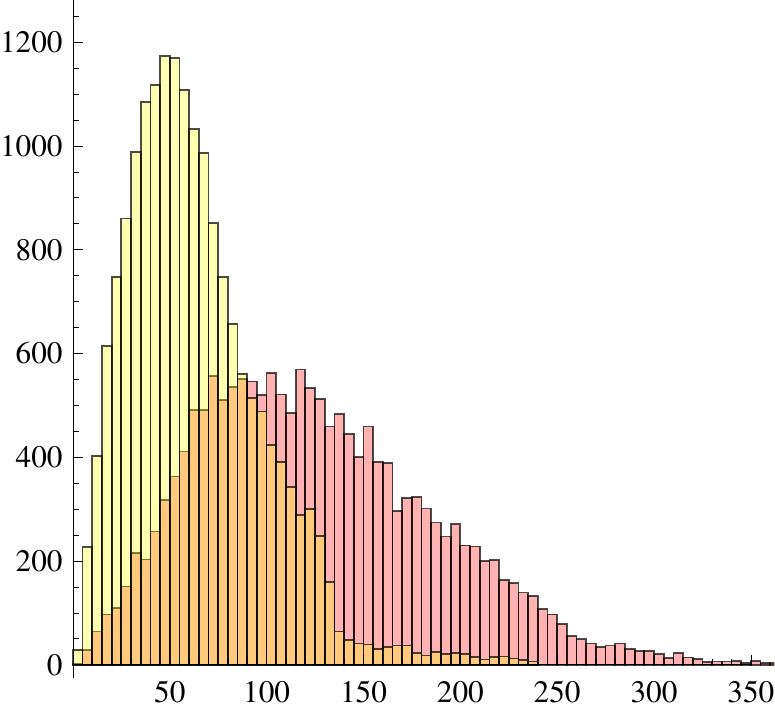}\\
\caption{Histograms of the singularity variables $D_i$ of
Eq.~(\ref{eq:Ds}). Top: $D_1$. Middle: $D_2$. Bottom: $D_3$.
Left: comparison of the signal for ${\cal M}=M_H=120$ GeV
(the distribution peaking closer to zero) with the $WW$
background. Right: comparison of the same
signal with the $t\bar t$ background. Abscissae in GeV units.
  \label{fig:ToZeros}}
\end{center}
\end{figure}

To construct true singularity variables that reflect a complete set
of singularity conditions we must exploit a measure of the distance
between a data point (its values of $\vec k$ and $\vec l\,$) and one
of the three center-of-mass singularities which, for ${\cal M}=M_H$,
are the points $\{0,0\}$ of the planes  $\{\Sigma_0,\Sigma_i \}$, $i=1$ to 3. 
These are the quantities $D_i$ defined in Eqs.~(\ref{eq:Ds}).
Distributions of these variables  are shown
in Fig.~\ref{fig:ToZeros}. The three choices appear to be comparably
efficient at telling signal from backgrounds.

\begin{figure}[htbp]
\begin{center}
\includegraphics[width=0.23\textwidth]{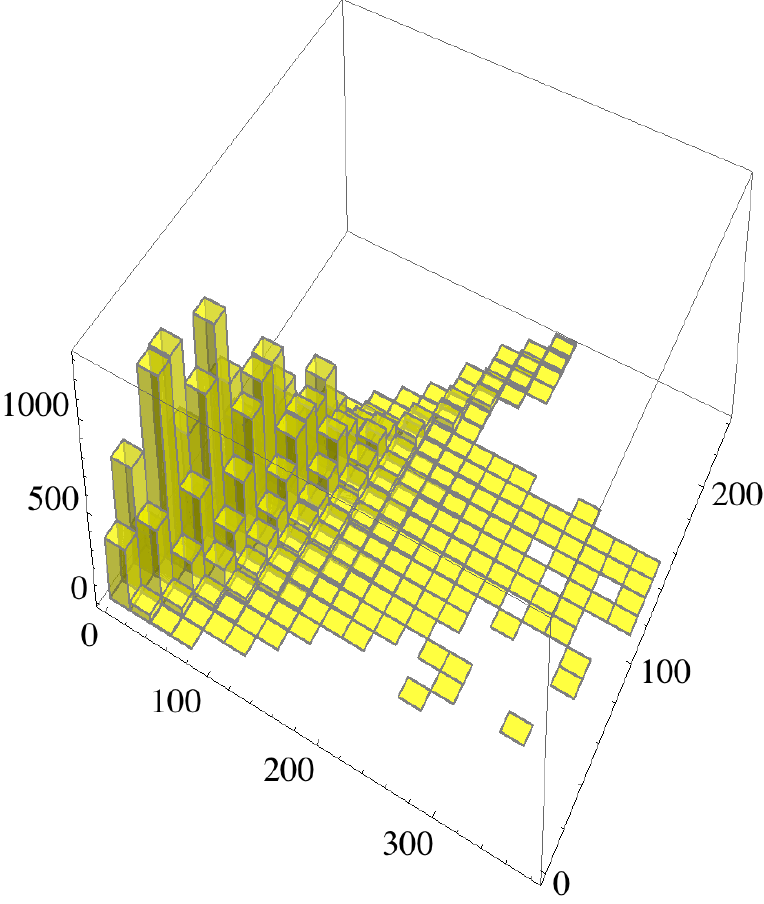}
\includegraphics[width=0.23\textwidth]{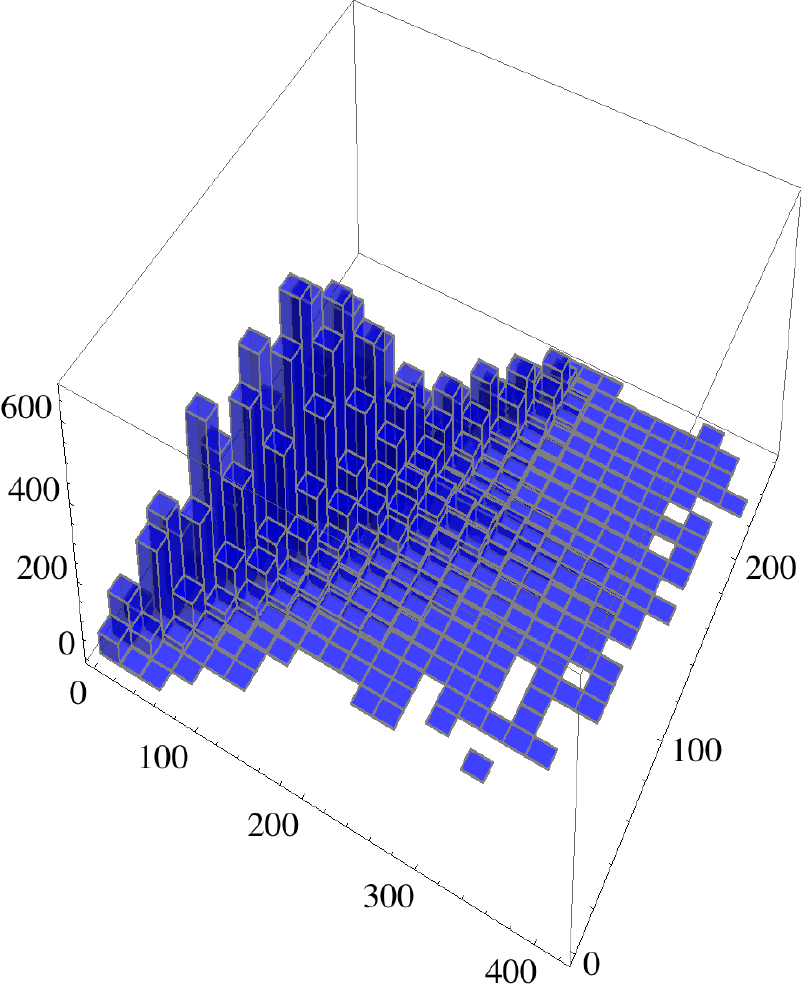}\\
\vspace{.3cm}
\includegraphics[width=0.23\textwidth]{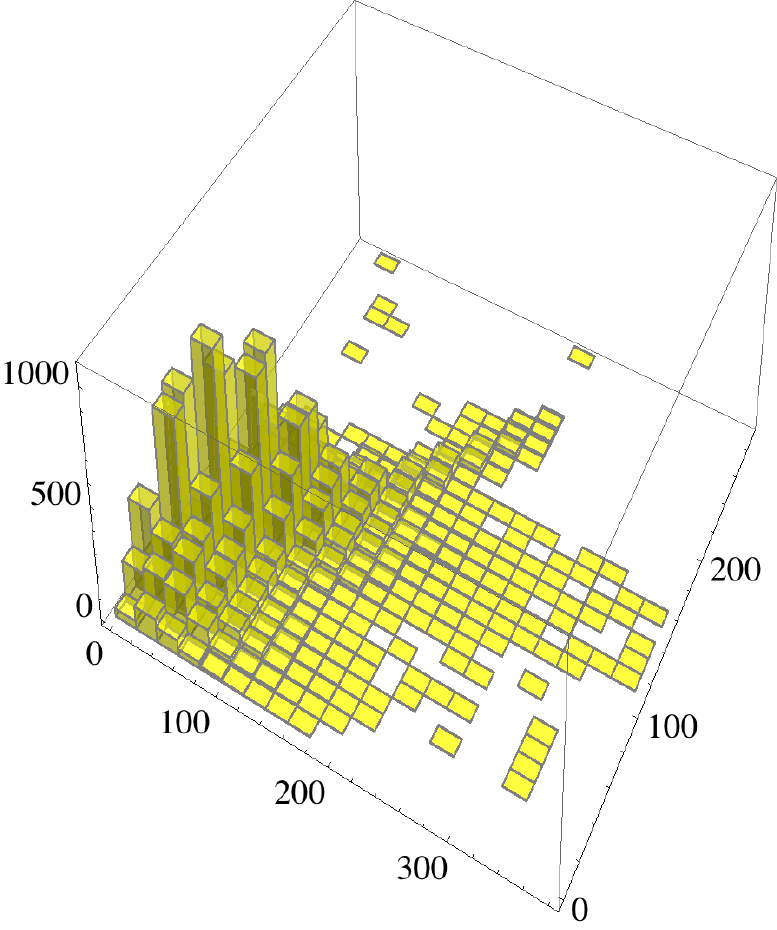}
\includegraphics[width=0.23\textwidth]{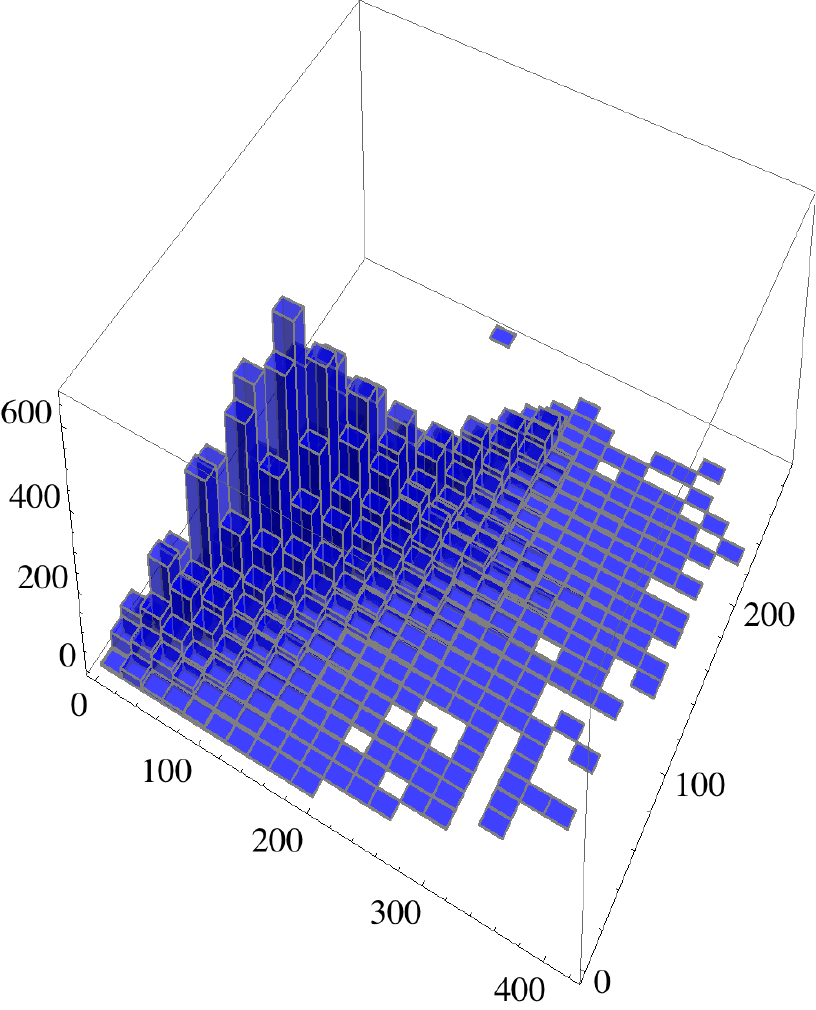}\\
\vspace{.2cm}
\includegraphics[width=0.23\textwidth]{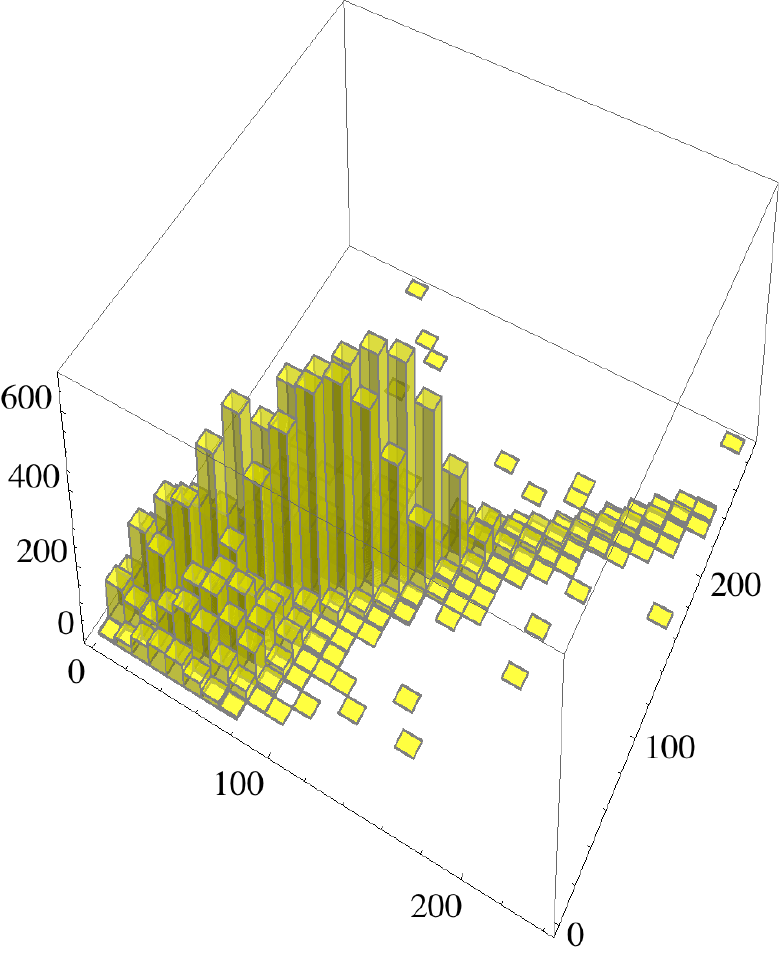}
\includegraphics[width=0.23\textwidth]{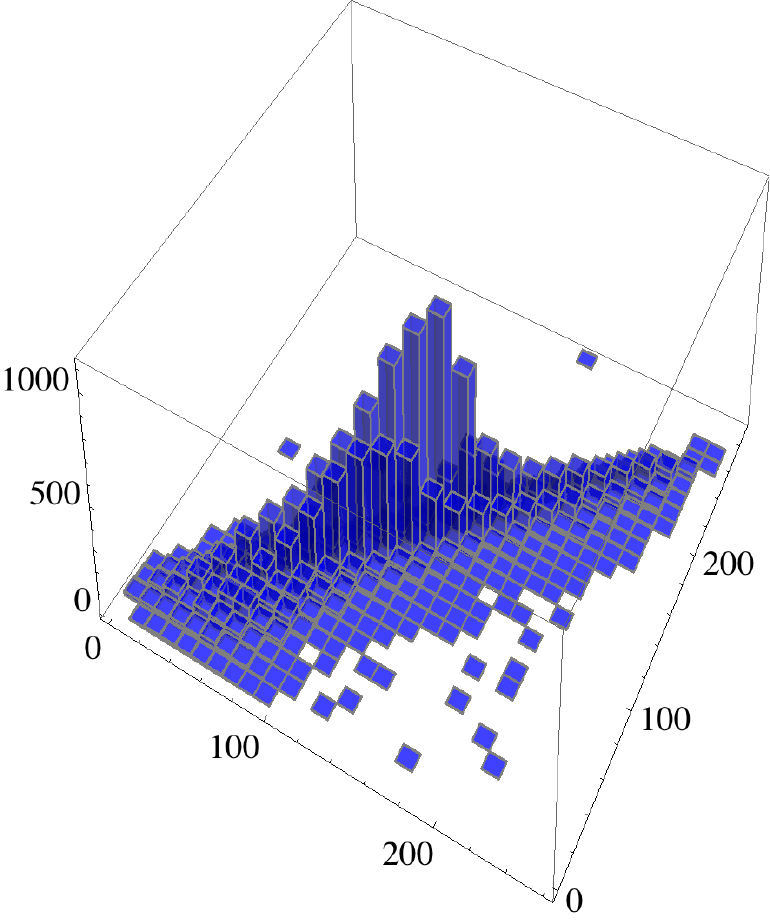}\\
\caption{Histograms of the correlations between the singularity variables $D_i$ of
Eq.~(\ref{eq:Ds}). Left: the signal for $M_H=120$ GeV. Right: $WW$ background.
Top: $\{D_1,D_2\}$. Middle: $\{D_1,D_3\}$. Bottom: $\{D_2,D_3\}$.
Horizontal axes is GeV units.
  \label{fig:D1D2D3Correlations}}
\end{center}
\end{figure}

The $\{D_i,D_j\}$ correlations are illustrated in Fig.~\ref{fig:D1D2D3Correlations}
for a signal with $M_H=120$ GeV and for the $WW$ background, all with 
 ${\cal M}=M_H$. Only the $\{D_2,D_3\}$ correlation is strong.
In all cases the signal and background results are fairly distinct. This is even
more so for the $t\bar t$ background, results for which we do not show.

\section{Data analysis beyond the CM approximation}
\label{BoostedData}

We have derived three longitudinally boost-invariant singularity variables.
The first of them, $\Delta_1$ is algebraically simple
and factorizable as $\Delta_1\!\propto\! M^2\,\Sigma$, see Eq.~(\ref{eq:Delta1}).
The mass dimension of $C'_1$ is 1 and its degree in $\xi$ is 2.
The corresponding numbers for $C'_0$ are 6 and 8, see Eqs.~(\ref{eq:CPrimes}).
The mass dimension of their resultant is $1\times 8+6\times 2=20$.
The mass dimensions of $M$ and $\Sigma$ are 4 and 12,
respectively. It is therefore convenient to discuss the results in terms of 
$M^{1/4},\,\Sigma^{1/12}$ and 
\begin{equation}
\tilde\Delta_1\equiv\Delta_1^{1/20},
\label{DeltaTilde1}
\end{equation}
where the root is always real, since $\Delta_1$ is always positive.

\begin{figure}[htbp]
\begin{center}
\includegraphics[width=0.23\textwidth]{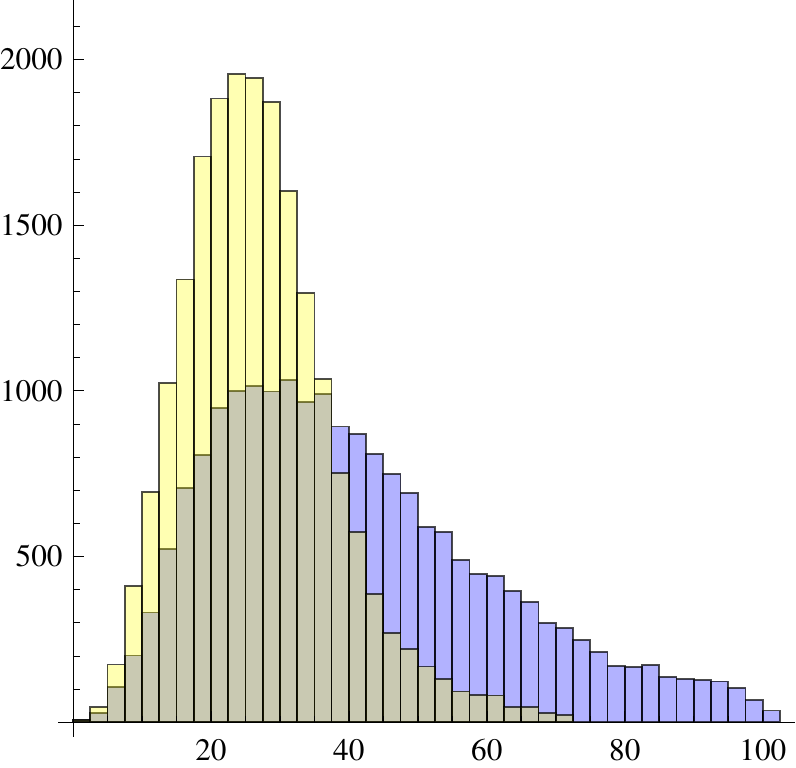}
\includegraphics[width=0.23\textwidth]{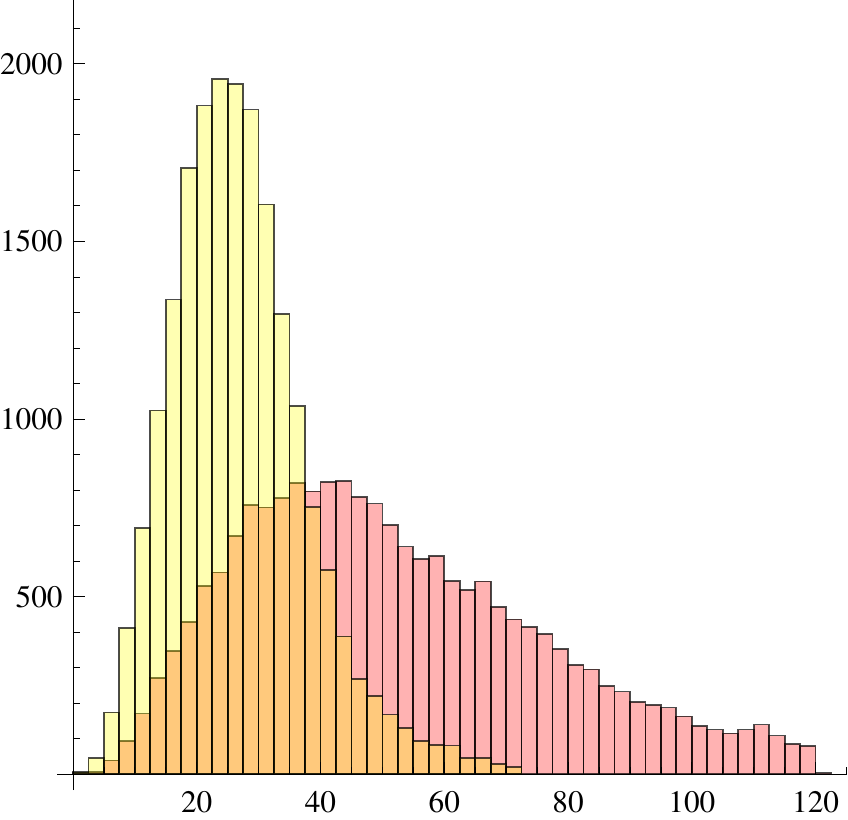}\\
\vspace{.3cm}
\includegraphics[width=0.23\textwidth]{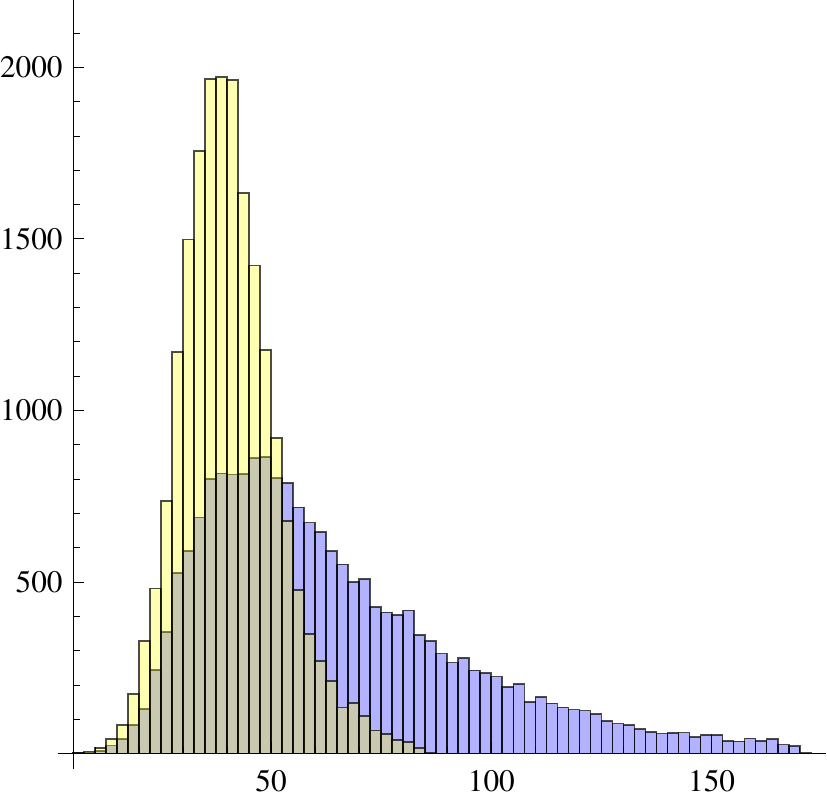}
\includegraphics[width=0.23\textwidth]{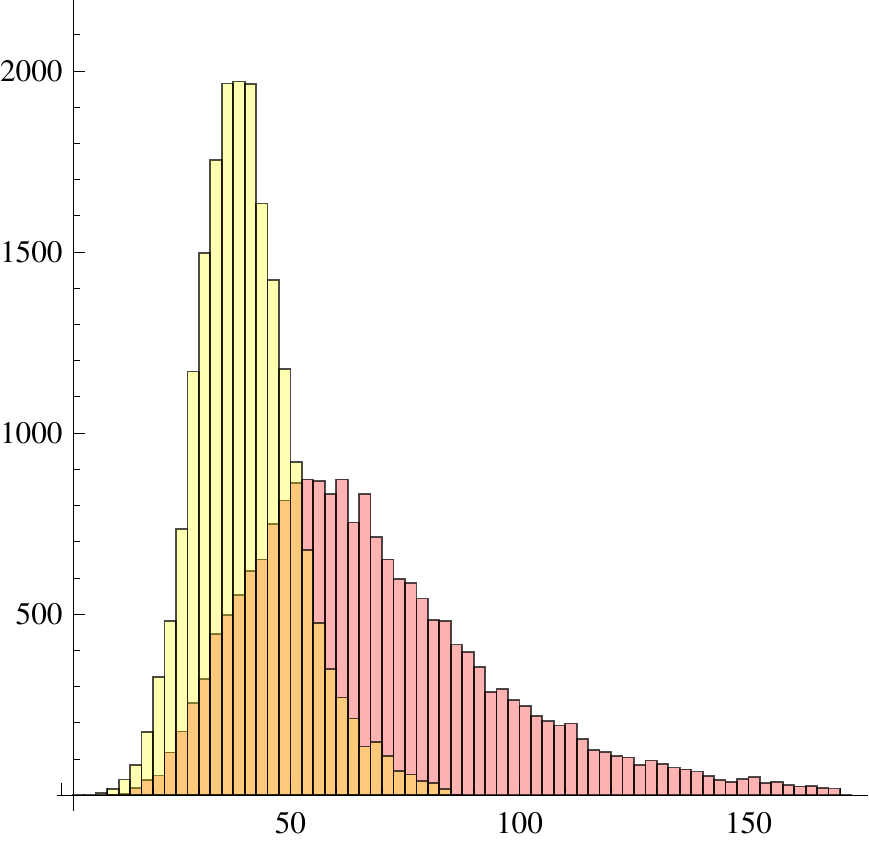}\\
\vspace{.2cm}
\includegraphics[width=0.23\textwidth]{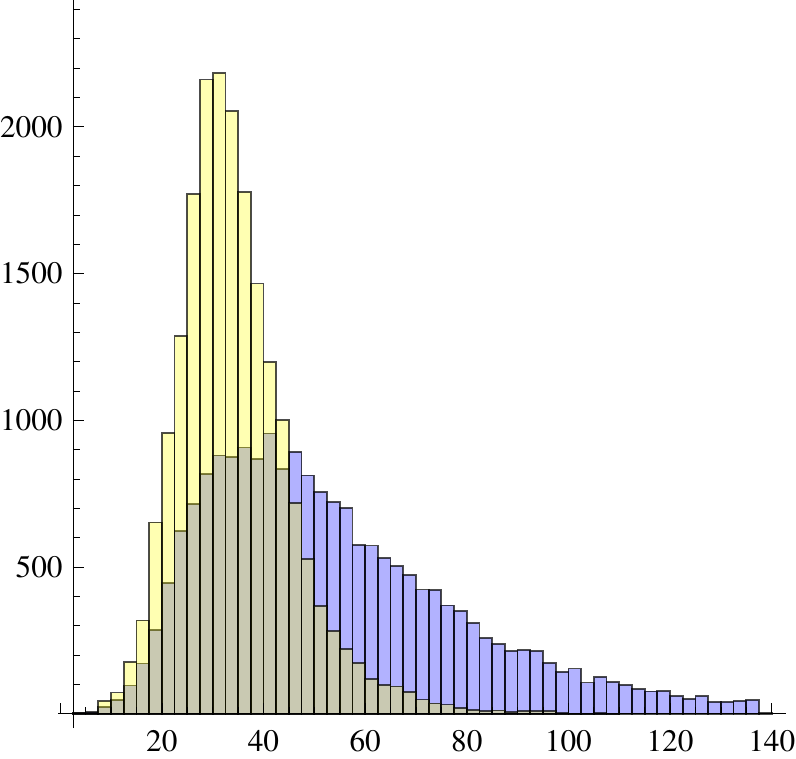}
\includegraphics[width=0.23\textwidth]{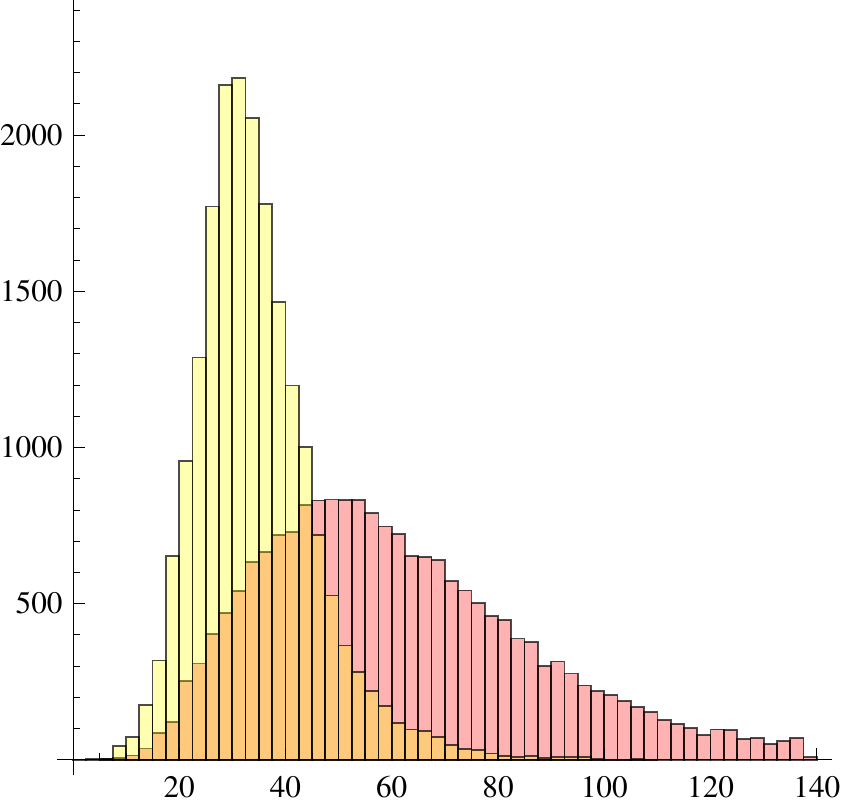}\\
\caption{Histograms of the singularity variables 
$M^{1/4},\Sigma^{1/12}$ and $\tilde\Delta_1$ of
Eq.~(\ref{eq:Delta1}). Top: $M$. Middle: $\Sigma$. Bottom: $\Delta_1$.
Left: comparison of the shape of the signal for ${\cal M}=M_H=120$ GeV
(the distribution peaking closer to zero) with the shape of the $WW$
background. Right: comparison of the same
signal shape with that of the $t\bar t$ background. Abscissae in GeV units.
  \label{fig:Dprime1}}
\end{center}
\end{figure}

\begin{figure}[htbp]
\begin{center}
\includegraphics[width=0.35\textwidth]{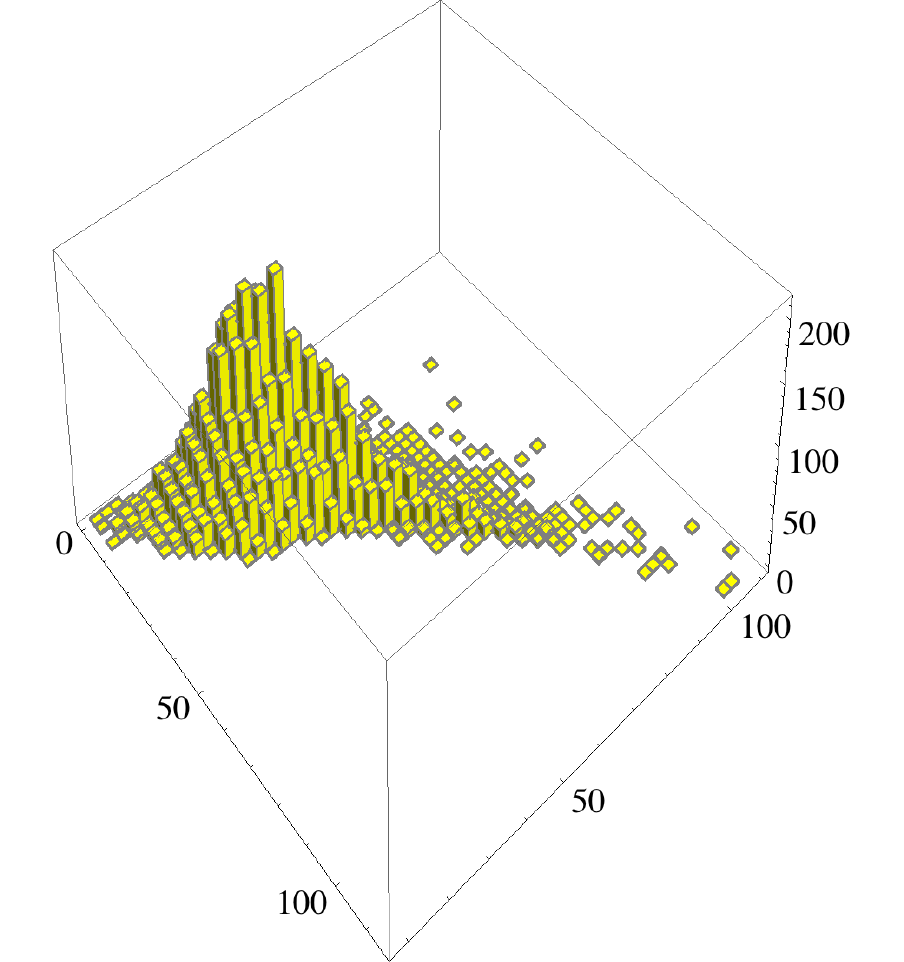}\\
\includegraphics[width=0.35\textwidth]{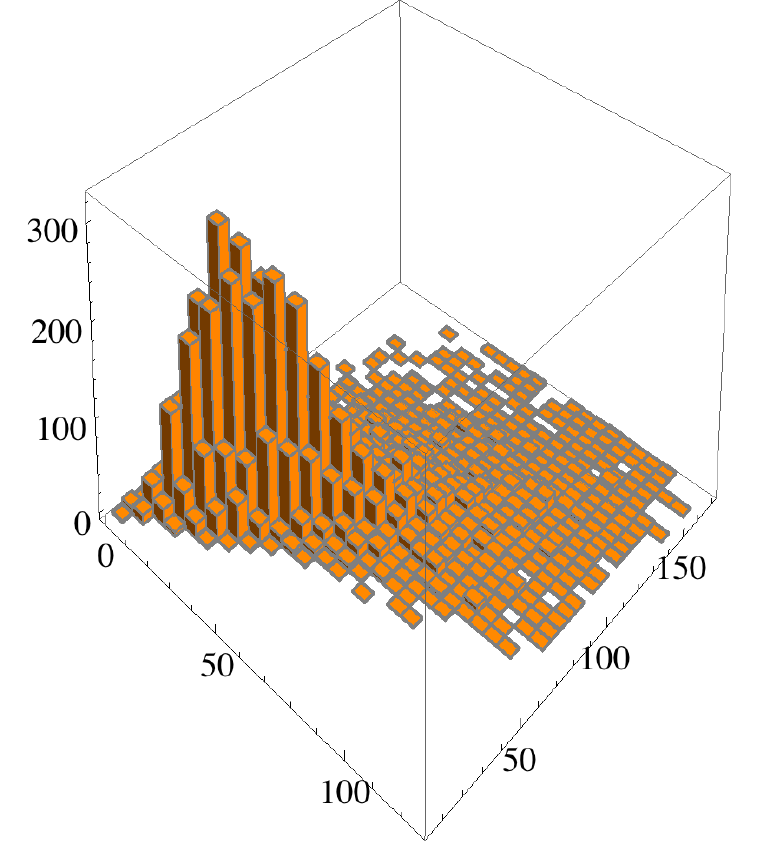}\\
\includegraphics[width=0.38\textwidth]{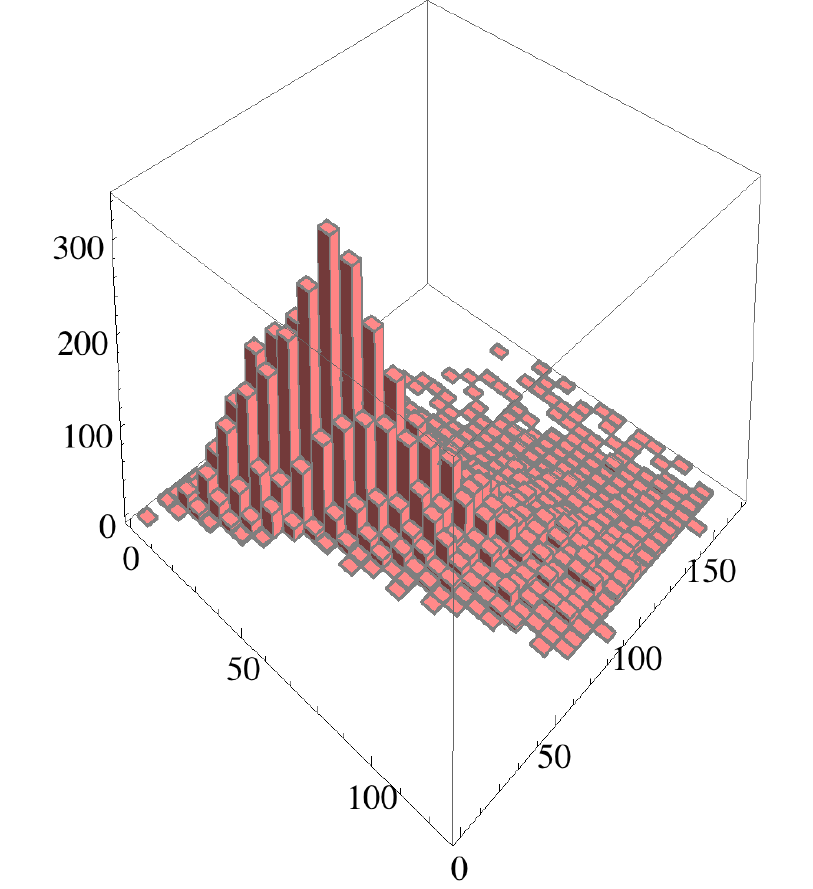}\\
\caption{Correlations between $M^{1/4}$ and $\Sigma^{1/12}$.
Top: The Signal for ${\cal M}=M_H=120$ GeV.
Middle:  $WW$ background. Bottom: $t \bar t$ background.
The horizontal scales are in GeV and do not have the same extent
in all figures.
  \label{fig:Dprime1Corr}}
\end{center}
\end{figure}

For the quoted variables we show in Fig.~\ref{fig:Dprime1} histograms comparing 
the quite distinct
shapes of the distributions for a signal of a Higgs boson of mass
$M_H=120$ GeV and the $WW$ and $t\bar t$ backgrounds, for
${\cal M}=M_H$ in the case of $\Sigma$ and $\Delta_1$. The distributions
of $M^{1/4}$ and $\Sigma^{1/12}$ --and consequently $\tilde\Delta_1$--
are similar, since the first two variables are correlated.
The correlations, shown in Fig.~(\ref{fig:Dprime1Corr}), are
not as strong as one might have suspected on the basis of
Fig.~\ref{fig:Dprime1} . The correlations are weaker for the signal than they
are for the background, except in the high-mass tails of the background
distributions. 

\begin{figure}[htbp]
\begin{center}
\includegraphics[width=0.23\textwidth]{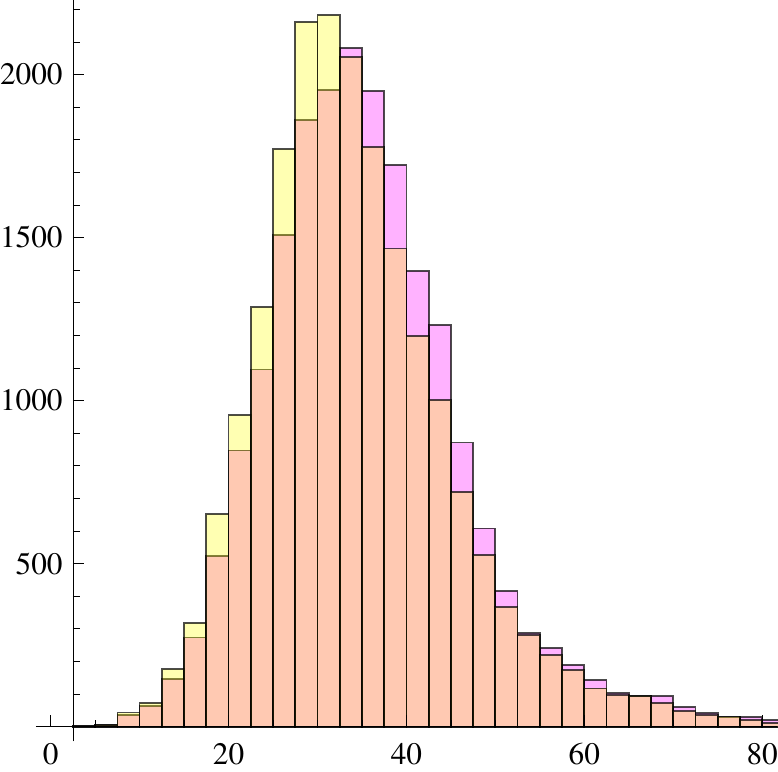}
\includegraphics[width=0.23\textwidth]{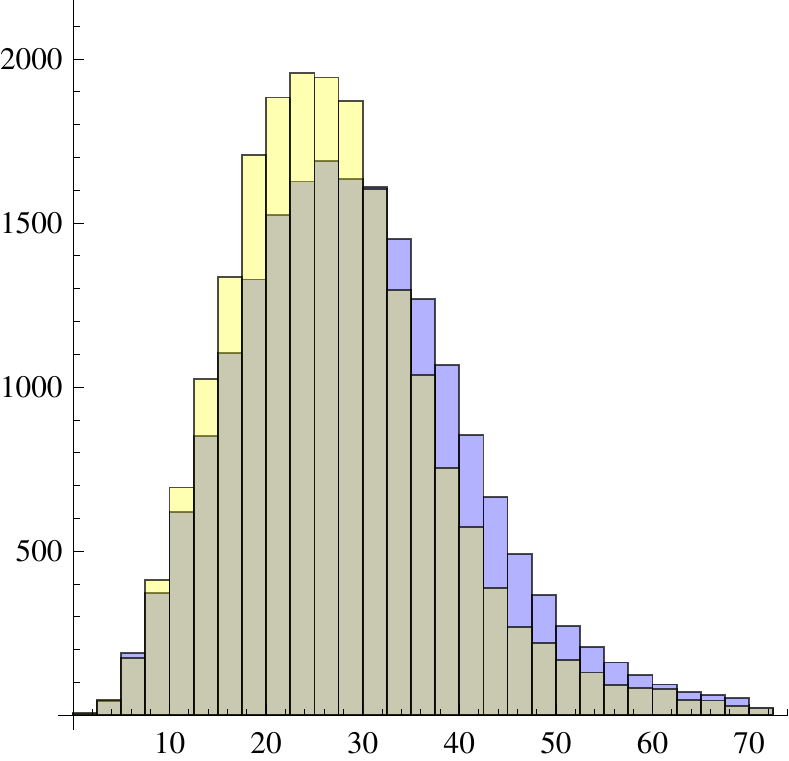}\\
\includegraphics[width=0.23\textwidth]{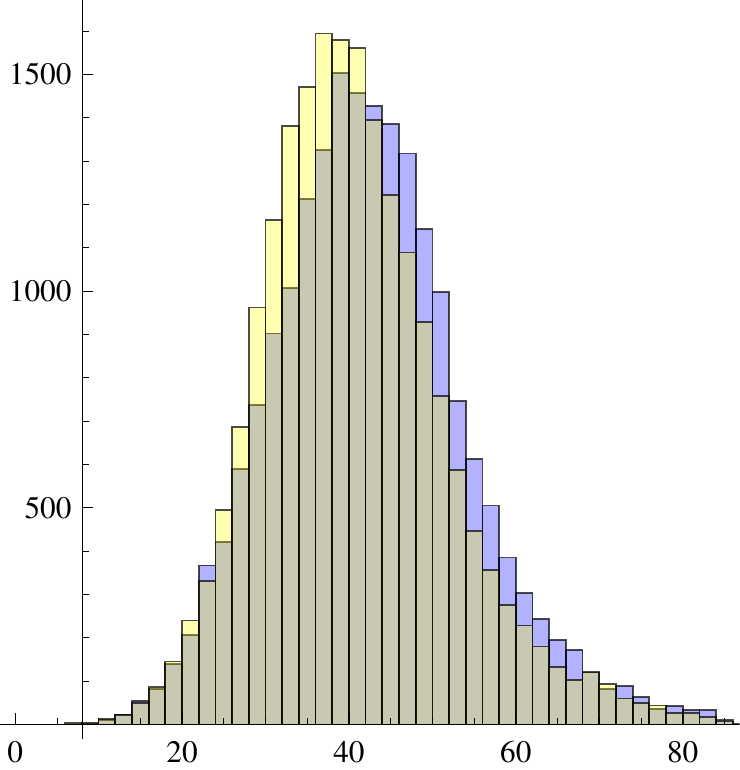}
\includegraphics[width=0.23\textwidth]{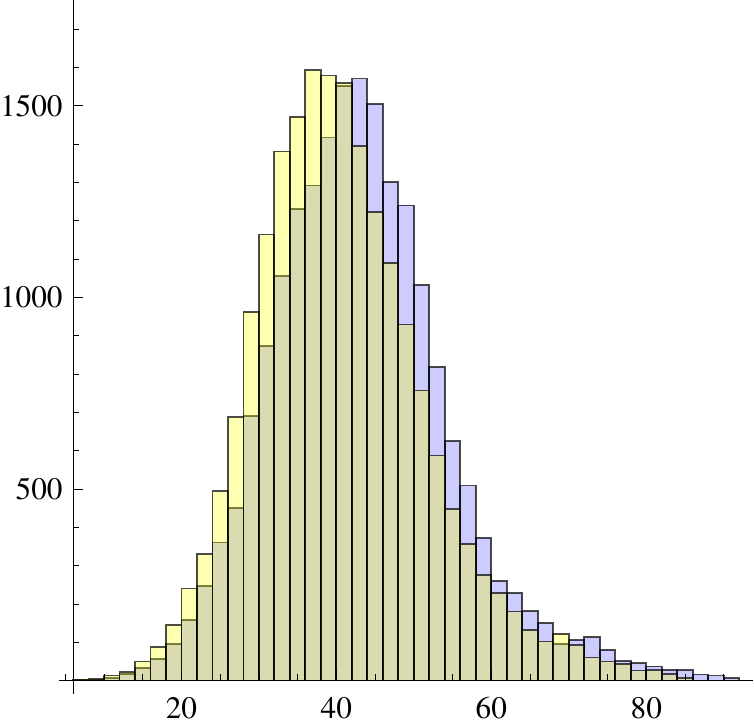}
\caption{Top left: The distribution of $\Sigma^{1/12}$ for a signal with $M_H=120$ GeV.
The (yellow) histogram peaking at a smaller value of $\Sigma$ 
 is for the correct ${\cal M}=M_H$. The other histogram is for ${\cal M}=140$ GeV.
 Top right: The distribution of $M^{1/4}$ for ``data" with $M_H=120$ and $M_H=140$ GeV.
 Bottom left: The distribution of $\Sigma^{1/12}$ for the same data, with ${\cal M}=M_H$
 in each case.
 Bottom right: The same as the last entry, for $\tilde\Delta_1$. In all cases
 the lower-peaking histogram his for $M_H=120$ GeV.
 The horizontal scales are in GeV.
\label{fig:SigmaandcalM}}
\end{center}
\end{figure}

The ability of $\Sigma$ and its factors
to tell apart diverse masses (120 and 140 GeV in the coming instance) 
is studied in Fig.~\ref{fig:SigmaandcalM}. The top left figure, histogramming
$\Sigma^{1/12}$, is for a fixed $M_H=120$
GeV, with ${\cal M}=M_H$ or 140 GeV. The other tree figures, for
$M^{1/4},\,\Sigma^{1/12}$ and $\tilde\Delta_1$ are for $M_H=120$, 140 GeV
with, in all cases, ${\cal M}=M_H$. The function $\Sigma({\cal M})$ is quadratic
in ${\cal M}$ so that its roots, in analogy with $M_\pm$ in Eq.~(\ref{eq:Sigma4})
can be made explicit. But they are not very efficient at telling apart signal from backgrounds,
nor at zooming into a value of $M_H$. Thus, we do not show results for them.

We conclude that the singularity variable $\Delta_1$ and its factors are strong 
boost-invariant tools to tell signal from backgrounds, and is
not very stringent in constraining the value of $M_H$.

\begin{figure}[htbp]
\begin{center}
\includegraphics[width=0.23\textwidth]{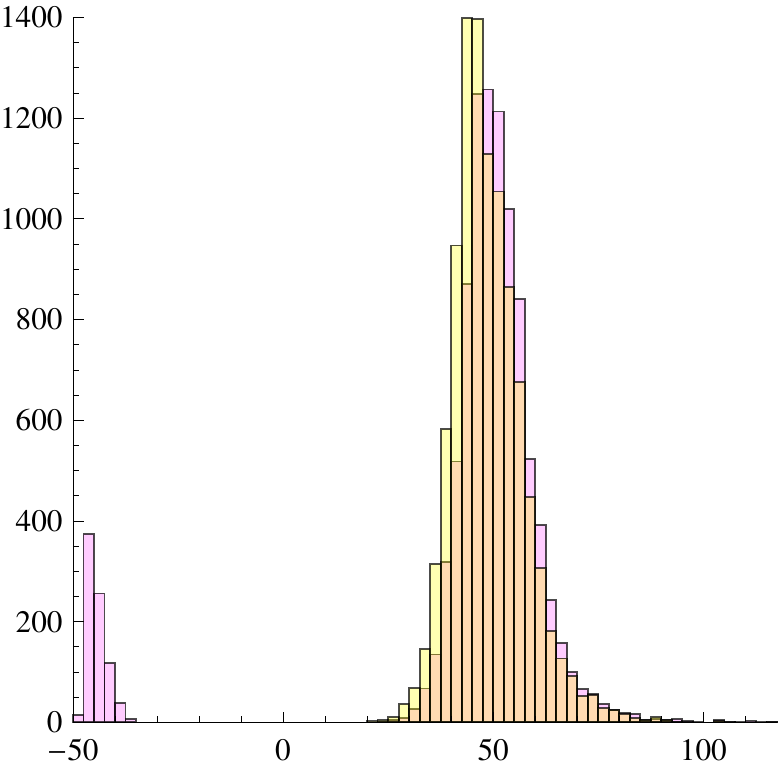}
\includegraphics[width=0.23\textwidth]{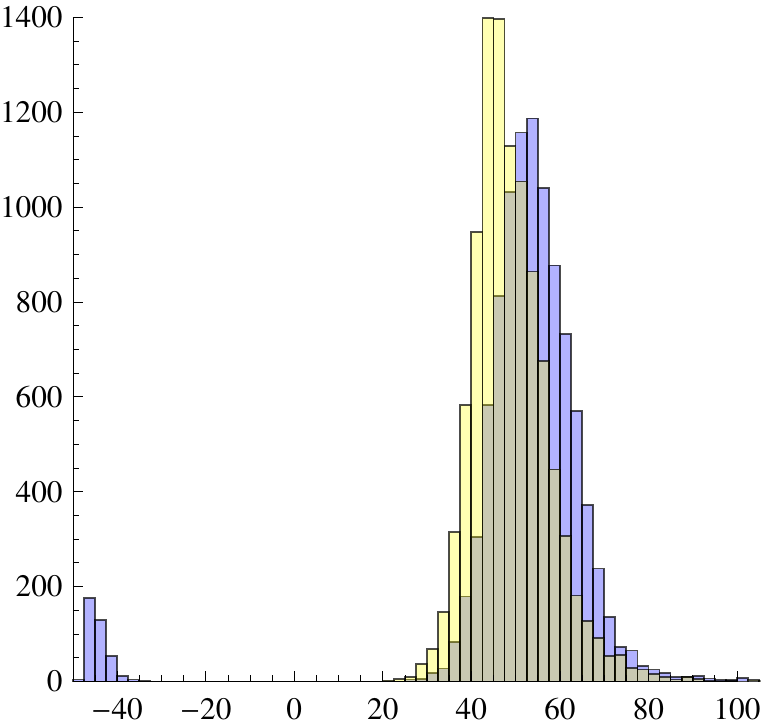}\\
\includegraphics[width=0.23\textwidth]{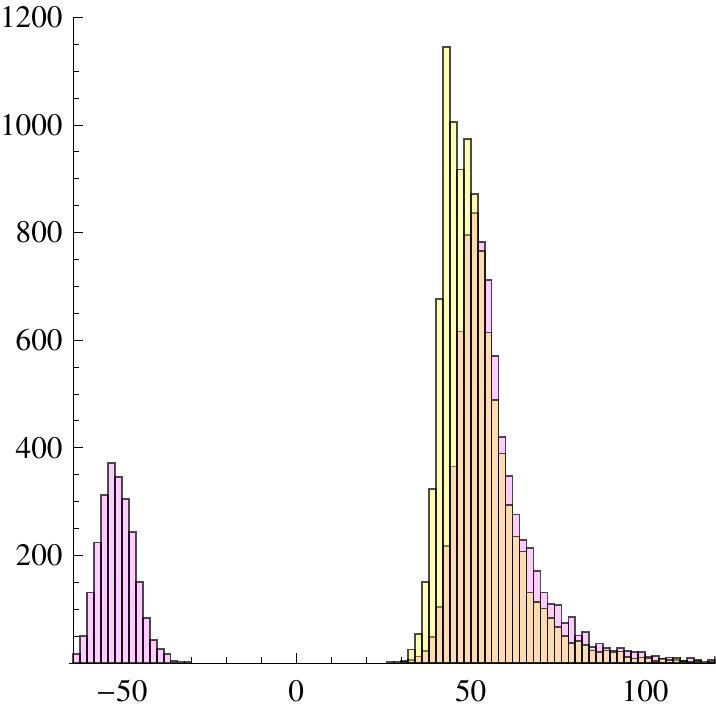}
\includegraphics[width=0.23\textwidth]{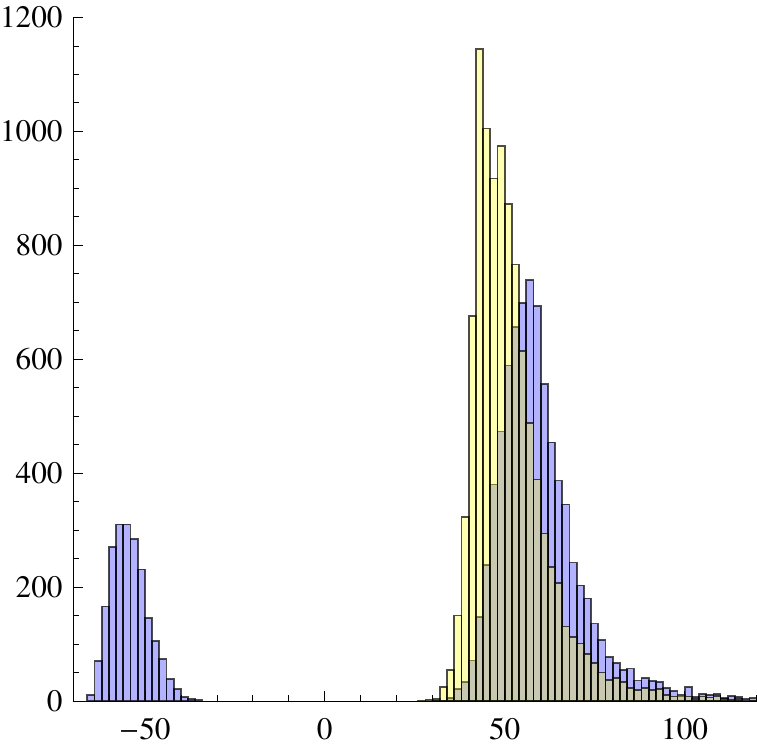}
\caption{Study of the sensitivity to mass of $\tilde\Delta_{2}$, 
Eq.~(\ref{DeltaTilde2}) (top row), and of $\tilde\Delta_{3}$, 
Eq.~(\ref{DeltaTilde3}) (bottom row).
The (yellow) histograms peaking at a smaller values  
 are, in all graphs, for the correct ${\cal M}=M_H=120$ GeV. 
Left: Comparison with an incorrect choice ${\cal M}=140$ GeV.
Right: Comparison with data for $M_H=140$ GeV, analized with the correct 
 ${\cal M}=M_H$.  The horizontal scale is in GeV.
\label{fig:Delta23Sensitivity}}
\end{center}
\end{figure}

\begin{figure}[htbp]
\begin{center}
\includegraphics[width=0.23\textwidth]{D1SigvsWW.pdf}
\includegraphics[width=0.23\textwidth]{D1Sigvstt.pdf}\\
\vspace{.3cm}
\includegraphics[width=0.23\textwidth]{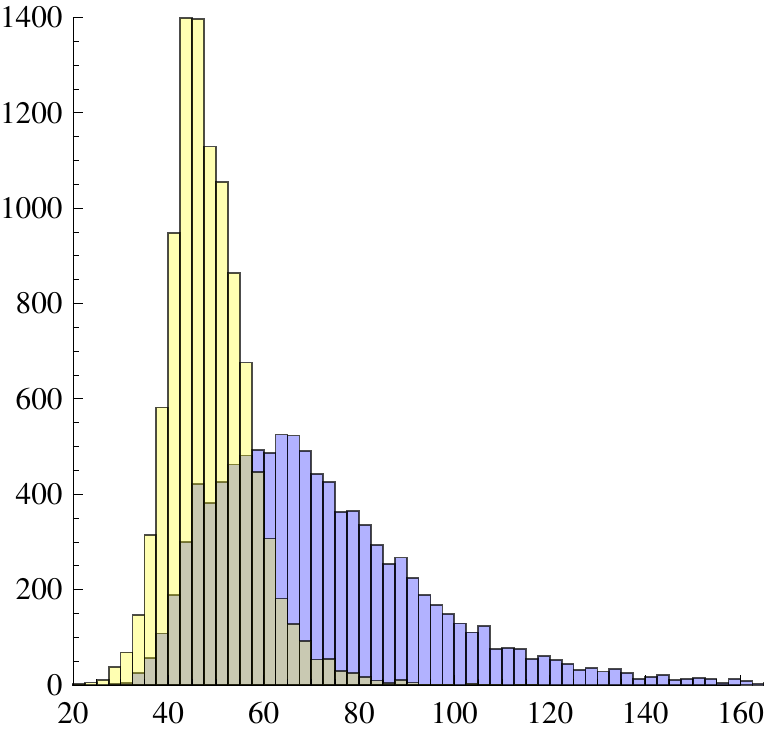}
\includegraphics[width=0.23\textwidth]{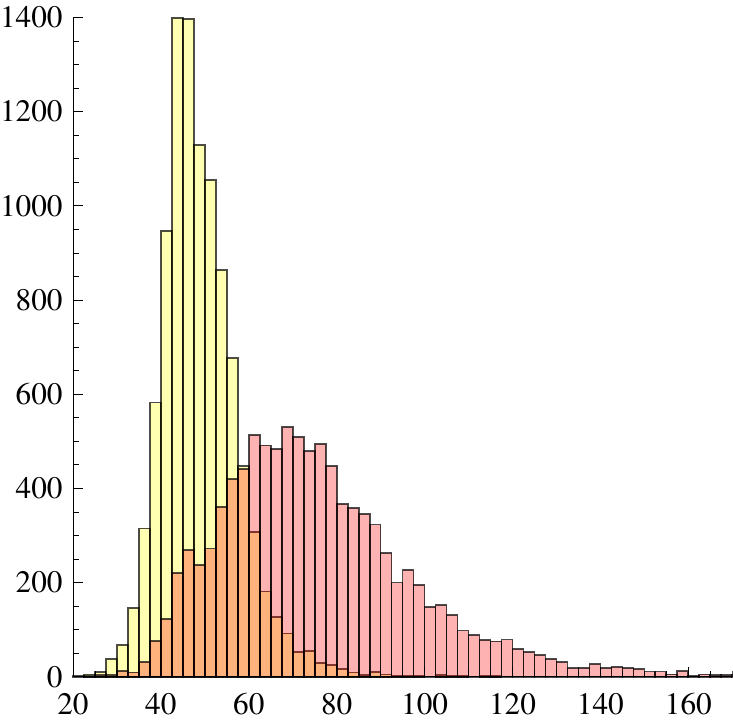}\\
\vspace{.2cm}
\includegraphics[width=0.23\textwidth]{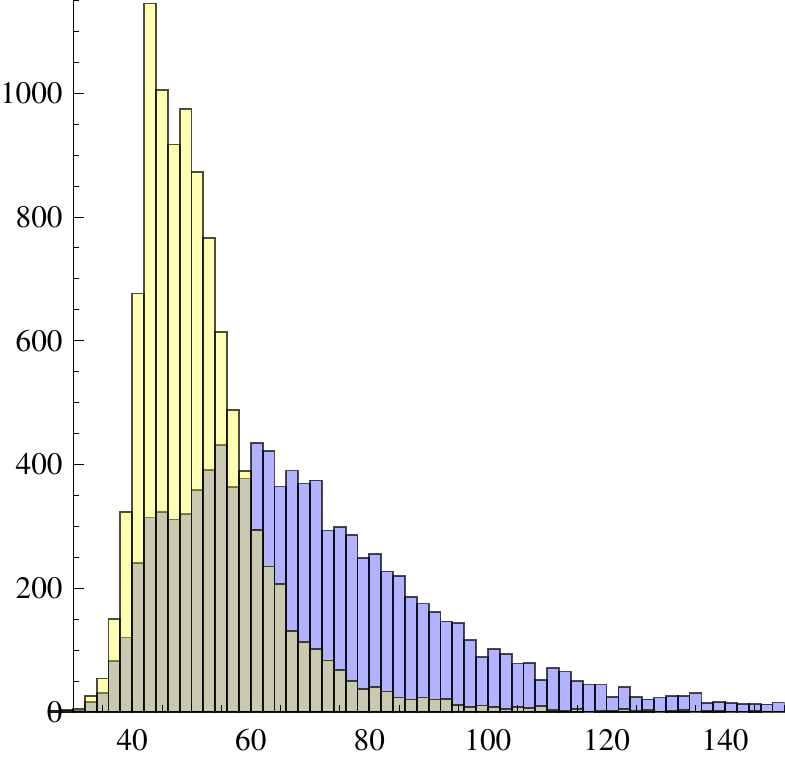}
\includegraphics[width=0.23\textwidth]{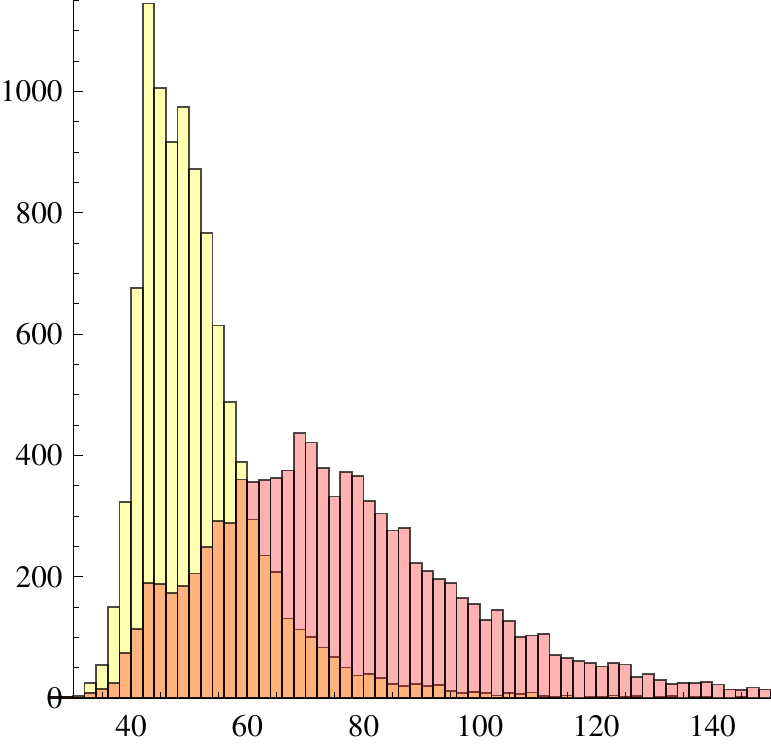}\\
\caption{Signal vs $WW$ (left column) and $t\bar t$ backgrounds
(right column) for the variables $\tilde\Delta_1$ (top row),
$\tilde\Delta_2$ (middle row) and $\tilde\Delta_3$ (bottom row),
respectively defined in Eqs.~(\ref{DeltaTilde1},\ref{DeltaTilde2},\ref{DeltaTilde3}).
In all graphs  ${\cal M}=M_H=120$ GeV. The horizontal scale is in GeV.
  \label{fig:DprimesvsBG}}
\end{center}
\end{figure}

The mass dimension of $C'_2$ is 3 and its degree in $\xi$ is 4.
The corresponding numbers for $C'_0$ are 6 and 8, see Eq.~(\ref{eq:CPrimes}).
Thus, the mass dimension of $\Delta_2\equiv {\rm Res}\{C'_2,C'_0\}$
is $3\times 8+4\times 6=48$. In analogy with Eqs.~(\ref{eq:Sigma123}), it is convenient
to define
\begin{equation}
\tilde\Delta_2 \equiv {\rm Sign}(\Delta_2)\,\vert \Delta_2\vert^{1/48}
\label{DeltaTilde2}
\end{equation}
Results for the distributions of this singularity variable are presented in
Figs.~\ref{fig:Delta23Sensitivity}, \ref{fig:DprimesvsBG} and commented later.

The mass dimension of $C'_3$ is 7 and its degree in $\xi$ is 12.
Recall that thee corresponding numbers for $C'_0$ are 6 and 8.
The mass dimension of $\Delta_3\equiv {\rm Res}\{C'_3,C'_0\}$
is $7\times 8+12\times 6=128$. 
In analogy with Eq.~(\ref{DeltaTilde2}), it is thus convenient
to define
\begin{equation}
\tilde\Delta_3 \equiv {\rm Sign}(\Delta_3)\,\vert \Delta_3\vert^{1/128}
\label{DeltaTilde3}
\end{equation}
Results for the distributions of this singularity variable are presented in
Figs.~\ref{fig:Delta23Sensitivity}, \ref{fig:DprimesvsBG}. 
The message of these figures
is that the variables $\tilde\Delta_ {1,2,3}$ are very good both
at distinguishing
signal and background events and $\tilde\Delta_ {2,3}$
are very good
at pinpointing the mass of a putative signal.

\begin{figure}[htbp]
\begin{center}
\includegraphics[width=0.45\textwidth]{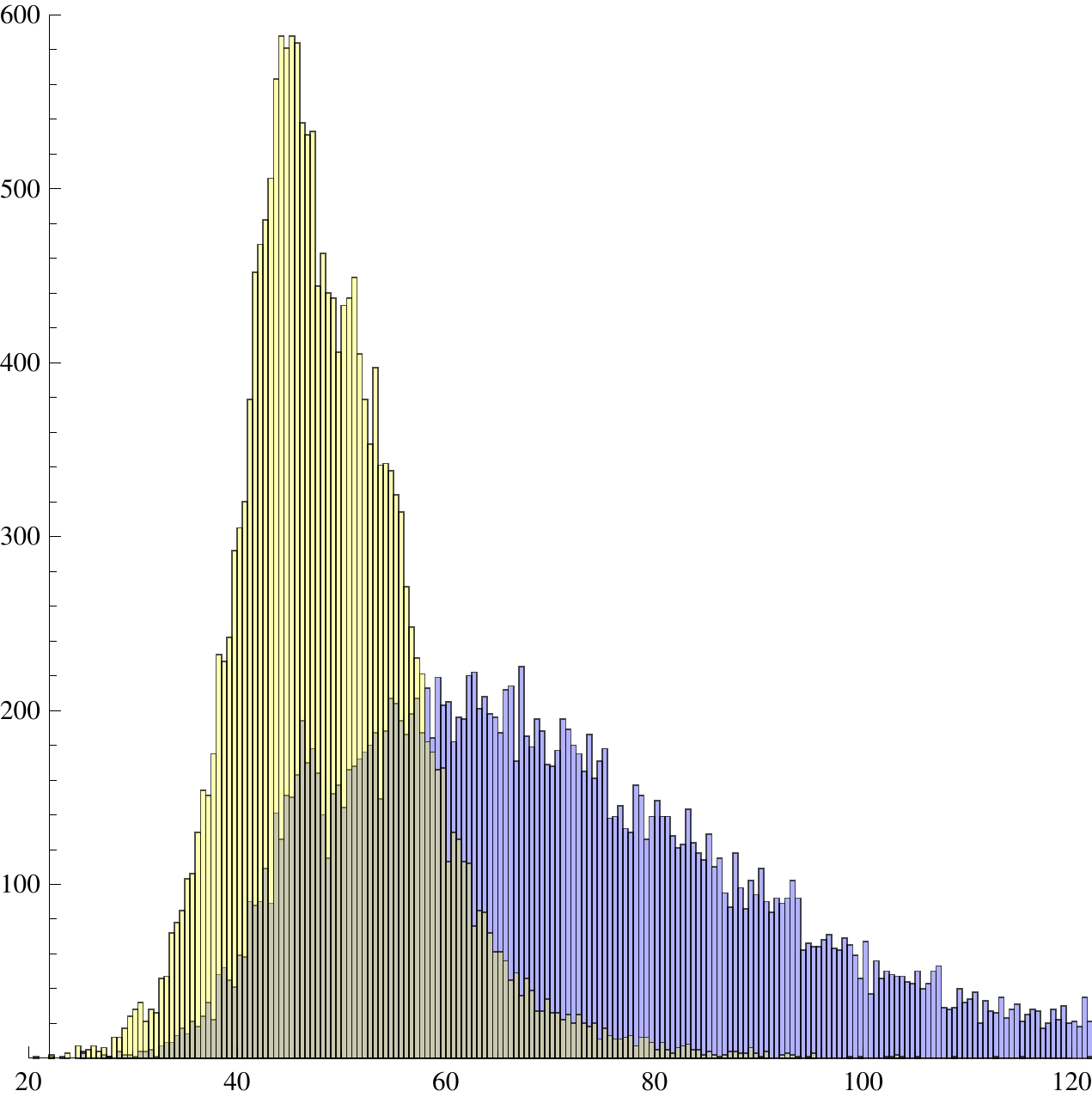}\\
\includegraphics[width=0.45\textwidth]{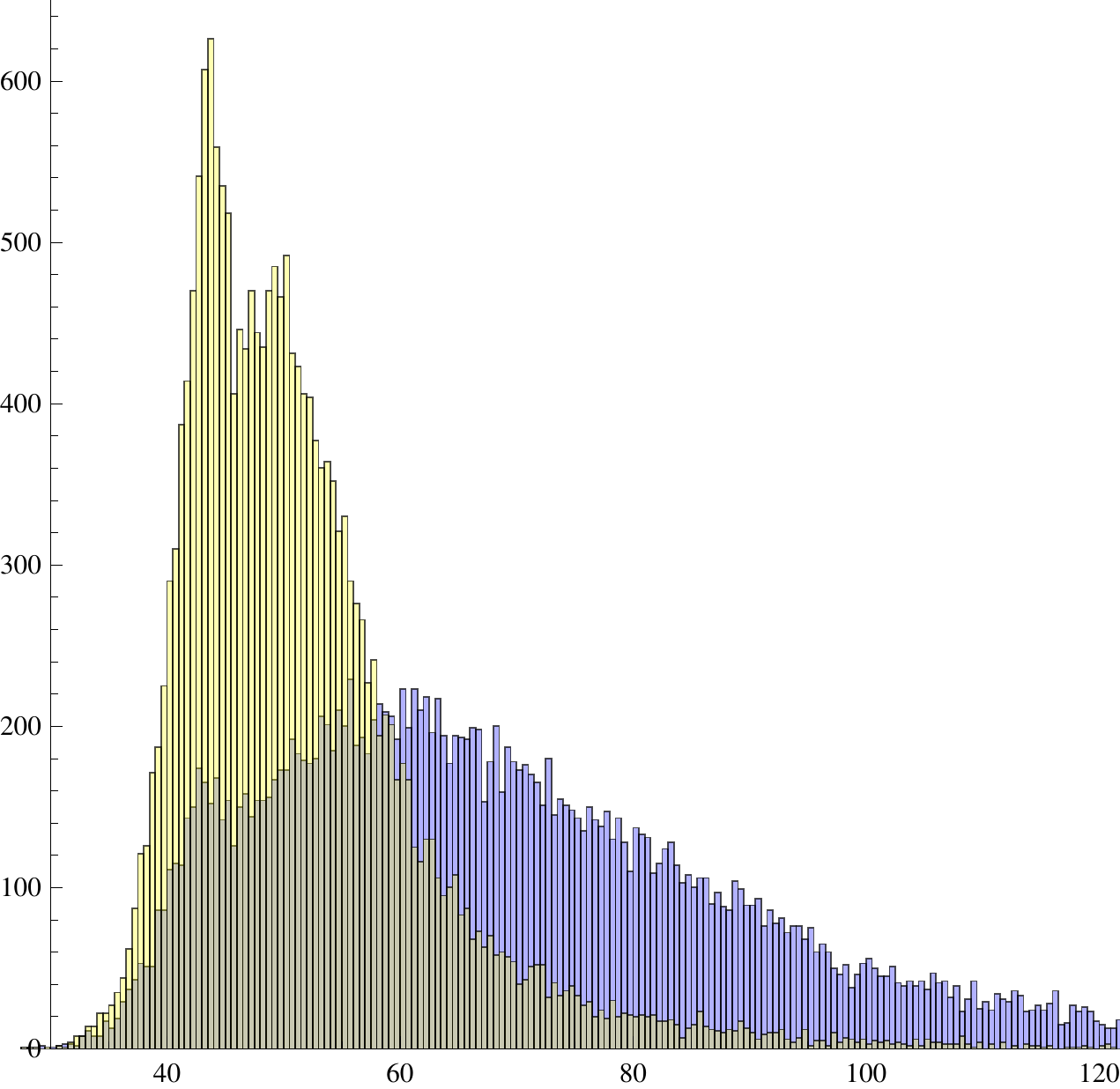}\\
\caption{High resolution histograms for $\tilde\Delta_2$ (upper panel)
and $\tilde\Delta_3$, showing the (narrower, yellow) signal and
the (wider, blue) $WW$ background. The abscissae are in GeV units.
  \label{fig:HRDelta23}}
\end{center}
\end{figure}

An interesting feature emerges when some of the histograms in Fig.~\ref{fig:DprimesvsBG}
are remade with higher statistics and resolution, concerning the singularity
functions $\tilde\Delta_{2,3}$, but not $\tilde\Delta_{1}$. This is shown in 
Fig.~\ref{fig:HRDelta23}. A very clear narrow double peak shape appears
for $\tilde\Delta_{3}$ (lower figure), and a hint  of a similar structure for $\tilde\Delta_{2}$
(upper figure). 

The peaks in Fig.~\ref{fig:HRDelta23}  reflect individual roots in $\cal M$ of 
$\tilde\Delta_{2,3}({\cal M})$. The function $\tilde\Delta_2$ in Eq.~(\ref{eq:Delta2})
is, after elimination of the overall $E^4=({\cal M}/2)^4$ factor, still a polynomial of sixth degree
in ${\cal M}^2$. The $\tilde\Delta_{3}({\cal M})$ resultant in Eq.~(\ref{eq:Res23and0})
has an even more intractable degree in ${\cal M}^2$: twenty-two. Thus, their roots can only be 
extracted numerically event by event, a rather laborious task, which we postpone. 

\subsection{Correlations}

\begin{figure}[htbp]
\begin{center}
\includegraphics[width=0.24\textwidth]{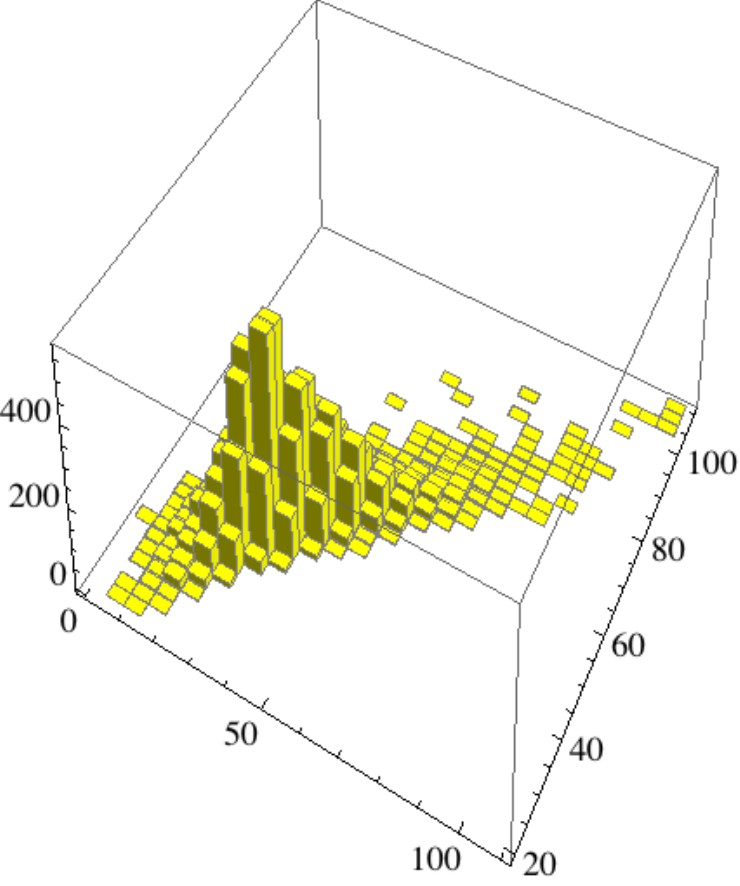}
\includegraphics[width=0.22\textwidth]{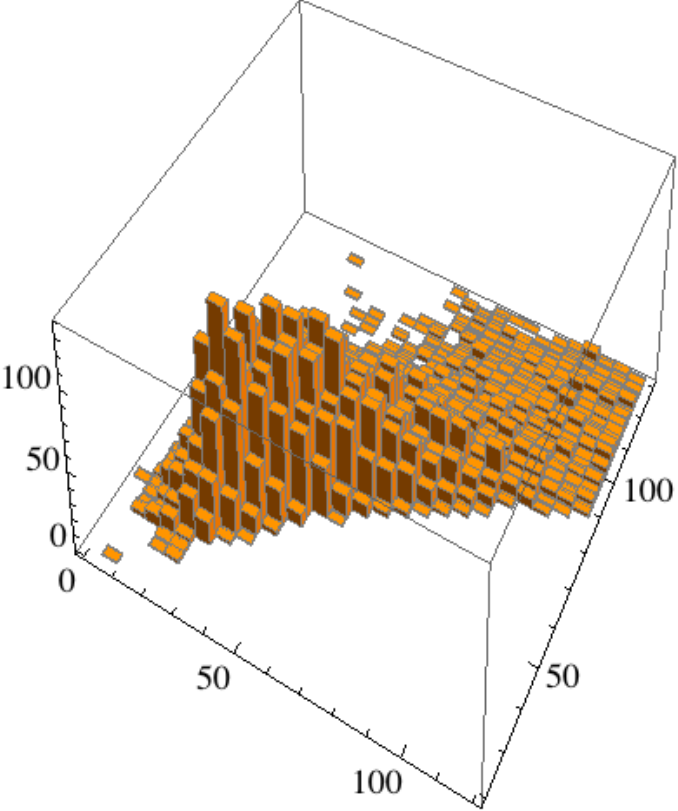}\\
\vspace{.3cm}
\includegraphics[width=0.23\textwidth]{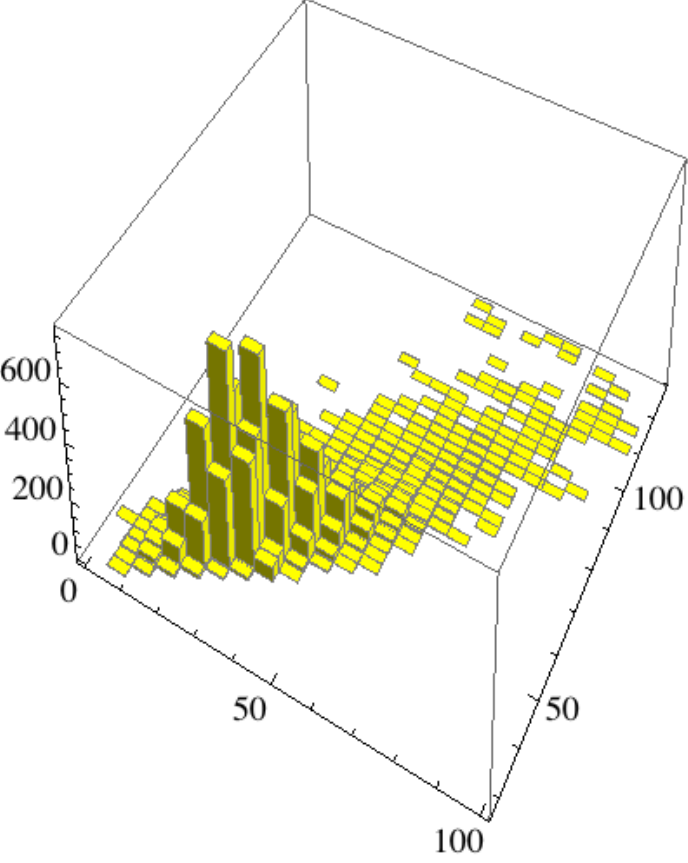}
\includegraphics[width=0.22\textwidth]{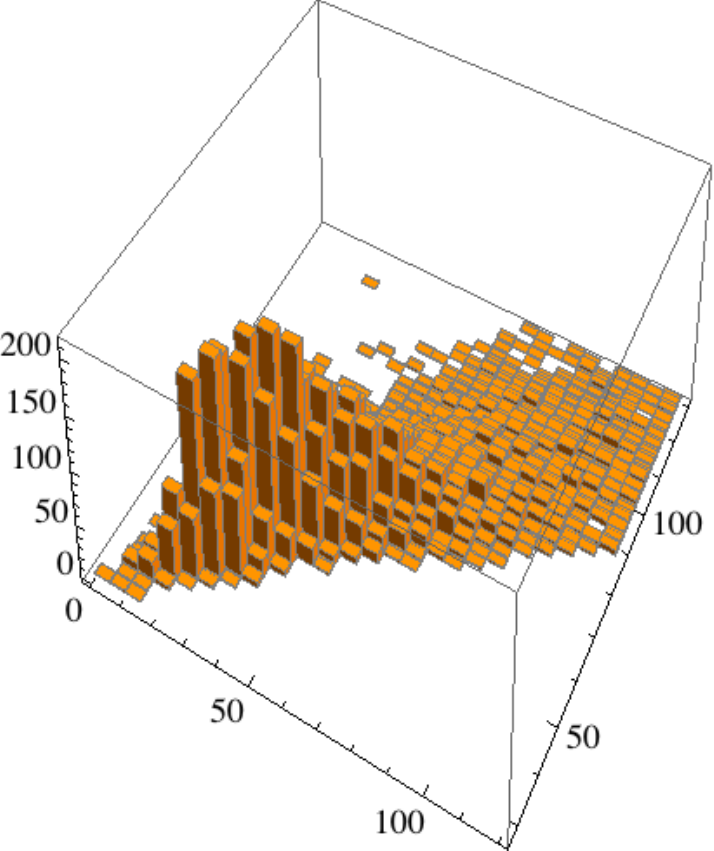}\\
\vspace{.2cm}
\includegraphics[width=0.23\textwidth]{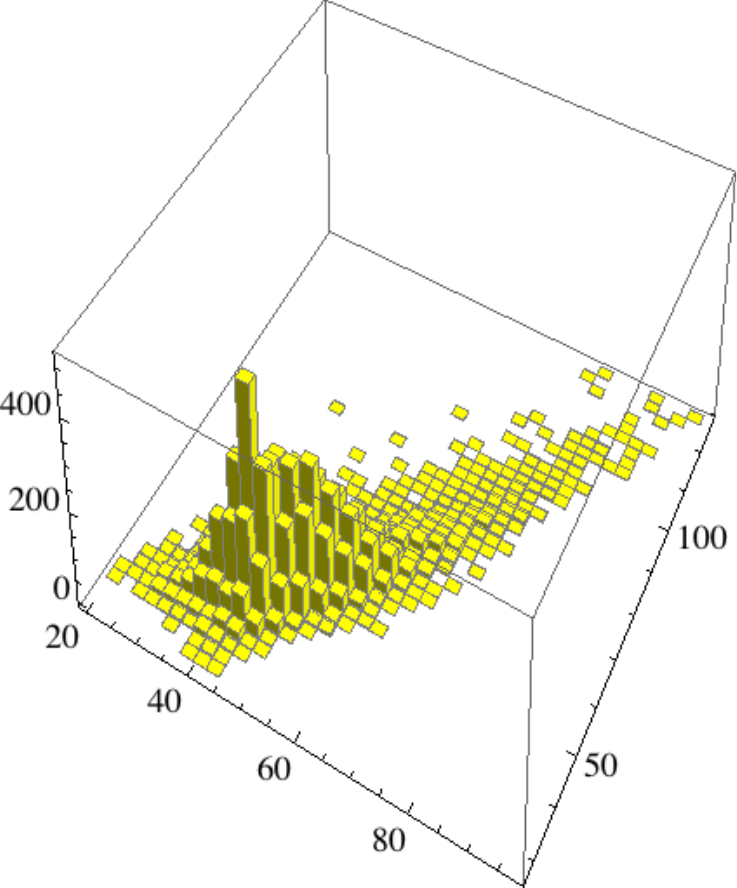}
\includegraphics[width=0.22\textwidth]{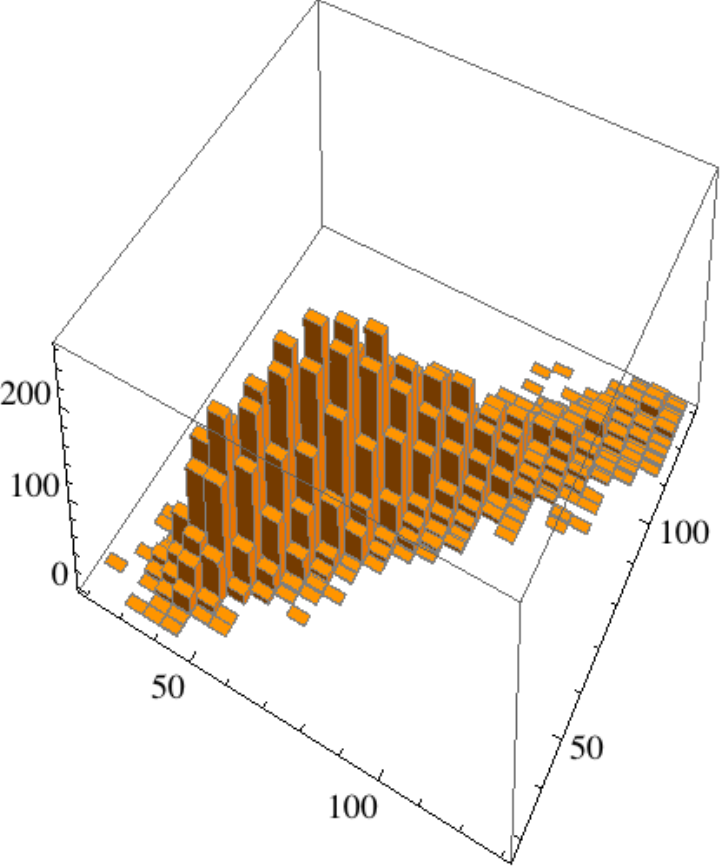}\\
\caption{Histograms of the correlations between the singularity variables $\tilde\Delta_i$ of
Eqs.~(\ref{DeltaTilde1},\ref{DeltaTilde2},\ref{DeltaTilde3}). 
Left: The $M_H=120$ GeV signal. Right: $WW$ background.
Top: $\{\tilde\Delta_1,\tilde\Delta_2\}$. Middle: $\{\tilde\Delta_1,\tilde\Delta_3\}$. 
Bottom: $\{\tilde\Delta_2,\tilde\Delta_3\}$. The $\tilde\Delta_i$ axes are in GeV.
 \label{fig:DeltaiDeltajCorrelations}}
\end{center}
\end{figure}

The longitudinally boost invariant singularity variables $\tilde\Delta_{1,2,3}$
have correlations similar to the ones between $D_{1,2,3}$ that we
showed in Fig.~\ref{fig:D1D2D3Correlations}. They are shown in 
Fig.~\ref{fig:DeltaiDeltajCorrelations}, for an $M_H=120$ GeV signal. 
Once again, there are significant
but not extreme correlations. Moreover the correlated histograms are
quite different for the signal and $WW$ background. This is even more
so for the $t\bar t$ background, which we do not show.

\begin{figure}[htbp]
\begin{center}
\includegraphics[width=0.24\textwidth]{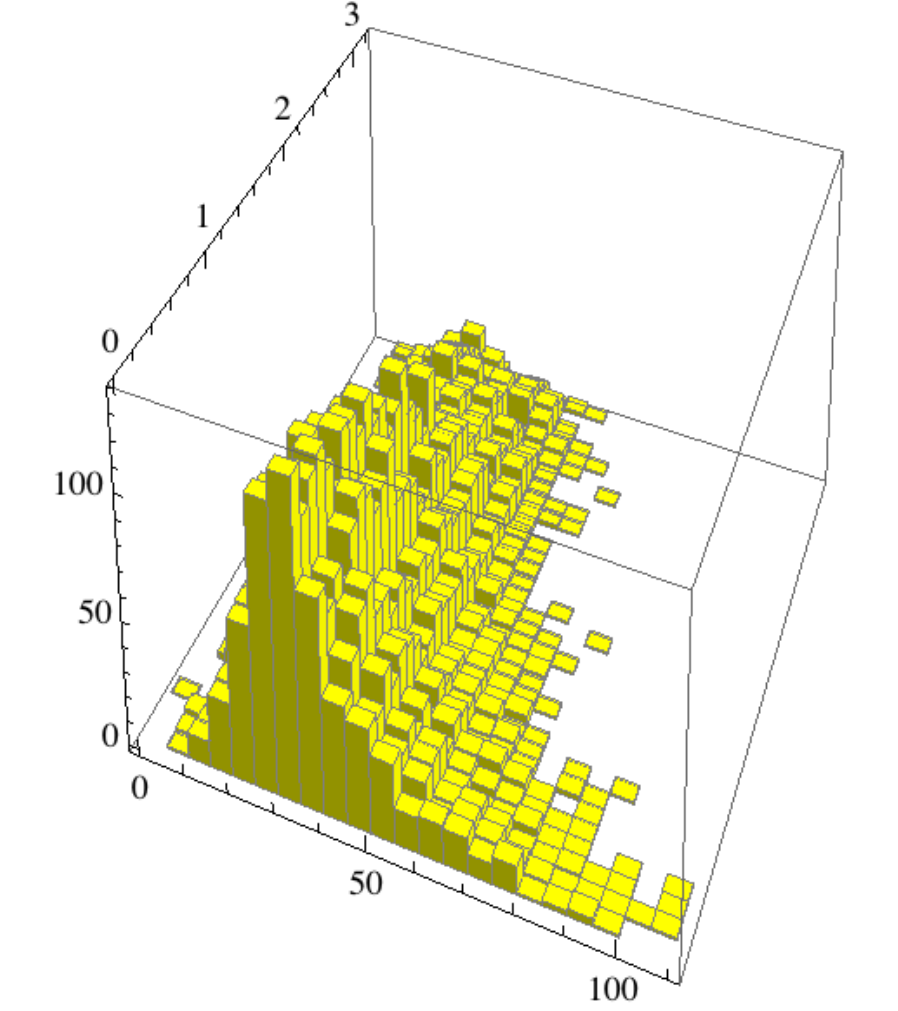}
\includegraphics[width=0.22\textwidth]{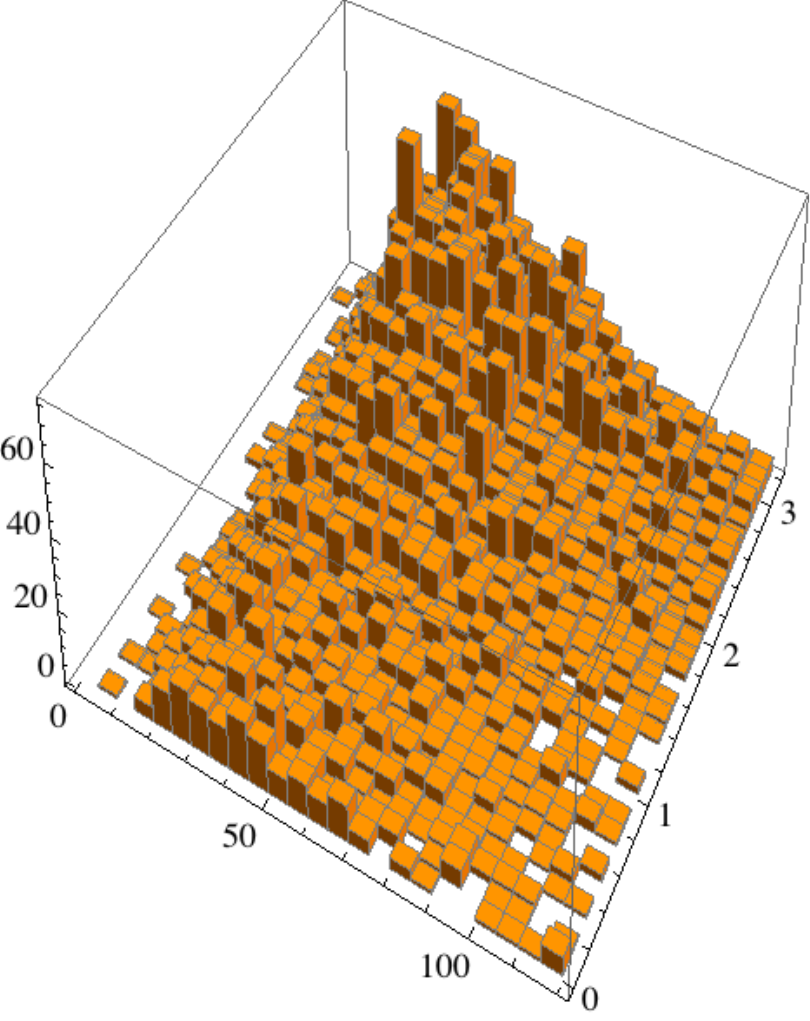}\\
\vspace{.3cm}
\includegraphics[width=0.23\textwidth]{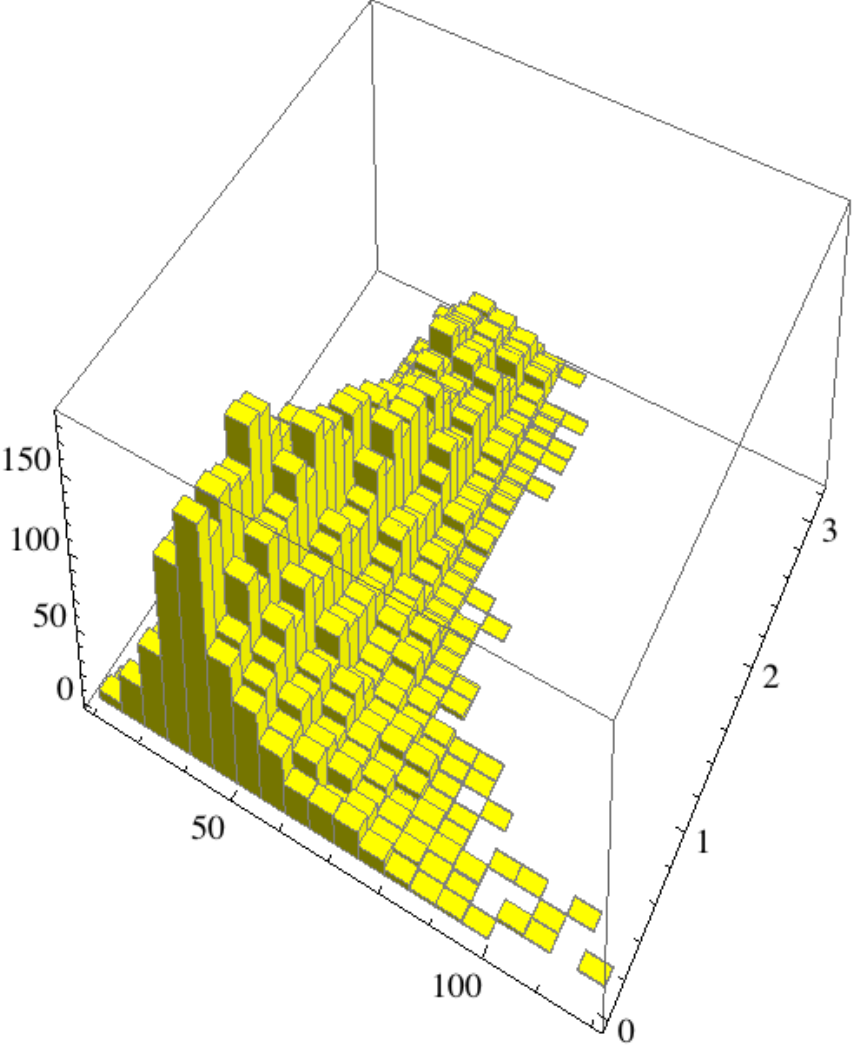}
\includegraphics[width=0.22\textwidth]{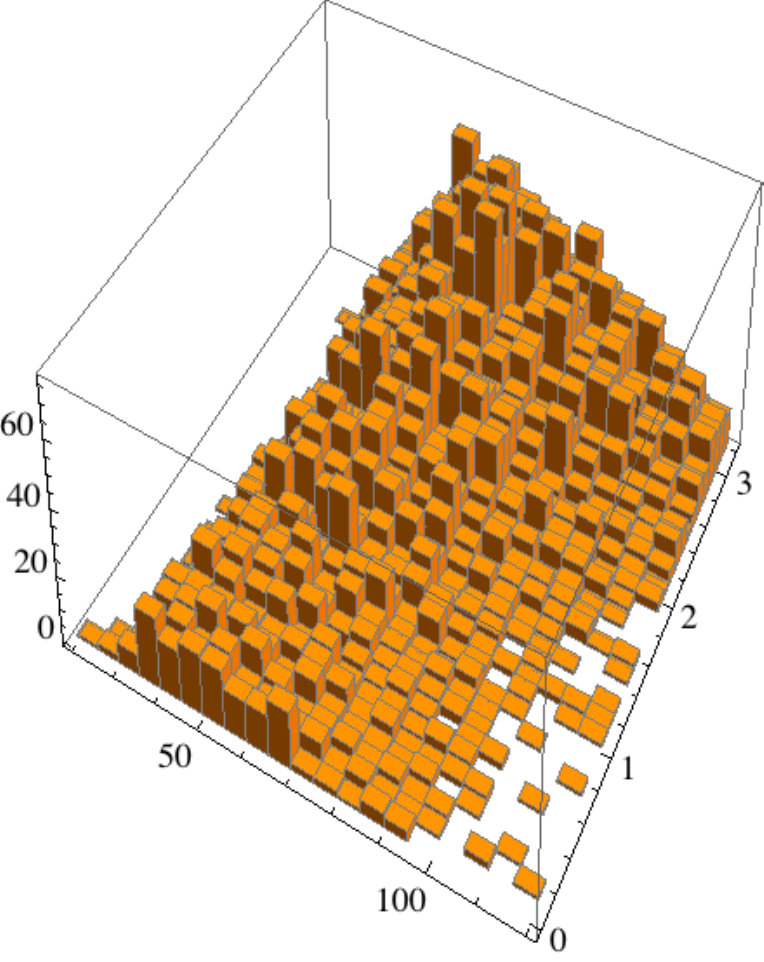}\\
\vspace{.2cm}
\includegraphics[width=0.23\textwidth]{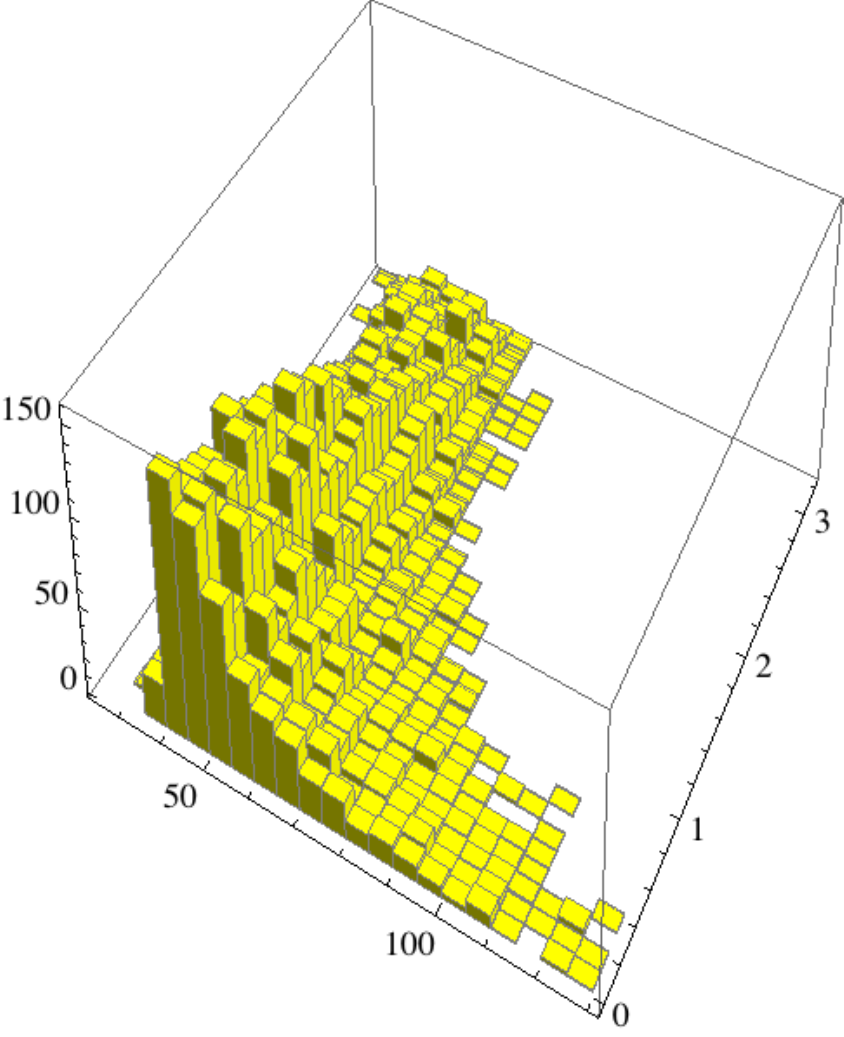}
\includegraphics[width=0.22\textwidth]{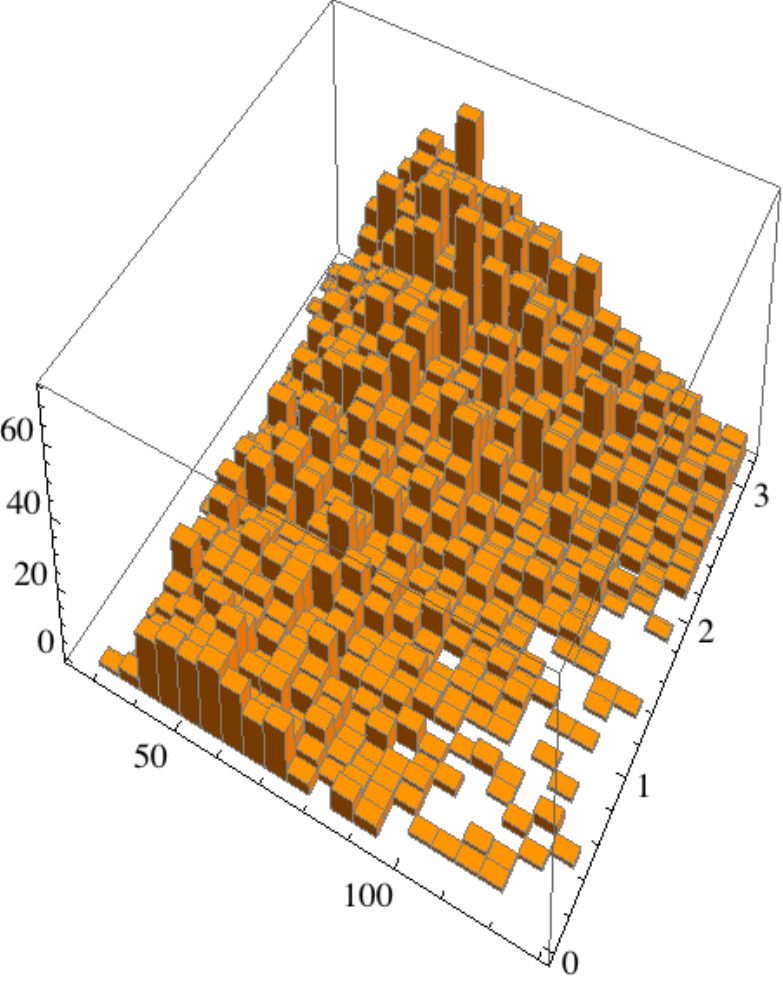}\\
\caption{Histograms of the correlations between the singularity variables $\tilde\Delta_i$ of
Eqs.~(\ref{DeltaTilde1},\ref{DeltaTilde2},\ref{DeltaTilde3}) and $\Delta\varphi$. 
Left: The $M_H=120$ GeV signal. Right: $WW$ background.
Top: $\{\tilde\Delta_1,\Delta\varphi\}$. Middle: $\{\tilde\Delta_2,\Delta\varphi\}$. 
Bottom: $\{\tilde\Delta_3,\Delta\varphi\}$. The $\tilde\Delta_i$ axes are in GeV.
 \label{fig:DiDphiCorrelations}}
\end{center}
\end{figure}

In Fig.~\ref{fig:DiDphiCorrelations}, we illustrate the correlations
between $\tilde\Delta_{1,2,3}$ and $\Delta\varphi$ for the signal and the
$WW$ background, to which the $t\bar t$ background again is in this sense
similar. For the relatively light $M_H=120$ GeV signal shown in the
figure, as expected,  signal and background densely populate very
different regions of phase space.

\section{Summary of results}
\label{Sec:Summary}

With an eye on potential practical usefulness, let us call ``good" the singularity variables that
do an efficient job at focusing on the correct value of $M_H$ and, more so, the ones 
that produce the most significant difference in the shape of their distributions for
a potential signal and the $WW$ and $t\bar t$ backgrounds.

Some of the singularity variables derived in the CM approximation of a motionless Higgs boson
are unexpectedly good. This is the case for $\Sigma_0$, associated with the CM condition of 
coplanarity, defined in Eq.~(\ref{eq:Sigma4}) and illustrated in 
Figs.~\ref{fig:C4Determinant} and \ref{fig:C4DeterminantCompare}.
Its two roots, $M_\pm$, defined in the same equation, are of the simpler kind that does
not involve a trial mass $\cal M$. Their correlations, shown in 
Figs.~\ref{fig:AbsPhaseCorr2},\ref{fig:RootsofC4} in two different ways, are moderate,
both for the signal and the background. The quantities $M_\pm$ and their product
$\Sigma_0^2({\cal M})\equiv ({\cal M}-M_+)({\cal M}-M_-)$ are good.

Still in the CM, we see in Fig.~\ref{fig:C2&C1} that the na\"ive variable $M_1$ (or $\Sigma_1$),
defined in Eqs.~(\ref{eq:C2},\ref{eq:Sigma123}), is not good. The variable 
$\Sigma_2\, (\tilde\Sigma_3)$, 
also defined in Eq.~(\ref{eq:Sigma123}) is good (not so good), as one can conclude from
Figs.~\ref{fig:C2&C1}, \ref{fig:C3}.
These limitations are lifted as we construct from these variables the quantities
$D_i$ defined in Eq.~(\ref{eq:Ds}), which are histogrammed in Fig.~\ref{fig:ToZeros}:
they are reasonably good at telling signal from backgrounds. Their
correlation plots, shown in Fig.~\ref{fig:D1D2D3Correlations}, are quite disimilar for
the signal and the $WW$ irreducible background.

The longitudinally boost-invariant analogs of $D_i$, $i=1$ to 3 are  the 
singularity variables $\Delta_1$ of Eq.~(\ref{eq:Delta1}) and $\Delta_{2,3}$ of
Eq.~(\ref{eq:Res23and0}). We have redefined them to have unit mass dimensionality
in Eqs.~(\ref{DeltaTilde1},\ref{DeltaTilde2},\ref{DeltaTilde3}). 
Only for $\Delta_{1,2}$ we have analytical expressions, which are factorizable.
The factors of $\Delta_{1}$,
 and the complete variable are very good, as illustrated in 
Figs.~\ref{fig:Dprime1},\ref{fig:Dprime1Corr},\ref{fig:SigmaandcalM}.
So are the variables $\tilde\Delta_{2,3}$, as shown in 
Figs.~\ref{fig:Delta23Sensitivity}, \ref{fig:DprimesvsBG}. We see in 
Fig.~\ref{fig:DeltaiDeltajCorrelations}
that the three $\tilde\Delta_i$ are quite correlated, but the correlation plots of signal
and background are populated in a significantly different way. 
A comparison of Fig.~\ref{fig:DprimesvsBG} with the corresponding result for the CM variables $D_i$,
Fig.~\ref{fig:ToZeros}, shows, once more, that the $\tilde\Delta_i$ 
are demonstrably better.

The confrontation of the results for the boost-invariant variables, $\Delta_i$,
and their siblings, $\D_i$, obtained in the approximation in which the boson
is at rest is very gratifying. The signal peaks are significantly narrower
and the correlations weaker for the $\Delta$s than for the $D$s. 

In the sense of their correlations with the function $\Delta\varphi$, shown in 
Fig.~\ref{fig:DiDphiCorrelations}, the singularity variables $\tilde\Delta_i$ 
are optimal tools to separate signal from backgrounds.

\section{Conclusions}
\label{sec:conclusions}

Recall that, as discussed in the Introduction, in the case of the CM singularity
variables one can construct up to five independent combinations of the relevant
observables. The best choice is the set $\{M_+,M_-,D_1,D_2,D_3\}$, whose ingredients
are defined in Eqs.~(\ref{eq:Sigma4},\ref{eq:Ds}).
The main appeal of these CM variables is that they are simple and
explicit functions of the relevant observables. Their main drawback is that
they are not as good as the boost-invariant variables, to be revisited next. 

The only imperfection of the boost-invariant singularity variables 
is that for one of them, $\Delta_{3}$, we are unable to derive its explicit
analytical expression. For $\Delta_{2}$ the analytical expression,
Eq.~(\ref{eq:Delta2}), is so complex that we have opted to compute it
 event by event as a numerical resultant, as we are forced to do in the
case of $\Delta_3$.  For the computer, this is fast and simplest.

The practical virtues of the variables $\Delta_i$ --in being able to pinpoint the actual
value of $M_H$ and to tell apart signal from backgrounds--
amply overcome their quoted single limitation. The theoretical toil required
to go beyond the Higgs-at-rest approximation pays. 

Recall that in terms of the four
boost-invariant observables one can construct up to four useful combinations.
The next-to-best choice is $\{M^{1/4},\Sigma^{1/20},\tilde\Delta_2,\tilde\Delta_3\}$,
where $M$ and $\Sigma$ are the factors building up $\Delta_1$, see Eq.~(\ref{eq:Delta1}).
The best choice is $\{\tilde\Delta_1,\tilde\Delta_2,\tilde\Delta_3,\Delta\varphi\}$,
combining our kinematical singularity variables with the good old ``dynamical"
(spin-dependent) angle, $\Delta\varphi$, of the charged leptons in the transverse plane.

The bell shapes of the signal histograms in Fig.~\ref{fig:DprimesvsBG}
are very satisfactory, even if obtained with theoretical expressions in the $\vec p_T\!=\! 0$
approximation for the produced hadrons. We have checked, by generating and analizing
events with $\vec p_T\!=\! 0$, that the improvement brought by theoretical variables 
that avoid the $\vec p_T\!=\! 0$ approximation is unlikely to be very significant.

We have only studied variables and their pair-wise correlations. We have not attempted
to quantify the absolute values of signals and backgrounds --as opposed to just the shape 
of their distributions. Thus, we are far from being able to show potential
``significance" results in terms of a full multi-dimensional analysis of all variables and 
their correlations. Yet our results for $\tilde\Delta_{1,2,3}$ in 
Fig.~\ref{fig:DprimesvsBG} are competitive in ``goodness"
with the $\Delta\varphi$ diagnosis recalled in Fig.~\ref{fig:DeltaPhi}. That was one of 
our goals.

For an experimentalist eager to test the tantalizing hints that $M_H=126$ or 124 GeV
\cite{LHC}, it should not be too streneous to prepare the relevant  
one- or multi-dimensional singularity-variable templates 
for the relatively copious channel $H\to WW\to$ leptons.

Our main aim was the theoretical derivation of a complete set of phase-space
singularity conditions and variables for the process $H\to W^+W^-$, $W^\pm\to \ell^\pm\nu$.
We have seen it is a rather laborious task. The origin of its difficulty is
many-fold. First, because of the elusiveness of neutrinos, the kinematical
constraints of  Eqs.~(\ref{alleqs}) are incomplete. Second, several of these equations are 
non-linear. Finally and most severely, the 7-th equation, the one reflecting that the
invariant mass of the four leptons is $M_H$, inextricably links the leptons resulting
from the decay of one $W$ to those from the other, very significantly complicating
the ensuing algebra.

In most processes relevant to a hadron-collider search 
for new physics involving unobservable particles, the
initial step is a non-resonant production of a pair of novel particles.
This means one cannot assume a fixed invariant mass for 
the pair and (approximately) boost each event to the pair's rest system. 
But the last difficultly mentioned in the previous paragraph is absent.
That is why, even for $\vec p_T\neq 0$ --and a surfeit of unknown masses--
the pertinent singularity variables are relatively simple, and analytical
\cite{AAbis}.

\section*{Appendix:  SO(1,1) invariance  of the resultants}

Let $C_i, C_0$ be the CM functions defined in Eqs.~(\ref{eq:Cs},\ref{eq:det}).
Let $C'_i, C'_0$ be the functions in Eq.~(\ref {eq:CPrimes}), boosted by
$L_\xi\in {\rm SO}(1,1)$, whose action on a longitudinal vector $v=\{v_0,v_3\}$ is
 $v \mapsto L_\xi(v)$. Very explicitly, with $\xi$ the boost parameter in Eq.~(\ref{eq:xi}),
 \begin{equation}
\begin{split}
& v_0 \mapsto (L_\xi v)_0 := \half(\xi+\xi^{-1}) v_0 + \half(\xi-\xi^{-1}) v_3,
\\
& v_3 \mapsto (L_\xi v)_3 := \half(\xi-\xi^{-1}) v_0 +
\half(\xi+\xi^{-1}) v_3,
\end{split}
\end{equation}
and
\begin{equation}
\begin{split}
& C'_i(l,k,\xi) = \xi^{r_i}C_i(L_\xi^{-1} l,L_\xi^{-1}k),
\\
&  C'_0(l,k,\xi) = \xi^{r_0}C_0(L_\xi^{-1}l,L_\xi^{-1}k),
\end{split}
\end{equation}
where $r_i,r_0$ are the minimal entire numbers required for $C'_i,C'_0$ 
to be polynomials in $\xi$. It is easy to check that 
\begin{equation}
\begin{split}
& r_i =  g_i/2,\;\;\; r_0 = g_0/2,
\end{split}
\label{eq:rr}
\end{equation}
with $g_i$ and $g_0$ the degrees in $\xi$ of  $C'_i$ and $C'_0$.

Let $R(l,k)$ be the resultant in $\xi$ of $C'_i(l,k,\xi),C'_0(l,k,\xi)$:
\begin{equation}
R(l,k) := {\rm Res}(C'_i(l,k,\xi),C'_0(l,k,\xi),\xi).
\end{equation}

We want to prove that
\begin{equation}
R(l,k)=R(L_\eta l,L_\eta k)
\end{equation}
for all $L_\eta\in {\rm SO}(1,1)$, that is, the resultant is invariant under longitudinal boosts.

Indeed, using the aforementioned definitions, we have:
\begin{equation}
\begin{split}
& R(L_\eta l,L_\eta k) = {}
\\
& {\rm Res}(C'_i(L_\eta l,L_\eta k,\xi),C'_0(L_\eta l,L_\eta k,\xi),\xi)= {}
\\
& {\rm Res}(\xi^{r_i}C_i(L_\xi^{-1}L_\eta l,L_\xi^{-1}L_\eta k),
\xi^{r_0}C_0(L_\xi^{-1}L_\eta l,L_\xi^{-1}L_\eta k),\xi)= {}
\\
& {\rm Res}(\xi^{r_i}C_i(L_{\xi/\eta}^{-1} l,L_{\xi/\eta}^{-1} k),
\xi^{r_0}C_0(L_{\xi/\eta}^{-1}l,L_{\xi/\eta}^{-1}k),\xi)= {}
\\
& {\rm Res}(\eta^{r_i}C'_i(l,k,\xi/\eta),
\eta^{r_0}C'_0(l,k,\xi/\eta),\xi).
\label{eq:proof1}
\end{split}
\end{equation}

Taking into account that, if $p_1(x),p_2(x)$ are two arbitrary polynomials
of degrees $g_1,g_2$, respectively,
\begin{equation}
\begin{split}
&{\rm Res}(a_1 p_1(x/b),a_2 p_2(x/b),x) ={}
\\
& a_1^{g_2}a_2^{g_1}b^{-g_1g_2}{\rm Res}(p_1(x),p_2(x),x),
\end{split}
\label{eq:identidad}
\end{equation}
it immediately follows, with use of  Eqs.~(\ref{eq:rr},\ref{eq:identidad}), that:
\begin{equation}
\begin{split}
& {\rm Res}(\eta^{r_i}C'_i(l,k,\xi/\eta),
\eta^{r_0}C'_0(l,k,\xi/\eta),\xi)= {}
\\
& {\rm Res}(C'_i(l,k,\xi),
C'_0(l,k,\xi),\xi)= R(l,k).
\end{split}
\label{eq:proof2}
\end{equation}
The desired ``Q.E.D." is simply reached by putting together Eqs.~(\ref{eq:proof1},\ref{eq:proof2}).
It is also simple and gratifying, in the case of the $\Delta_{3}$ resultant
that we were unable to derive explicitly, to check its boost invariance numerically.

\section*{Acknowledgments}

We are indebted to Rakhi Mahbubani, Emanuele Di Marco, Maurizio Pierini and Chris Rogan for discussions and suggestions.

\end{document}